\documentclass[aps,prd,nofootinbib,floatfix]{revtex4}
\usepackage{amsmath}
\usepackage{amssymb}
\usepackage{graphicx}
\usepackage{hyperref}
\usepackage{color}
\usepackage{float}
\usepackage{slashed}
\usepackage{float}
\usepackage{bbold}
\usepackage[caption=false]{subfig}
\usepackage{multirow}
\definecolor{darkblue}{rgb}{0.0, 0.0, 0.62}
\definecolor{darkred}{rgb}{0.55, 0.0, 0.0}
\definecolor{violetryb}{rgb}{0.55, 0.0, 0.50}
\definecolor{royalpurple}{rgb}{0.47, 0.32, 0.66}
\definecolor{regalia}{rgb}{0.32, 0.18, 0.5}
\definecolor{purpleheart}{rgb}{0.41, 0.21, 0.61}
\definecolor{plum}{rgb}{0.56, 0.27, 0.52}
\definecolor{deeppurple}{rgb}{0.41, 0.16, 0.38}
\hypersetup{colorlinks=true,citecolor=darkred,linkcolor=darkblue}
\hypersetup{breaklinks=true,citecolor=darkblue,linkcolor=darkblue,linktoc=page,urlcolor=violetryb}
\newcommand{\g}{\gamma}

\newcommand{\A}{\mathcal{A}}

\newcommand{\Wt}{{\tilde{W}}}
\newcommand{\ETmiss}{\slashed{E}_T}

\DeclareMathOperator{\Br}{Br}

\newcommand{\tr}{{\rm Tr}}
\renewcommand{\to}{\rightarrow}

\newcommand{\de}{\partial}

\renewcommand{\to}{\rightarrow}

\newcommand{\beq}{\begin{equation}}
\newcommand{\eeq}{\end{equation}}

\def\pt{p_\textrm{T}}

\begin{document}

\title{Probing the Interactions of Axion-Like Particles with Electroweak Bosons and \\ the Higgs Boson in the  High Energy Regime at LHC}

\author{Tisa Biswas} 
\email{tibphy\_rs@caluniv.ac.in}

\affiliation{\\~\\Department of Physics, University of Calcutta,  
	92 Acharya Prafulla Chandra Road, \\ Kolkata 700009, India}

\begin{abstract}
	We study the interactions of axion-like particles (ALPs) with the Standard Model particles, aiming to probe their phenomenology via non-resonant searches at the LHC. These interactions are mediated by higher dimensional effective  operators within two possible frameworks of linearly and non-linearly realised electroweak symmetry breaking. We consider the ALPs to be light enough to be produced on-shell and exploit their derivative couplings with the SM Higgs boson and the gauge bosons. We will use the high momentum transfer processes, namely $hZ, Z\gamma, WW$ and $WW\gamma$ production from $pp$ collisions. We derive upper limits on the gauge-invariant interactions of ALPs with the electroweak bosons and/or Higgs boson that contribute to these processes, from the re-interpretation of the latest Run 2 available LHC data. The constraints we obtain are strong for ALP masses below 100 GeV. These allowed effective interactions in the ALP parameter space yield better significance at HL-LHC and thus, offer promising avenues for subsequent studies. Furthermore, we augment our cut-based analysis with gradient-boosted decision trees, which improve the statistical significance distinctly across these interaction channels. We briefly compare the results with the complementary probe of these couplings via direct production of ALPs in association with the Higgs boson or a vector boson.
\end{abstract}

\vskip 1.0cm

\maketitle

\section{Introduction}
\label{introduction}
Originally motivated by the efforts to solve the strong CP problem~\cite{Peccei:1977hh,Peccei:1977ur,Weinberg:1977ma,Hook:2014cda,Wilczek:1977pj}, pseudo-Nambu-Goldstone bosons (pNGBs) generically arise in a variety of new physics (NP) scenarios. Their implications are many, including the dynamic generation of small neutrino masses (Majorons)~\cite{Chikashige:1980ui},  attempting to solve the flavor problem (Flavons)~\cite{Froggatt:1978nt}, contributing to composite Higgs models and extra-dimensional theories~\cite{Izawa:2002qk}. The pNGBs also play a role in addressing the long standing anomaly of  muon magnetic moment~\cite{Chang:2000ii}, the hierarchy problem~\cite{Graham:2015cka} and electroweak baryogenesis~\cite{Jeong:2018jqe}.  In addition, they can serve as potential dark matter candidates or provide a portal connecting the Standard Model (SM) particles to dark sector~\cite{Abbott:1982af,Dine:1982ah,Preskill:1982cy}. Typically, pNGBs exhibit symmetry under a continuous or, in some cases, a discrete shift of the field. These pNGBs, which enjoy a variety of origins, interactions and masses, are often grouped together in a much broader class of the axion-like particles (ALPs).

Owing to diverse origins, ALPs connect different sectors in high-energy physics. Studies aiming to detect ALPs in a range of masses and interactions have guided the current and future direction of experiments for {\em{beyond}} the Standard Model (BSM) physics, as discussed in recent reviews~\cite{Irastorza:2018dyq,Graham:2015ouw,Bruggisser:2023npd}.  ALPs can manifest through a variety of traces in experiments running at different energy scales.  At the LHC, ALP interactions are probed through signatures including new resonances or missing energy~\cite{Georgi:1986df,Mimasu:2014nea,Brivio:2017ije,Bauer:2017ris}, or interactions with top quarks~\cite{Esser:2023fdo} and via Higgs decays~\cite{Bauer:2016zfj,Biekotter:2022ovp}. Meson decay experiments typically provide the conventional constraints on ALP-QCD couplings. In the context of flavor experiments, ALPs with masses below a few GeV can be resonantly produced via meson~\cite{Bjorkeroth:2018dzu,Dobrich:2018jyi,MartinCamalich:2020dfe,Bandyopadhyay:2021wbb} and lepton decays~\cite{Bauer:2019gfk,Calibbi:2020jvd} or directly in $e^+e^-$ interactions~\cite{Dolan:2017osp,Acanfora:2023gzr}. Fixed-target settings~\cite{Dobrich:2015jyk,Dobrich:2019dxc} further enable the search of sub-GeV ALPs. In cosmological and astrophysical probes, still lighter ALPs manifest through observable phenomena~\cite{Cadamuro:2011fd,Millea:2015qra,Depta:2020wmr}.

In this work, we aim to probe the effects of non-resonant ALP-mediated production processes involving SM final states only, at the LHC. The ALP serves as an off-shell propagator in these $s$-channel scattering processes. We analyze the behavior of differential cross-sections at high energies for scatterings that produce electroweak gauge bosons and the Higgs boson from $pp$ collisions. As we will see, the enhanced high-energy sensitivity of the LHC enables us to impose significant constraints on ALP-interactions, from such processes as their rates grow with energy. This deviates from the SM scenarios, which exhibit a decrease in the production rates with the collisional  center-of-mass energy($\sqrt{\hat{s}}$), as  $1/\hat{s}$. The explicit dependence of the derivative interactions of the ALPs with SM particles lead to enhanced cross-sections at increasing energies, especially when $\sqrt{\hat{s}} \gg m_a$, the mass of the ALP.

While presently, ALP interactions with massive gauge bosons are accessible primarily through indirect channels, such as loop corrections to low-energy processes~\cite{Izaguirre:2016dfi,Bauer:2017ris,Alonso-Alvarez:2018irt,Ebadi:2019gij,Gavela:2019wzg,Ertas:2020xcc,Gori:2020xvq,Kelly:2020dda,Galda:2021hbr,Bonilla:2021ufe,Bauer:2021mvw}, or direct detection at colliders, based on the ALP mass and decay width, these interactions might be observed in associated production processes involving a gauge boson and an ALP. The ALP may either go undetected~\cite{Mimasu:2014nea,Brivio:2017ije} or participate in resonant decays~\cite{Craig:2018kne,Wang:2021uyb}. In the context of the kinematics discussed above in presence of the low-lying ALPs, the LHC stands out as a key facility for exploring events that yield SM bosons in the final state, including electroweak gauge bosons $(W,Z,\gamma)$, gluons and the Higgs boson. The earlier investigations on processes where the ALP acts as an off-shell propagator included  $ZZ~({\rm semileptonic~decay~channel}),WW$ and $Z\gamma$ in  leptonic channel, $\gamma\gamma$ and di-jet production~\cite{Gavela:2019cmq,Carra:2021ycg,Bonilla:2022pxu}. In this work, we will consider that the ALP is too light to be produced resonantly and study the gluon-initiated scatterings such as $pp(gg) \to Zh,Z\gamma$ (hadronic final states from Z), $WW$ (semi-leptonic final states) and $WW\gamma$ (leptonic final states) that proceed through an intermediate ALP. The search strategy to probe the ALP interactions with the SM particles entails identifying deviations from the SM expectations within the high-energy tails of the invariant mass distributions of the boosted final states. 

Focusing on a particular ALP-mass range of below $100$ GeV, we highlight the novel features of our study:
\begin{itemize}
	\item  We carry out a comprehensive analysis of non-resonant ALP-mediated $Zh, Z\gamma,WW$ and the $WW\gamma$ production processes at the LHC, aiming to probe at high energy regime. In each of the $2\to2$ processes, atleast one of the heavy SM particles undergoes hadronic decay.  We study these processes for ALP interactions by contrasting the results with the latest experimental data. By taking advantage of the derivative ALP couplings and the boosted state of heavy bosons in the final state, we investigate for new physics effects in the $Zh$ process spanning the final state invariant mass spectrum  from 320 GeV to 2.8 TeV, the $WW$ process from 900 GeV to 4 TeV and the $Z\gamma$ process between 800 GeV to 2.1 TeV. Previous studies, to our knowledge, have focused on leptonic channels in lower energy ranges, particularly for the $Z\gamma$ and $WW$ processes~\cite{Carra:2021ycg}. Our study improves  the ALP limits obtained in these production processes. Prospects at HL-LHC are also discussed.
	
	\item  For the Higgs-strahlung and $Z\gamma$ production processes, where the Higgs and $Z$ boson, respectively, are boosted, result in a `fat-jet' comprising two b-tagged subjets. Similarly, in the semi-leptonic channel of the $WW$ production, one of the $W$ bosons is manifested as a fat-jet. We explore final state topologies featuring such fat-jets in combination with a hard photon or a $Z$ boson to search for signatures of new interactions differing from the SM. These heavy particles, decaying hadronically, are abundantly produced at the LHC but are challenging to detect through conventional searches due to overwhelming QCD backgrounds. Tagging these jets and employing jet-substructure techniques enhances the sensitivity of LHC to previously unexplored kinematic regions for axion-like particle mediated processes.
	
	\item This is the first detailed analysis of non-resonant ALP-mediated $WW\gamma$ production at the LHC. This $2\to 3$ process is dominantly produced from gluon-fusion under the allowed constraints of ALP couplings from available data. We explore how this process could probe $g_{a\gamma\gamma}$, $g_{aZ\gamma}, g_{aWW}$ couplings, either through $WWZ,WW\gamma$ triple gauge coupling or via a four-point interaction of $a\gamma WW$.
	
	\item We also briefly compare the limits on the ALP couplings obtained from the aforementioned non-resonant production processes with those from direct probe of ``mono-$X$'' signatures $(X=Z,W^\pm,h)$ through the production of ALP in association with a Higgs or a vector boson.
	
	\item To enhance the distinction between signal and background in the processes under study, we employ a multivariate analysis using the Boosted Decision Tree (BDT) technique. This approach, going beyond the conventional cut-based method, exhibits a marked improvement in signal significance, as will be explicitly demonstrated in the subsequent sections.
\end{itemize}

In this work, we adopt a model-independent effective field theory (EFT) approach. Considering that the Higgs boson observed at LHC is still part of an $SU(2)_L$ doublet, as in the SM, then any electroweak (EW) physics that extends beyond the SM can be systematically examined using a linear EFT expansion~\cite{Buchmuller:1985jz,Grzadkowski:2010es}. The setup of a linear EFT  includes the SM and an ALP~\cite{Georgi:1986df,Choi:1986zw, Brivio:2017ije} and contrasts with the framework of a chiral EFT when considering interactions involving the ALP and the Higgs boson~\cite{Brivio:2017ije,Bauer:2017ris}. The  current experimental results do not preclude the existence of a Higgs component that deviates from this doublet structure, at least within a 10\% uncertainty margin~\cite{ATLAS:2019slw}, thus making the non-linear EFT methodology equally pertinent for exploration \cite{Feruglio:1992wf,Alonso:2012jc,Azatov:2012bz,Alonso:2012px,Alonso:2012pz,Buchalla:2013rka,Brivio:2013pma}. We will mainly focus on the linear EFT framework in this paper, while also consider the chiral EFT context to assess the ALP-Higgs interactions.

In future LHC runs, the non-resonant ALP searches are set to become increasingly competitive. This improvement is expected not just because of the significant growth of available data on the high luminosity frontier but also due to progress inspired by the SMEFT studies which encourage a generalised, systematic approach to probing new physics~\cite{Cepeda:2019klc}. While the SMEFT presumes that new physics manifests through particles that are too heavy to be produced on-shell~\cite{Biswas:2021qaf,Biswas:2022fsr}, non-resonant ALP searches aim to look for ALPs too light to undergo resonant decays. This distinct approach enables non-resonant ALP searches to explore complementary areas of the parameter space, depending on minimal assumptions about the ALP decay width.

The plan of the paper is the following. In section~\ref{alp_eft_framework}, we describe the ALP effective theory and set the framework for our analysis. This has been followed by a discussion on the general features of non-resonant ALP EW processes considered in this study in section~\ref{alp_mediated_processes}. In section~\ref{collider_probe}, we undertake a detailed collider analysis studying the kinematical features of the signal and background processes. We present the constraints derived on the parameters of the ALP Lagrangian using measurements from the latest available Run 2 LHC data. We discuss the validity range of our  analysis. Thereafter, we define some benchmark scenarios for ALP signals and discuss the projected sensitivities to the effective  couplings  in the upcoming HL-LHC run. We also discuss the constraints arising from direct probe of these couplings through the production of ALP in association with a Higgs boson or a vector boson.  In section~\ref{sensitivity_reach}, numerical results and their interpretations, along with detailed discussions on cross-section parameter dependencies, are covered. In section~\ref{multivariate_analysis}, the use of boosted decision trees to improve the cut-based results is explored. In section~\ref{existing_constraints}, we summarise the existing constraints from other experiments on ALP mass and couplings. Finally, we draw our conclusions in section~\ref{summary}.
\vspace{-0.5cm}
\section{ALP Effective Lagrangian}
\label{alp_eft_framework}
\vspace{-3mm}
We consider an ALP, denoted by $a$, which is a pseudo-scalar state. Its interactions are constructed to respect the invariance under shifts $a(x)\to a(x)+\alpha$, where $\alpha$ is a constant (reflecting to be of the form $J^{\mu}\partial_{\mu}\alpha$, consistent with its Goldstone nature). Within the EFT framework, we express all ALP interactions with suppression factors which are inversely proportional to the characteristic scale $f_a\gg m_a$ (mass of the ALP), that is unknown and naturally close to the mass scale of the heavy sector the ALP originates from. Also, it is implicitly assumed that $f_a\gg v$ where $v$ denotes the EW scale. We require all ALP interactions to be invariant under the full SM gauge group.  For linear EWSB realization,  the most general  linear bosonic Lagrangian, incorporating next-to-leading order (NLO) effects related to $a$, is given by 
\beq
\mathcal{L}_\text{eff}^\text{linear}= \mathcal{L}_{\text{LO}}+\, \delta \mathcal{L}_{\mathrm{eff}}^\text{bosonic}\,, 
\label{Lbosonic-lin}
\eeq
where the leading order Lagrangian now comprises the SM Lagrangian along with the ALP kinetic term,
\beq
\mathcal{L}_{\text{LO}}= \mathcal{L}_\text{SM}\,+\frac{1}{2}(\partial_\mu a)(\partial^\mu a) -\frac{1}{2}m_{a}^{2} a^2\,,
\label{Lbosoniclin1}
\eeq
while the NLO bosonic corrections  due to the  ALP interactions with the SM fields are included in the  effective Lagrangian : 
\begin{align}
\begin{split}
\delta \mathcal{L}_{\mathrm{eff}}^\text{bosonic} &\supset 
c_{\tilde{G}}\, \mathcal{O}_{\tilde{G}}+c_{\tilde{B}}\, \mathcal{O}_{\tilde{B}}+c_{\tilde{W}} \,\mathcal{O}_{\tilde{W}}\, +{c_{a\Phi}} \,\mathcal{O}_{a\Phi}\,,
\end{split}
\label{eq:lagrangian}
\end{align}
\noindent Eqn.~\eqref{eq:lagrangian} contains a complete and non-redundant set of dimension-5 bosonic operators which are given by: 
\begin{align}
\begin{split}
\mathcal{O}_{\tilde{G}} \equiv - \dfrac{a}{f_a}G_{\mu\nu}^A \widetilde{G}^{A\mu\nu}\,, \quad &  
\mathcal{O}_{\tilde{W}} \equiv - \dfrac{a}{f_a}W_{\mu\nu}^i \widetilde{W}^{i\mu\nu} \,,\\[0.4em]
\mathcal{O}_{\tilde{B}} \equiv - \dfrac{a}{f_a}B_{\mu\nu} \widetilde{B}^{\mu\nu} \,, \quad &
\mathcal{O}_{a\Phi}  \equiv i\dfrac{\partial^\mu a}{f_a}\, \Phi^\dagger \overleftrightarrow{D}_\mu \Phi 
\,.
\label{operators}
\end{split}
\end{align}
%
\noindent 
Here, 
$G_{\mu\nu}$, $W_{\mu\nu}$ and $B_{\mu\nu}$ are the generic field strength tensors corresponding to the SM gauge groups $SU(3)_c$, $SU(2)_L$ and $U(1)_Y$ respectively. The dual field strength tensors $\widetilde{X}^{\mu\nu}$ are defined by $\tilde{X}^{\mu\nu}\equiv \frac12 \epsilon^{\mu\nu\rho\sigma}X{\rho\sigma}$, with $\varepsilon^{0123}=1$.  
The associated operator coefficients  $c_i$ in Eqn.~\eqref{eq:lagrangian} are real constants. $\Phi$ is the SM Higgs doublet,
with $\Phi \overleftrightarrow{D}_\mu \Phi \equiv \Phi^\dagger \big{(}D_\mu\Phi\big)-\big{(}D_\mu \Phi\big{)}^\dagger \Phi {}$.
The first three operators in Eqn.~\eqref{eq:lagrangian} induce ALP couplings to the gluon, the photon and the $Z$ and $W$ bosons as given by :
%
\begin{eqnarray}
\delta \mathcal{L}_{\mathrm{eff}}^\text{bosonic} &\supset& - \frac{g_{agg}}{4}a\,G_{\mu\nu}^A \widetilde{G}^{A\mu\nu}  -
\frac{g_{a\gamma\gamma}}{4}a\,F_{\mu\nu} \widetilde{F}^{\mu\nu} - \frac{g_{aZ\gamma}}{4}a\,F_{\mu\nu} \widetilde{Z}^{\mu\nu} \nonumber \\
&-& \frac{g_{aZZ}}{4}a\,Z_{\mu\nu} \widetilde{Z}^{\mu\nu} - \frac{g_{aWW}}{4}a\,W_{\mu\nu}^+ \widetilde{W}^{-\mu\nu}\,,
\label{eq:lagrangian_gs}
\end{eqnarray}
The coupling strengths are defined as:
\begin{subequations}
\begin{align}
g_{agg} & =\frac{4}{ f_{a}}\,c_{\tilde{G}} \,, \quad
g_{a\gamma\gamma} = \frac{4}{f_{a}} \, \big(s_w^2 \, c_{\tilde{W}}+ c_w^2 \, c_{\tilde{B}} \big) \label{eq:gagluon}\\
g_{aWW} &= \frac{4}{ f_a}  \, c_{\tilde{W}} \,,\quad 
g_{aZZ} = \frac{4}{ f_a} \,({c_w^2}\, c_{\tilde{W}}+ {s_w^2}\, c_{\tilde{B}}) \label{eq:gaZZ}\\
g_{a\gamma Z} &= \frac{8}{ f_a}  \,s_wc_w ( c_{\tilde{W}}-  c_{\tilde{B}} )
\label{eq:gaZgamma}
\end{align}
\label{eq:alp_coupling}
\end{subequations}
with $s_w$ and $c_w$ denoting the sine and cosine of the Weinberg angle, respectively. 

After electroweak symmetry breaking, the last operator in Eqn.~(\ref{eq:lagrangian}), $\mathcal{O}_{a\Phi}$ induces a contribution to a two-point function involving longitudinal gauge fields and can be removed via a Higgs field redefinition. To assess its effect on observables, one approach is to substitute it with a fermionic vertex~\cite{Georgi:1986df}. This substitution can involve a vertex that either conserves or flips chirality, or a combination of both.  For illustration, the Higgs field redefinition:
\begin{equation}
\Phi\to e^{ic_{a\Phi}\, a/f_a}\Phi
\label{Oaphiequiv}
\end{equation}
when applied to  the bosonic Lagrangian in Eqn.~\eqref{Lbosonic-lin}, leads to a modification originating from the Higgs kinetic energy term in the  SM. This modification precisely negates $\mathcal{O}_{a\Phi}$ up to $\mathcal{O}(a/f_a)$. Meanwhile, the Yukawa terms in the SM generate a new Yukawa-axion coupling, allowing for a complete substitution of  $\mathcal{O}_{a\Phi}$. The overall effect  is, the  replacement in Eqn.~(\ref{eq:lagrangian})   by:
\begin{align}
i \dfrac{a}{f_a} \Big{[}\overline{Q} Y_u \widetilde{\Phi} u_R -
\overline{Q} Y_d \Phi d_R - \overline{L} Y_\ell \Phi \ell_R\Big{]}+\text{h.c.},
\label{eq:fermion_lagrangian}
\end{align}
where $Y_{u,d,\ell}$ are the SM Yukawa matrices.
In this work, we focus on experimental signatures that involve ALPs and SM bosons ($W,\,Z,\,\gamma$ and $h$). We do not consider the CP-violating terms and direct ALP-fermion interactions (stemming from the $\mathcal{O}_{a\Phi}$ operator) since such interactions are markedly suppressed at tree-level due to their proportionality to the involved fermion Yukawa couplings \footnote{For light fermions as initial states, the effect of these contributions are effectively negligible.}.

\vspace{3mm}

Within the framework of non-linear (chiral) electroweak theory, the interactions of the ALP with SM fields at leading order are captured by the following expression:
\begin{equation}
\label{eq:lag}
\mathcal{L}=\mathcal{L}^{\text{HEFT}}_{\text{LO}}+ \mathcal{L}^{\text{ALP}}_{\text{LO}}\,.
\end{equation}
Here, $\mathcal{L}^{\text{HEFT}}_{\text{LO}}$ denotes the chiral Lagrangian within the Higgs Effective Field Theory (HEFT)~\cite{Alonso:2012px,Brivio:2016fzo,Herrero:2020dtv,Herrero:2021iqt} framework. In this model, the Higgs boson is treated as a singlet field, while the Goldstone bosons $\pi^{a}$ are introduced in a non-linear representation, through the exponential parametrization by means of a unitary matrix $U$ given by :
\begin{equation}
U(\pi^{a}) \equiv \exp\left({i \pi^{a} \tau^{a}/v}\right)\,,
\end{equation}
with $\tau^{a}$,  $a=1,2,3$ are the Pauli matrices. The $U$ matrix   which  transforms as
a bi-fundamental  under   $SU(2)_L \times SU(2)_R$:
\vspace{-3mm}
\begin{equation}
\label{eq:Umatrix_transform}
U(\pi^{a}) \ = \ \exp\left( \frac{i\pi^a \tau^a}{v}  \right)   \quad \xrightarrow[L\times R]{}  \quad
L \: U(\pi^{a}) \: R^\dag,
\end{equation}\vspace{-3mm}
The series expansion of $U$ is as follows:
\begin{equation}
U(\pi^{a}) =  {\mathbb{1}}_{2} + i \frac{\pi^{a}}{v} \tau^{a} - \frac{2G^{+}G^{-} + G^{0} G^{0}}{2 v^2}  {\mathbb{1}}_{2} + \dots \,,
\end{equation}
where $G^{\pm}$ and $G^{0}$ are defined as $G^{\pm} = (\pi^{2} \pm i \pi^{1})/\sqrt{2}$ and $G^{0} = -\pi^{3}$, respectively. This peculiarity implies that there are multiple Goldstone boson interactions possible in the HEFT formalism, not just among themselves but also with the other fields. We work under this framework to study novel  ALP-Higgs interactions that can probe the unique singlet nature of the Higgs boson as described by the HEFT Lagrangian. Now the leading order Lagrangian for ALP interactions is expressed as:
\begin{equation}
\mathcal{L}^{\text{ALP}}_{\text{LO}} =  \frac{1}{2} \partial_{\mu} a \partial^{\mu}a -\frac{1}{2}m_{a}^{2} a^2 + c_{2D}\A_{2D}(h),~~{\rm{with}}~ \A_{2D}(h)= \left[iv^2\tr[\mathcal{T}\mathcal{V}_\mu]\de^\mu\frac{a}{f_a}\mathcal{F}_{2D}(h) \right]\,,
\label{eq:Higgs_operator}
\end{equation}
where the fields $\mathcal{V}_\mu(x)$ and $\mathcal{T}(x)$ are defined by the relations :  
$\mathcal{V}_\mu(x)\equiv \left(D_\mu U(x)\right)U(x)^\dag$ and $\mathcal{T}(x)\equiv U(x)\sigma_3U(x)^\dag$.
In this framework, as stated, the Higgs boson is introduced as a gauge-singlet scalar  field. There are no limitations from symmetry arguments on the implementation of this field and its interactions with itself and with the other fields. Its  interactions incorporated by polynomial functions such as:
\begin{equation}
\mathcal{F}_{2D}(h)=1+ a_{2D} h/v + b_{2D}(h/v)^2+ \mathcal{O}(h^3/v^3),
\label{eq:HEFT_operator}
\end{equation}
where coefficients $a_{2D}$ and $b_{2D}$ are independent constants. The term $\mathcal{A}_{2D}$ serves as the chiral analogue to the linear operator $\mathcal{O}_{a\Phi}$, with a distinct feature: it facilitates not only ALP-fermion interactions comparable to those in Eqn.~\ref{eq:fermion_lagrangian}, but it also induces new interactions at leading order between the ALP, electroweak gauge bosons and the Higgs, such as the trilinear $aZh$, $a\gamma h$ coupling. Exploring these interaction phenomenology yield an  understanding of the process of electroweak symmetry breaking, distinct from the linear approach and its interplay with axion-like states. Also, the other induced interactions in the polynomial function~\eqref{eq:HEFT_operator} can be important compared to the effects from other possible  operator involving interactions of Higgs and the gauge boson at the same order. Within the linear paradigm, such interactions emerge at the next-to-next-to-leading order (NNLO),  corresponding to mass dimension seven and thus, their effects are expected to be relatively subdominant. Furthermore, within the chiral framework,  the operators $\mathcal{O}_{\tilde{G}}$, $\mathcal{O}_{\tilde{W}}$ and $\mathcal{O}_{\tilde{B}}$ (in Eqn.~\eqref{operators}) also become relevant at NLO.
\vspace{-0.5cm}
\section{ALP mediated processes}
\label{alp_mediated_processes}
\begin{figure}[h]
	\begin{center}
		\subfloat{
			\begin{tabular}{cccc}
				\includegraphics[width=0.2\textwidth]{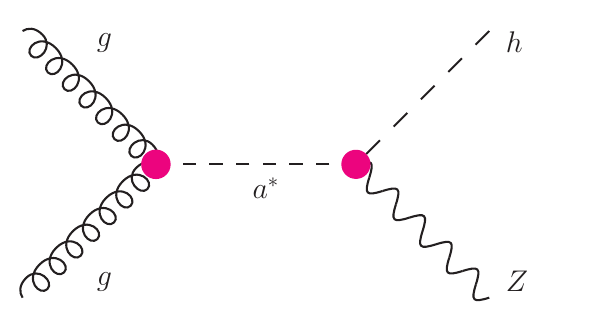}&
				\includegraphics[width=0.2\textwidth]{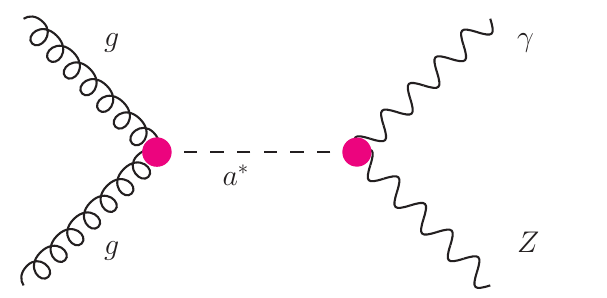}&
				\includegraphics[width=0.2\textwidth]{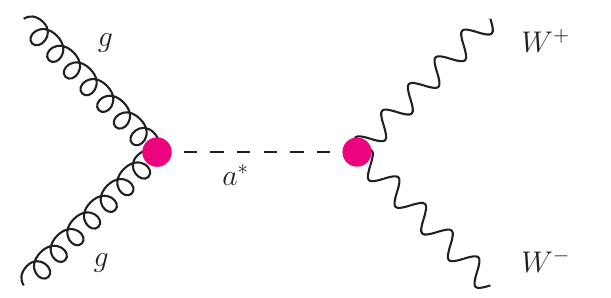}&
				\includegraphics[width=0.2\textwidth]{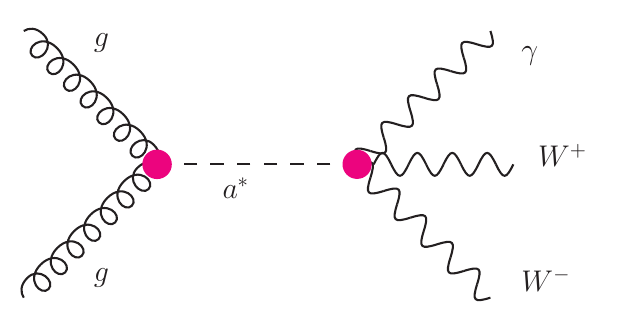}\\
				(a)&(b)&(c)&(d)
		\end{tabular}}\\
		
		\subfloat{
			\begin{tabular}{ccccc}
				\includegraphics[width=0.2\textwidth]{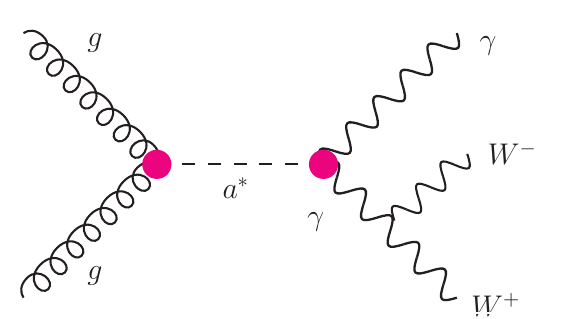}&
				\includegraphics[width=0.2\textwidth]{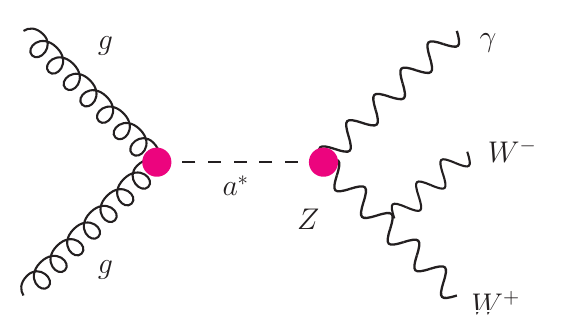}&
				\includegraphics[width=0.2\textwidth]{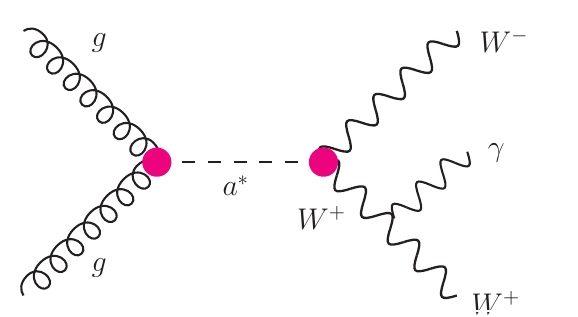}&
				\includegraphics[width=0.2\textwidth]{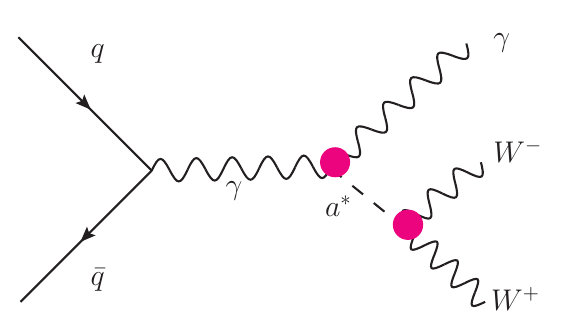}&
				\includegraphics[width=0.2\textwidth]{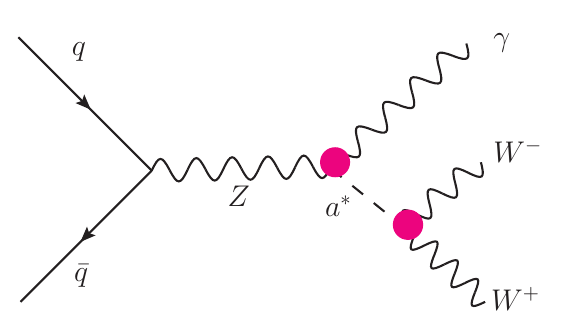}\\
				(e)&(f)&(g)&(h)&(i)
		\end{tabular}}
		\caption{\small Representative Feynman diagrams depicting the production of (a) $gg \to Zh$, (b) $gg \to Z \gamma$, (c) $gg \to W^+ W^-$  and  (d)-(i) $gg(q\bar{q}) \to W^+W^-\gamma$  mediated by an off-shell ALP. Each of the diagram consistently involves a double insertion of ALP operators.}
		\label{Fig_Feyndiag}
	\end{center}
\end{figure}

\vspace{-3mm}
We focus exclusively on processes with off-shell production of ALP into SM final states only. These processes are production of $Zh$, $Z\gamma$, $W^\pm W^\mp$ and $W^\pm W^\mp\gamma$ from $pp$ collisions. They all probe different operator combinations within the ALP EFT parameter space. To facilitate our discussion, we present in Fig.~\ref{Fig_Feyndiag}, the Feynman diagrams  which, by virtue of higher dimensional operators, contribute to the aforementioned processes. The blobs on  the vertices of diagrams (a)-(i)  stand for possible inclusion of one of the higher dimensional operators listed  in Eqns.~\eqref{eq:lagrangian_gs} and \eqref{eq:Higgs_operator}. ALP production in these processes is dominated by gluon-gluon fusion as the $q\bar{q}$ induced process for these final states is proportional to the quark masses from the operator $\mathcal{O}_{a\Phi}$ (See Eqn.~\eqref{eq:fermion_lagrangian}) and thus, highly suppressed. These channels have been studied for heavy resonant searches in the differential measurements of the invariant mass of the final state system by the CMS and ATLAS collaborations. No excess of events have been found and we shall reinterpret these measurements for the ALP mediated processes. We particularly aim to probe the boosted regime with at least one of the weak bosons or the Higgs boson decaying hadronically. This ensures that we have a large fraction of events and reduce uncertainties and yet maintain a balance with clean environment, using the jet substructure techniques for tagging the heavy bosons. Such boosted regimes with improved techniques are useful for identifying lighter ALPs that get rejected by the selection criteria of the cross-section measurements.  The $WW\gamma$ channel is an exception for which we will study the fully leptonic final state. All the four processes receive contributions from {\em{s-channel}} mediated non-resonant ALP. The $WW\gamma$ process receives additional contribution from initial quark states. We included these diagrams in our calculation for consistency. We have, however, checked that their contribution is significantly lower compared to those initiated by gluons. We investigate into the non-resonant triboson production, mediated through ALP. It is known that the resonant triboson production puts stringent constraints on ALP couplings for $m_a>100$ GeV~\cite{Craig:2018kne}. The non-resonant ALP mediated $WW\gamma$ process can be induced by the couplings  $\{g_{agg},g_{aZ\gamma},g_{aWW},g_{aWW\gamma},g_{a\gamma\gamma}\}$. The couplings $g_{aWW}$ and $g_{aWW\gamma}$ depend on one parameter $c_{\tilde W}$. However, a four point interaction of $aWW\gamma$, with a different Lorentz structure, can leave distinct kinematic effects in the process than the $aWW$ interaction. Both the couplings $g_{aWW}$ and $g_{aWW\gamma}$ lead to an amplitude growing with energy. In the case of $g_{aWW}$, the energy growth arises because of the extra powers of momenta in the $aWW$ vertex, whereas for the contact interaction, $g_{aWW\gamma}$, the energy growth is also due to the fact that there is absence of one propagator in the diagram involving this vertex (Fig.~\ref{Fig_Feyndiag} (d)).

As the ALP is always off-shell, its propagator acts as a suppression in the hadronic scattering amplitudes. However, due to the presence of the  explicit momentum dependence of the ALP interactions under discussion, the ALP couplings lead to higher energy growth with the invariant mass of the event final states as compared to that in  the corresponding SM  backgrounds.

All the diagrams in Fig.~\ref{Fig_Feyndiag} must arise with double  insertions of ALP operators. This results for the amplitude to scale as $f_a^{-2}$ and cross-sections in the order $f_a^{-4}$. In all generality, the contributions from bosonic ALP couplings in Eqn.~\eqref{eq:lagrangian_gs}  interfere with the SM  amplitudes.
Thus, a generic cross-section when expressed as polynomial functions of Wilson coefficients $c_i\over f_a$, including both SM and EW ALP contributions, has the structure
\begin{equation}
\label{eq.polynomial}
\begin{aligned}
\sigma_{\rm {ALP}} &= \sigma_{\rm{SM}} 
+ \frac{1}{f_a^2}\, \sigma_{\rm int}
+ \frac{1}{f_a^4}\, \sigma_{\rm NP}\,.
\end{aligned}
\end{equation}
\begin{equation}
{\rm where}~~\sigma_{\rm int} = \sum_{i}^{m} c_i^2 \sigma^{a}_{\rm int}+\sum_{i>j=1}^{m} c_i c_j\sigma^{b}_{\rm int}~ {\rm and}~ \sigma_{\rm NP}=\sum_{i}^{m} c_i^4 \sigma^{a}_{\rm NP}+\sum_{i>j=1}^{m} c_i^2 c_j^2\sigma^{b}_{\rm NP} + \sum_{i>j=1}^{m} c_i c_j  (c_i^2+c_j^2)\sigma^{c}_{\rm NP}\nonumber
\end{equation}

When the ALP couplings are relatively small, their interference with the SM background may become comparable with the pure ALP-signal and thus must be considered in evaluation of the process. The coupling value at which this interference becomes significant varies based on the specific final state being analyzed. In processes with electroweak diboson final states, the ALP signal interferes with the SM ones occurring at one-loop.  The nature of interference could be constructive or destructive and it depends on the relative sign of the couplings $g_{agg}$ and $g_{a V_1 V_2}$ (new vertices in the diagram, Fig.~\ref{Fig_Feyndiag}).  Currently, however, the magnitude of ALP-gluon couplings accessible at LHC are loose and the interference effect is suppressed in the total cross-section estimation~\cite{Gavela:2019cmq}. The quartic dependence from the pure ALP interactions dominate and  result in  large-$\hat{s}$ enhancement in the cross-section, $\sigma_{\rm {ALP}} \sim {\hat{s}}/{f_a^4}$. Such energy scaling is valid only as long as the energies involved in the scattering process remain below the cutoff scale of EFT, $\sqrt{\hat{s}} < f_a$. On the other hand, the SM backgrounds usually scale as $1/\hat{s}$  well above the resonance of the  s-channel.  In hadronic collisions, the calculation of any cross-section involves a convolution of this partonic cross-section with the parton distribution functions (PDFs). These PDFs exhibit a declination with the increase in energy. Taking this effect into account, the ALP mediated rates show a slower decrement  with the invariant mass of the system compared to the SM background.  This allows to distinguish ALP-mediated processes from the SM background as discussed in the following sections.

\vspace{-0.5cm}
\section{Collider Analysis}
\label{collider_probe}
\vspace{-3mm}
The effective Lagrangian has been implemented  into {\tt{FeynRules}}~\cite{Alloul:2013bka} to generate the {\tt{UFO}} model file~\cite{Brivio:2017ije} for the event generator {\tt{Madgraph5\_aMC@NLO}}~\cite{Alwall:2014hca}. MadGraph was employed for producing all signal and background sample events. These events are   generated at leading order (LO) and subsequently processed by {\tt{Pythia}} (v8)~\cite{Sjostrand:2014zea} for parton showering and hadronization. For event generation, {\tt{NNPDFNLO}} parton distribution functions~\cite{pdf} are utilized, setting both factorization and renormalization scales  dynamically to half the sum of all final state transverse energies in the scattering processes. The matching parameter,  {\tt QCUT}, was specifically determined for the different processes as discussed  in Ref. \cite{MLMMerging}. Detector effects are incorporated by passing the events through {\tt Delphes-v3.4.1}~\cite{deFavereau:2013fsa}. Jets are reconstructed using  {\tt{Fastjet-v3.3.2}}~\cite{Cacciari:2011ma}.  We impose a set of cuts at the generator level on the final state particles, namely, 
\begin{equation}
\begin{aligned}
p_{T_{(l,j)}}	& > 30~\rm{GeV}\,, & p_{T_{\gamma}} & > 20~\rm{GeV}\,, & \Delta R_{(\gamma,j)} & >0.4\,,& \Delta R_{(j,j/ll/jl)} & >0.2\,,\\
\eta_j & <5\,, &\eta_{(l,\gamma)} &<2.5\,,   &  m_{ll}& >50~\rm{GeV}\,&&
\end{aligned}
\label{preselection}
\end{equation}
all processes, except for the $WW\gamma$ channel where we require $m_{ll'}>10~\rm{GeV}$. The angular separation between two particles is defined as $\Delta R = \sqrt{\Delta \eta ^2 + \Delta \phi^2}$, where $\eta$ is the pseudorapidity and $\phi$ is its azimuthal angle of each particle. The ALP signal events are generated fixing $m_a=1~\rm{MeV}$, treating the ALP as effectively massless at LHC energies. The ALP width $\Gamma_a$ is assumed to be considerably smaller than $m_a$. The specific choices of the ALP mass and its decay width have negligible impact in the non-resonant regime. We generate signal samples with pure ALP-mediated production and the interference between the ALP and the SM processes. However, we have checked that the estimation of the total rate of the process is numerically dominated by the $\sigma_{\rm NP}$ (Eqn.~\eqref{eq.polynomial}).

\subsection{13 TeV LHC probes}
\label{13tev_collider_probe}
In this section, we will present the details of the process analyses. All of these processes are sensitive to the product of the ALP-gluon coupling $g_{agg}$ and the relevant ALP-bosonic couplings. We derive constraints on these ALP interactions via the non-resonant ALP-mediated signals mentioned above, utilizing publicly available data from the ATLAS and CMS collaborations at Run II 13 TeV LHC as listed in Table~\ref{tab:data}.
\begin{table}[h]
	\begin{center}
		\begin{tabular}{ |l | l | c | l | l|}
			\hline
			Channel  & Distribution & \# bins   & \hspace*{0.2cm}Int.  Lum. &\hspace*{0.2cm} Data set
			\\ 
			\hline
			$Zh\rightarrow \ell^+\ell^{-} b \bar{b}$ &  $m_{VH}$ Fig.~4(d) & 16&139 fb$^{-1}$
			& ATLAS 13 TeV~\cite{ATLAS:2022enb}
			\\[0mm]
			$Z\gamma\rightarrow b \bar{b} + \gamma$ & $m^{J\gamma}$, Fig.~5a & 18&36.1 fb$^{-1}$
			& ATLAS 13 TeV~\cite{ATLAS:2018sxj}
			\\[0mm]
			$WW\rightarrow \mu^\pm+J+\slashed{E}_T\;$ & $m_{WW}$, Fig.~4 (middle, right) &  16 &35.9 fb$^{-1}$
			& CMS 13 TeV~\cite{CMS:2019ppl}
			\\[0mm]
			$WW\rightarrow e^\pm+J+\slashed{E}_T\;$ & $m_{WW}$, Fig.~4 (middle, left) &  16 &35.9 fb$^{-1}$
			& CMS 13 TeV~\cite{CMS:2019ppl}
			\\[0mm]
			$WW\gamma\rightarrow e^\pm\mu^{\mp}+\gamma+\slashed{E}_T\;(0j)$ & $m_{T}^{WW}$, Fig.~3 (left) & 12 & 138 fb$^{-1}$
			& CMS 13 TeV~\cite{CMS:2023rcv}
			\\[0mm]
			\hline
		\end{tabular}
		\caption{Experimental data used in our study to constrain the ALP couplings.
			The third column indicates the number of bins used in our analysis, always starting from the highest.}\label{tab:data}
	\end{center}
\end{table}
\vspace{-0.5cm}
\subsubsection{$pp \to Zh$}
\label{13tev_zh}
\vspace{-3mm}
This process yields a powerful probe of the ALP-Higgs coupling through the operator $\A_{2D}$ in Eqn.~\eqref{eq:Higgs_operator} and also assumes the additional presence of $g_{agg}$. It may be expected among the leading signals for ALP-Higgs interactions and a conclusive evidence if the underlying EWSB enjoys a non-linear character.  There can be further probes of this operator contribution in double Higgs production. In fact, this operator  also induces $ah\gamma$ interaction and thus, one can question for a process of $pp \to h \gamma$ signal mediated by the ALP. However, as the ALP forces the interaction to be derivative and the photon being transverse and on-shell in $pp\to h\gamma$, leading to a vanishing cross-section.

In order to study the current reach of the LHC in constraining this coupling through $pp \to Zh$, we optimize a hadron-level analysis to obtain the sensitivity to the BSM signal, which is well-pronounced in the high energy bins. To achieve this, we consider the $Z(\ell^+\ell^-)h$ production and scrutinize the $h \to b\bar{b}$ decay channel. The dominant backgrounds consist of $Zb\bar{b}$ and the irreducible SM production of $Zh$. Reducible contributions arise from $Z+$ jets production ($c$-quarks included but not explicitly tagged), where the light jets can be misidentified as $b$-jets and  $t\bar{t}$ production in the fully leptonic decay mode. Rather than performing a resolved analysis with two distinct b-tagged jets, our method focuses on a single fat-jet with a cone-radius  $R=1.0$. We apply the BDRS  method~\cite{Butterworth:2008iy} with some minor modifications to enhance sensitivity. This technique merges jets using the CA algorithm, from a significantly large cone radius to encapsulate all decay products of a resonance (like the Higgs boson). The process involves breaking the primary jet $J$ into two subjets,  $j_{1}$ and $j_{2}$ with $m_{j_{1}} > m_{j_{2}}$. We impose a mass drop condition such that $m_{j_{1}}<\mu m_J$ where $\mu=0.66$ ($m_{J}$ is mass of the fat-jet), along with a symmetry criterion between the subjets requiring  $\frac{min(p_{T,j_1}^2,p_{T,j_2}^2)}{m_J^2}\Delta R^2_{j_1,j_2} > 0.09$. If the condition fails, the lighter subjet, $ j_2 $, is removed and the process repeats with $ j_1 $. This iteration continues until a final jet $ J $  is obtained that satisfies the mass drop condition. This selection is fairly efficient in filtering out QCD jets but can still be impacted by the underlying events at the high energies and luminosities of the LHC. To further eliminate rare QCD events and effects from hard gluon emissions or underlying events, we refine the Higgs vicinity by recombining the components of $ j_1 $ and $ j_{2} $ using the CA algorithm with a reduced radius $R_{\text{filt}} = \min(0.2, R_{b\bar{b}}/2) $\footnote{We find that this choice of $R_{\text{filt}} $ is effective in background reduction.}. We keep only the three strongest filtered subjets for resonance (Higgs boson) reconstruction. Overall, this approach effectively distinguishes boosted electroweak-scale resonances from significant QCD backgrounds.

The event selection criteria are based on Ref.~\cite{ATLAS:2022enb}. We constructed fat-jets with a minimum transverse momentum, $p_T>$ 100 GeV and a rapidity cut of $|y| < 2.5$. Leptons are isolated within a $R=0.3$ radius, with their $p_T > 25$ GeV and $|\eta| < 2.5$. Events with exactly two isolated, opposite-charge, same-flavor leptons, conforming to the $Z$-peak with invariant mass between max$[40,87-0.030.m_{Zh}]$ GeV and $[97+0.013.m_{Zh}]$ GeV (as a function of $m_{Zh}$) and a leptonic separation of $\Delta R>$ 0.2 are selected. For Higgs reconstruction, we required at least one fatjet with a minimum of two $B$-meson tracks ($p_T > 15$ GeV) and a fatjet $p_T > 250$ GeV. After mass-drop and filtering criteria, events with exactly two $b$-tagged subjets, well-separated from isolated leptons are selected. The Higgs invariant mass is required to be between 75 and 145 GeV. To minimize the backgrounds, both reconstructed $Z$ and Higgs were required to have $p_T > 200$ GeV and the $t\bar{t}$ background was significantly reduced by setting a $\slashed{E}_T/\sqrt{H_T} < (1.15+(8\times10^{-3}).m_{Zh})/1$ GeV limit.  The $p_T^{ll}$ was also optimised to be greater than $\left[20+9.\sqrt{m_{Vh}/(1~{\rm GeV})-320}\right]$ GeV, where all the events are required to have a minimum invariant mass of $Z$ and Higgs of 320 GeV. The ATLAS provides a measurement of invariant mass of the $Zh$ system in the 2 leptons+2 b-jets final state~\cite{ATLAS:2022enb}. The bins extend in varying steps from 320 GeV to 2.8 TeV.  These cuts are relaxed for higher-energy tails to account for resolution effects and smaller backgrounds and  lead to a higher signal acceptance upto energies of multiple TeV. The corresponding signal and background distributions with the ATLAS data are shown in Fig.~\ref{13tev_validation} (a).  The SM background and the experimental data have been obtained from~\cite{hepdata}.

Cross-sections for each of the background processes simulated, are summarised in Table~\ref{tab:xsec_signal_backgrounds}. All the aforementioned background processes are generated at LO and multiplied with appropriate K-factors to obtain the higher order in QCD cross-sections.

\begin{table}[h]
	\centering
	\setlength{\arrayrulewidth}{.3mm}
	\setlength{\tabcolsep}{0.8 em}
	\renewcommand{\arraystretch}{1.2}
	\begin{tabular}{|c|c|c|c|c|c|}
		\hline\hline
		\multicolumn{2}{|c|}{Process $pp \to h(\to b\bar{b})Z(\to \ell^+\ell^-)$} 			& $\sigma^{bc}$ (fb) at 13 TeV & Efficiency$(10^{-4})$& $\sigma^{ac}$ (fb) at 13 TeV 	&K-factor\\ \hline\hline
		$Z+$ jets [N$^2$LO]& $ b-tagged$  & $20.5 \times 10^4$	&0.11&1.1275&1.41\\ \cline{2-6} 
		\cite{Catani:2009sm,Balossini:2009sa}& $ untagged $ 				& $30.39 \times 10^5$ 	&0.0573&0.174 & 1.23\\ \hline
		$V V + jets$ [NLO]  & $WW + jets$ & $1918$	&0.102&0.00977&1.66\\ \cline{2-6} 
		\cite{Campbell:2011bn} & $W Z + jets$  		& $1537$	&4.005&0.3078 &1.87\\ \cline{2-6} \hline
		Top pair [N$^3$LO]\cite{Muselli:2015kba} & $t\bar{t}  + jets$  &$10.24\times 10^{5}$ 	&$1 \times 10^{-4}$&0.00512&1.62\\ \cline{2-5}  \hline
		Electroweak~\cite{LHCWGxsec} & $h Z$ & $37.96$  	&34.5&0.0655&1.42\\ \hline \hline

		\multicolumn{2}{|c|}{Process $pp \to Z(\to b\bar{b})\gamma$} 			& $\sigma^{bc}$ (fb) at 13 TeV & Efficiency$(10^{-4})$& $\sigma^{ac}$ (fb) at 13 TeV 	&K-factor\\ \hline\hline
		{$J \gamma $} [NLO] & $ b-tagged$  & $47.20 \times 10^3$	&$0.5885$&$2.7778$&1.32\\ \cline{2-6} 
		~\cite{Alwall:2014hca}& $ untagged $ 				& $188.9 \times 10^3$ 	&$0.03184$&$0.6016$&1.32 \\ \hline
		{$V \gamma + jets$} [NLO]   & $Z \gamma + jets$ & $140.80$	&$22.38$&$0.3352$&1.19\\ \cline{2-6} 
		~\cite{Krause:2017nxq,Lombardi:2020wju}& $W \gamma + jets$  		& $541.89$ 	&$0.861$&$0.0467$&1.28\\ \cline{2-6} \hline
		{Top $\gamma + jets$} [NLO] & $t\bar{t} \gamma + jets$  & $6.69$ 	&$1.27$&$8.516\times 10^{-4}$&1.54\\ \cline{2-6} 
		~\cite{Pagani:2021iwa}& $tj \gamma + jets$  		& $5.418$ 	&$0.6652$&$3.604\times 10^{-4}$&1.17\\ \cline{2-6} \hline
		Electroweak~\cite{Gabrielli:2016mdd} & $h \gamma$ & $0.01377$ 	&$194.1$& $2.67\times 10^{-4}$&-\\ \hline \hline

		\multicolumn{2}{|c|}{Process $pp \to W(\to \ell\nu)W(\to jj)$} 			& $\sigma^{bc}$ (fb) at 13 TeV & Efficiency$(10^{-3})$& $\sigma^{ac}$ (fb) at 13 TeV 	&K-factor\\ \hline\hline
		\multicolumn{2}{|c|}{$W+$ jets [N$^2$LO]~\cite{Catani:2009sm,Balossini:2009sa}}& $18.92\times 10^{4}$	&$6.68\times 10^{-3}$&$126.455$&1.37\\  \hline
		$V V + jets$ [NLO]  & $WW + jets$ & $12.431\times 10^{4}$	&$0.964$&$11.985$&1.66\\ \cline{2-6} 
		\cite{Campbell:2011bn} & $W Z + jets$  		& $5.182\times 10^{4}$ 	&$0.0565$&$0.293$&1.87\\ \cline{2-6} \hline
		\multirow{2}{*}{Single $t + jets$~\cite{Pagani:2021iwa}} & $tW$  & $83.1\times 10^{3}$	&$0.135$&$11.231$ &1.21\\ \cline{2-6} 
		& $tj$  		& $12.35\times 10^{3}$	&$0.124$&$11.224$ &1.12\\ \cline{2-6} \hline
		Top pair [N$^3$LO]\cite{Muselli:2015kba} & $t\bar{t}  + jets$  & $98.857\times10^{4}$	&$0.0124$&$10.992$& 1.62\\ \cline{2-6} \hline\hline

		\multicolumn{2}{|c|}{Process $pp \to W(\to \ell\nu)W(\to \ell\nu) \gamma$} 			& $\sigma^{bc}$ (fb) at 13 TeV & Efficiency& $\sigma^{ac}$ (fb) at 13 TeV	&K-factor \\ \hline\hline
		\multicolumn{2}{|c|}{$WW\gamma$ [NLO]~\cite{CMS:2023rcv}}& $4.274 \times 10^3$ 	&$0.438$&$1.8734$& 2.10\\  \hline
		\multicolumn{2}{|c|}{$WZ\gamma$ [NLO]~\cite{CMS:2023rcv}}& $447.5 \times 10^3$ 	&$8.23\times 10^{-4}$&$0.3685$&2.10\\  \hline
		\multicolumn{2}{|c|}{$Z\gamma$ [NLO]~\cite{Krause:2017nxq}}& $60.12 \times 10^3$ 	&$0.02$&$1.2024$ &1.19\\  \hline
		{Top} [NLO] & $t\bar{t} \gamma + jets$  & $59.4$ 	&$0.0442$&$2.625$&1.54\\ \cline{2-6} 
		~\cite{Pagani:2021iwa,CMS:2023rcv}& $tW + jets$  		& $83.1\times 10^{3}$ 	&$5.122\times 10^{-6}$&$0.4256$&1.21\\ \cline{2-6} \hline
		\multicolumn{2}{|c|}{Non-prompt leptons}& $6.008 \times 10^2$ 	&$1.067\times 10^{-3}$&$0.641$&-\\  \hline
		\multicolumn{2}{|c|}{Non-prompt photons}& $4.45 \times 10^3$ 	&$8.51\times 10^{-5}$&$0.378$&-\\  \hline\hline
	\end{tabular}
	\caption{The cross-sections for the background processes used in this analysis are shown with the order of QCD corrections provided in brackets. $\sigma^{bc}$'s and $\sigma^{ac}$'s are cross-sections before the cuts and after the cuts discussed in the text are applied. The last column presents the K-factors for the higher order corrections of the  processes with respect to the leading order cross-sections.}
	\label{tab:xsec_signal_backgrounds}
\end{table}
In Ref.~\cite{CMS:2021xor}, the CMS collaboration has performed a search analysis for the non-resonant ALP-mediated production of $Zh$ in the semileptonic channel. The analysis requires the leading (sub-leading) lepton from the event to have $p_T > 40 (30)$ GeV and $|\eta| < 2.1 (2.4)$. The invariant mass of the dilepton pair is required to be in the range  $70~{\rm GeV} < m_{\ell \ell} < 110$ GeV and have $p_T^{\ell \ell} > 200$ GeV. In addition, the events contain an anti-kT jet with radius $R = 0.8$ and $p_T^J > 200$ GeV. The merged jet mass is required to be in the range $65~{\rm GeV} < m_J < 105$ GeV. The analysis also makes use of the N-subjettiness variable and requires events with $\tau_{21} < 0.4$ for the fat-jet. This study spans $m_{Zh}$ bins from 450 GeV to 2 TeV.  Overall, the CMS analysis translates into an average ALP signal selection efficiency of $\sim 7\%$ (Ref.~\cite{CMS:2021xor})\footnote{The observed 95\% C.L. limit on $c_{\tilde G}a_{2D}/f_a^2$ is obtained as $0.0269$ TeV$^{-2}$ in the CMS analysis.}.
\vspace{-3mm}
\subsubsection{$pp \to Z\gamma$}
\label{13tev_za}
\vspace{-3mm}
We then consider the signal of $Z\gamma$ production mediated by an off-shell ALP and the Z decays hadronically. This process receives contributions from bosonic operator coefficients $c_{\tilde{B}}$ and $c_{\tilde{W}}$, apart from the ALP-gluon coupling $c_{\tilde G}$. These coefficients also affect $a\gamma\gamma$ and $aZZ$ vertices. Hence, to fully understand $aZ\gamma$ vertex modification, assumptions on $g_{a\gamma\gamma}$ and $g_{aZZ}$ are necessary as we elaborate later. In this process, we consider the regime where both the $Z$ boson and the photon are significantly boosted, leading to the total hadronic decay products of $Z$ being contained within a large radius jet. Consequently, the final state features a fat-jet in recoil against a hard photon. We employ jet substructure techniques to reconstruct the $Z$ jet from its invariant mass, with the fat-jet radius estimated by the relation $R \sim \frac{2 m_h}{p_{T_{h}}}$. The following SM processes can mimic the $Z \gamma$ signal. Continuum $\gamma j$ process emerges as the most dominant background. The  $Z/W\gamma+$jets process, while having a similar topology to the signal, is less prevalent due to lower cross-section. Production of  $t\bar{t}\gamma$ with hadronic decays of the top quarks also contribute to the background. However, demanding a high $p_T$ photon and $Z$ tagging can suppress these backgrounds. Similarly, single top productions like $tj\gamma,tb\gamma$ also contribute in the background.   The $pp \to h(\to b\bar{b}) \gamma$ associated production in the SM has a nominal rate, either due to the  very small couplings of Higgs with the initial state quarks or because the process predominantly receives contribution at one-loop.

The ATLAS~\cite{ATLAS:2018sxj} Collaboration has searched for a resonance decaying into  $Z$ and photon. No significant excess over the SM expectation has been reported. In the signal from $800~\rm GeV < m_{J \gamma} < 2~\rm TeV$, the ATLAS has collected 55  events with $\int \mathcal{L}dt = 36.1~\rm fb^{-1}$. We reinterpret this analysis for deriving constraints on ALP interactions. With the SM background expectation, we compare Fig. 5 (a) of Ref.~\cite{ATLAS:2018sxj} as shown in Fig.~\ref{13tev_validation} (b). The selection criteria based on Ref.~\cite{ATLAS:2018sxj} and the corresponding cut efficiency are presented in Table~\ref{tab:zgamma_cutflow}.
\begin{table*}[h!]
	\centering
	\begin{tabular}{||c||c|c|c|c||}
		\hline
		\cline{1-5} 
		Cut flow & jets+ $\gamma$ & $W/Z+\gamma$ & $t\bar{t}\gamma+tj\gamma$ & $\rm Signal$   \\
		\cline{1-5} 			
		
		Atleast one fat-jet with two B-meson tracks  with $p_T > 200$ GeV  & $0.034$ & $0.13$ & $0.79$  & $0.46$  \\\hline
		
		Atleast 1 isolated photon and lepton veto & $0.79$ &$0.84$& $0.83$ & $0.92 $\\\hline
		
		Photon transverse momentum $> 200$ GeV  & $0.45$ &$0.55$& $0.58$ & $0.96$ \\\hline
		
		Atleast 1 fat-jet with two B-meson tracks ($p_T > 250$ GeV)  & $0.58$ &$0.74$& $0.78$ & $0.94$ \\\hline
		
		Two mass drop subjets and at least two filtered subjets & $0.32$ &$0.48$& $0.71$ & $0.45$ \\\hline
		
		Two b-tagged subjets& $0.17$ & $0.59$& $0.18$  & $0.78$   \\\hline
		
		$70.0 < m_{Z} < 110.0$ GeV & $0.43$ & $0.25$& $0.45$   & $0.26$ \\\hline
		
		$\Delta R(\gamma, b_i) > 0.4, \slashed{E_T}<30~\rm GeV, |\eta_{h}|<2.5$  & $0.38$ &$0.68$& $0.38$  & $0.42$ \\\hline
		
		$\Delta R(\gamma, h) > 2.4$  & $0.97$ & $0.99$ & $0.98$ & $0.98$\\\hline
		
	\end{tabular}
	\caption{The selection criteria applied for $Z(\to b\bar{b})+$photon production at $\sqrt{s}=13$ TeV. The signal corresponds to an ALP mediated process of $Z(\to b\bar{b})+$photon production, with $c_{\tilde{G}}=1.25, c_{\tilde{W}}=-c_{\tilde{B}}=1$  and $f_a = 5$ TeV.}
	\label{tab:zgamma_cutflow}
\end{table*}
\vspace{-3mm}
\subsubsection{$pp \to WW$}
\label{13tev_ww}
\vspace{-3mm}
The ALP mediated production of  $WW$ via the gluon-gluon fusion depends on only one bosonic operator $\mathcal{O}_{\tilde{W}}$ and the ALP-gluonic operator $\mathcal{O}_{\tilde{G}}$.  We consider final states where one $W$ decays leptonically $(e\nu$ or $\mu \nu)$ and the other $W$ decays hadronically. The fully leptonic decay channel has been recently studied in Ref.~\cite{Carra:2021ycg}. Although the hadronic decay channel of a vector boson is overwhelmed by the presence of background processes with significantly large cross-sections, it has a larger branching fraction than the leptonic decay channel. It also allows a full kinematic reconstruction of the diboson system $(W^{\rm lep}+W^{\rm had})$, using the $W$ mass to constrain the combined four-momentum of the lepton and neutrino. The semileptonic final state, therefore, offers a good balance between efficiency and purity.  Since the effects of the ALPs are most dramatic at high  momenta of vector boson, we consider highly Lorentz-boosted vector bosons where the hadronization products of the two final state quarks overlap in the detector to form a single, large-radius jet. Dominant backgrounds to this signal come from SM processes: $W +$ jets (and the W decaying leptonically), $t \bar{t}$ (semi-leptonic mode),  single top quark production ($t (\bar{t})\, j, \,\, tW$),  $W^+ W^- + $ jets ($W \rightarrow l \nu$, $ W \rightarrow j j $),  $t \bar{t} W +$ jets (when both top quarks decay hadronically and $W \rightarrow l \nu$) and  $W Z$ (with $W \rightarrow l \nu$, $Z \rightarrow j j )$. The event reconstruction and event selection criteria are based on Ref.~\cite{CMS:2019ppl}. To reject other subdominant backgrounds from Drell-Yan and fully leptonic $t \bar{t}$ events, we reject events that contain more than one lepton. Jets are clustered by the anti-kT algorithm with radius parameter $R=0.8$ and required to have a hard $p_T > 200$ GeV. The $\vec{p}_{T}^{miss}$ is required to be larger than $110~\rm{GeV}$ to reject QCD multijet background events.

\begin{figure*}[ht!]
\begin{center}
\includegraphics[width=8.9cm,height=6.4cm]{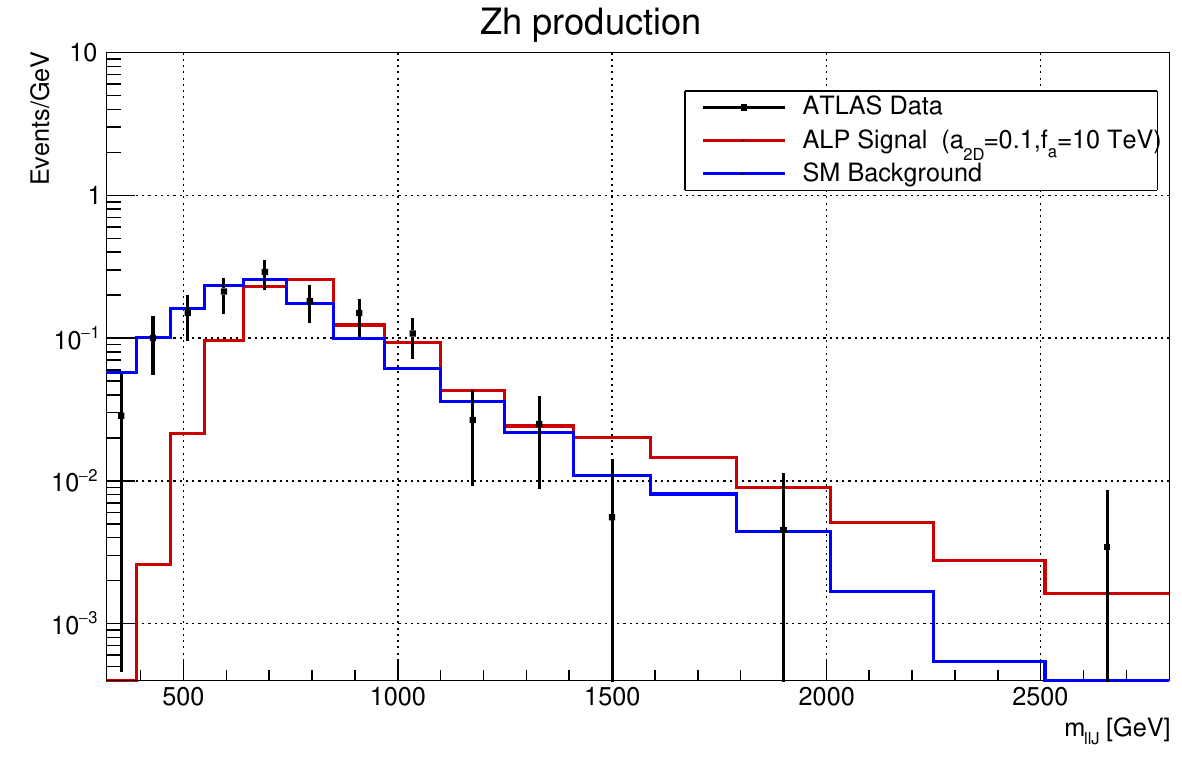}
\includegraphics[width=8.9cm,height=6.4cm]{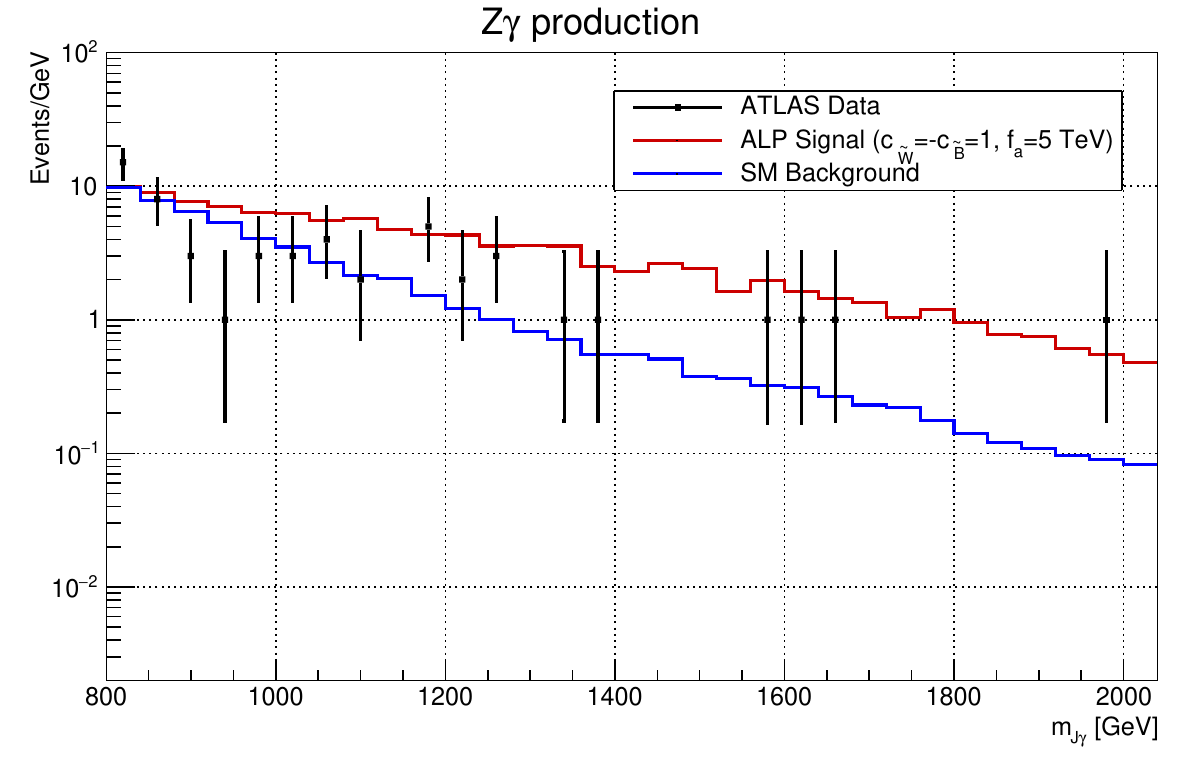}\\
\vspace{-2mm}
(a)\hspace{85mm}(b)\\
\includegraphics[width=8.9cm,height=6.4cm]{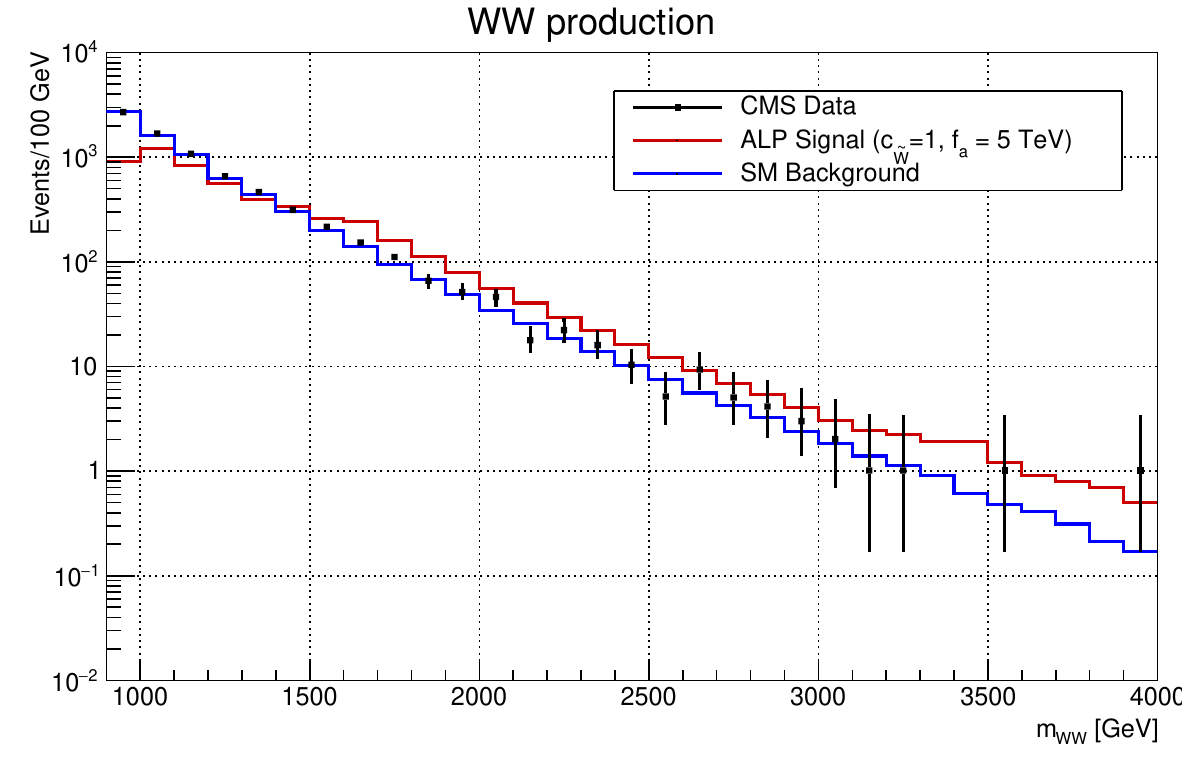}
\includegraphics[width=8.9cm,height=6.4cm]{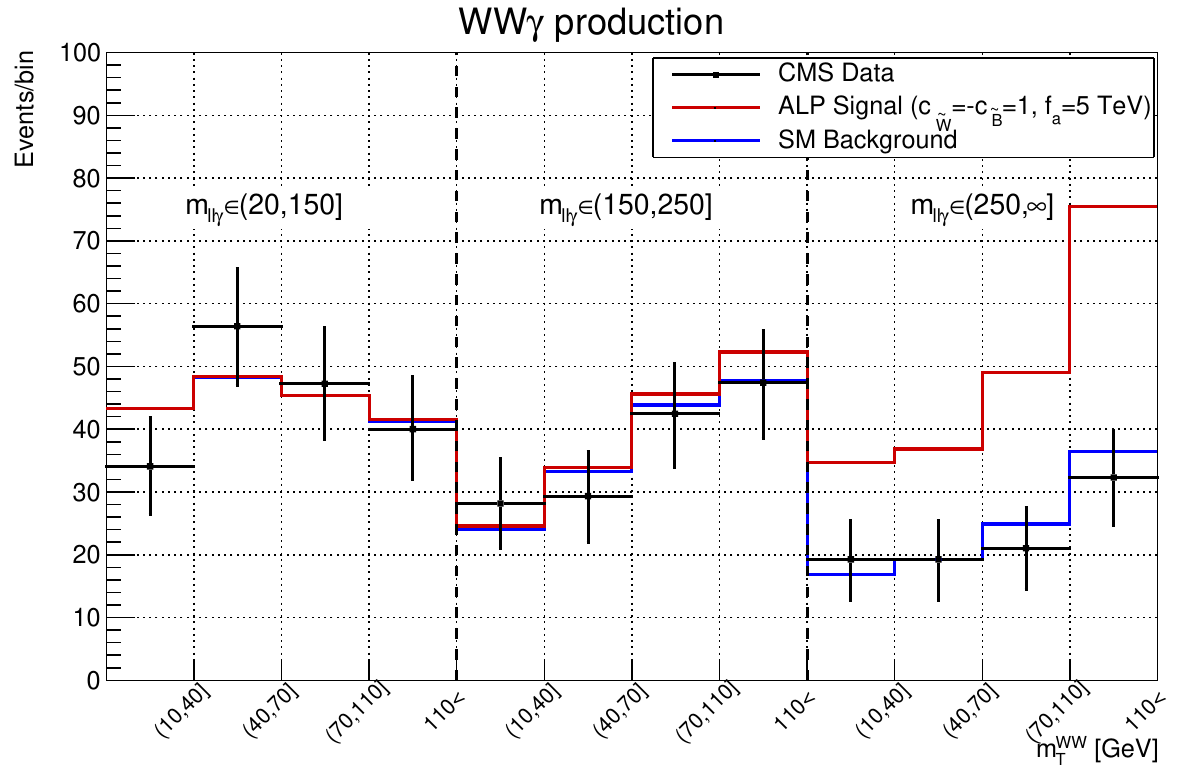}\\
(c)\hspace{85mm}(d)
\caption{(a) The differential distribution of events at 13 TeV LHC and an integrated luminosity of 139 fb$^{-1}$ with respect to reconstructed $m_{Zh}$ for the SM+ALP signal (red line) as well as the total SM background and the data (black dots with error bars) as given by the ATLAS measurement in Ref~\cite{ATLAS:2022enb}. The signal corresponds to coefficients $a_{2D}=0.1 $ and $f_a = 10$ TeV with $g_{agg}=1$ TeV$^{-1}$ (b)  Invariant mass distribution of $J\gamma$ with 36.1 fb$^{-1}$ data at 13 TeV run of LHC. The total SM prediction (blue line) and the data are taken from the analysis by the ATLAS collaboration in Ref.~\cite{ATLAS:2018sxj}. The signal (red line) correspond to coefficients $c_{\tilde{W}}=-c_{\tilde{B}}=1$  and $f_a = 5$ TeV  with $g_{agg}=1$ TeV$^{-1}$, (c) $m_{WW}$ distribution in $1 \ell+J+\slashed{E}_T$ channel, incorporating data points and total SM background from Ref.~\cite{CMS:2019ppl} by the CMS measurement at 13 TeV and an integrated luminosity of 35.9 fb$^{-1}$. The solid red line represents the ALP signal for $c_{\tilde{W}}=1 $ and $f_a = 5$ TeV  with $g_{agg}=1$ TeV$^{-1}$  and (d) Comparison of ALP signal events ($c_{\tilde{W}}=-c_{\tilde{B}}=1 $ and $f_a = 5$ TeV  and $g_{agg}=1$ TeV$^{-1}$) and the total SM expectation along with the CMS measurement data points for the transverse mass distribution of $WW$ system from the production of $WW\gamma$ in $2 \ell+ \slashed{E}_T+ \gamma$ channel~\cite{CMS:2023rcv} at 13 TeV and 138 fb$^{-1}$ integrated luminosity.}
\label{13tev_validation}
\end{center}
\end{figure*}

The leptonic $W$ boson candidate is reconstructed from the lepton and the $\vec{p}_{T}^{miss}$. The longitudinal momentum of the neutrino can be solved for by appying the $W$ boson mass constraint, assuming that the neutrino is the sole contributor to  ${p}_{T}^{miss}$. Here, we follow the CMS analysis method~\cite{CMS:2019ppl}. The transverse component of the neutrino momentum comes directly from the $\vec{p}_{T}^{miss}$. Fixing the mass of the $W$ boson candidate to its pole mass value, one can relate the four-momentum of the $W$ boson to those of the lepton and neutrino via a quadratic equation, which can have two real or complex solutions. In  case of two real solutions, the solution with the smaller absolute value is assigned as the neutrino longitudinal momentum, whereas in  case of two complex solutions, the real part common to both is instead assigned. The leptonic and hadronic boson candidates are combined into a diboson system by adding their four-momenta. Because the signal events are expected to have a back-to-back topology in the detector, we require events in the signal region to satisfy the following criteria: $\Delta R (J, \text{lepton}) > \pi/2$, $\Delta \phi (J, \vec{p}_{T}^{miss}) > 2$ and $\Delta \phi (J, W^{\rm lep}) > 2$, where $W^{\rm lep}$ denotes the reconstructed leptonic $W$ boson candidate. Additionally, we require $m_{WW} > 900~\rm{GeV}$ to isolate the signal events. The CMS collaboration presents a measurement of the $m_{WW}$ distribution in the 1 lepton+1 fat-jet+missing energy channel, employing a dataset of $35.9$ fb$^{-1}$ integrated luminosity from the  Run II LHC~\cite{CMS:2019ppl}. This analysis spans $m_{WW}$ bins up to 4 TeV.  The invariant mass of the reconstructed diboson system, $m_{WW}$, is the chosen event variable for the signal extraction. The comparison of the ALP signal with CMS data is illustrated in Fig.~\ref{13tev_validation} (c).
\subsubsection{$pp \to WW\gamma$}
\label{13tev_wwgamma}
\vspace{-3mm}

  We now consider the non-resonant ALP mediated production of triboson states of $W^+W^- \gamma$ from $pp$ collisions and both the $W$ bosons decaying leptonically. We find that even for an elusive ALP mass of $m_a < 100$ GeV, the process $W^+W^- \gamma$ deviates from the SM case as it gets modified due to the presence of ALP-gluon coupling and ALP-bosonic couplings $\{g_{agg},g_{aZ\gamma},g_{aWW},g_{aWW\gamma},g_{a\gamma\gamma}\}$. Both the couplings $g_{aWW}$ and $g_{aWW\gamma}$ depend on one parameter $c_{\tilde W}$ while couplings $g_{aZ\gamma}$ and $g_{a\gamma\gamma}$ depend on $c_{\tilde B}$ also along with $c_{\tilde W}$. 
The event reconstruction and event selection criteria are based on Ref.~\cite{CMS:2023rcv}. We look into final states with two different flavour, opposite sign (DFOS) leptons and  one photon along with $\slashed{E}_{T}$. Amongst the existing analyses for the same final state carried by the experimental collaborations, the CMS analysis has recently reported the first observation of SM $W^+W^-\gamma$ production in the leptonically decay channel~\cite{CMS:2023rcv} and hence, we reinterpret this measurement for our analysis to constrain the new physics parameter space. Although the cross-section for the  ALP signal in the $2\to3$ process is small ($\sim \mathcal{O}(1)$ fb for $f_a\sim \mathcal{O}(1)$ TeV and $m_a<100$ GeV), but the SM backgrounds for this channel are also small. The main SM backgrounds arise from $WW\gamma,WZ\gamma,Z\gamma$ and $t\bar{t}\gamma$ and processes with non-prompt leptons and photons. The final state events comprise of a photon having a transverse momentum of $p_{T}^\gamma > 20$ GeV and $|{\eta^\gamma}|<2.5$. There should be exactly one pair of DFOS leptons requiring $|{\eta^{l}}|< 2.5$  and $p_{T}^{l} > $ 20 $\rm{GeV}$. We also require  ${p}_{T}^{\rm{miss}} > 20~\rm{GeV}$. To minimise backgrounds from $WZ\gamma$ and relevant top quark processes, events are rejected that contain an additional lepton with $p_{T} > 10~\rm{GeV}$ or  at least one $b$-jet. The photon and the lepton must be well separated, such  that $\Delta R(l, \gamma) > 0.5$. To further suppress background contributions, we impose specific criteria:  the dilepton invariant mass ($m_{ll}$) $>10$~\rm{GeV}, the dilepton transverse momentum ($p_{T}^{ll}$) $>15$~\rm{GeV} and the transverse mass, $m_{T}^{WW} = \smash[b]{\sqrt{2 p_{T}^{ll} \vec{p}_{T}^{miss} [1 - \cos\Delta\phi(p_{T}^{ll}, \vec{p}_{T}^{miss})]}} >10~\rm{GeV}$. The observed distributions of $m_{T}^{WW}$ in the bins of the invariant mass of dilepton-photon system ($m_{ll\gamma}$)  are compared with the ALP signal (as shown in Fig.~\ref{13tev_validation} (d) for one such benchmark case of ALP scenario) to derive constraints on its couplings.

\vspace{-3mm}
\subsection{Fits to EFT coefficients}
\label{fitting_procedure}
\vspace{-3mm}
We take the experimental measurements in Table~\ref{tab:data} as input and our theoretical expectations for the observables in the ALP model. For the $Zh$ and $Z\gamma$ channels, we quantify the effects of the Wilson coefficients in the ALP EFT from a simplified binned likelihood ratio analysis. The likelihood function, constructed as a product of binned Poisson probabilities can be expressed as :
\begin{eqnarray}
L(\mu) = \prod_k \,e^{-(\mu s_k +\, b_k)}\, \frac{(\mu s_k + b_k)^{n_k}}{n_k !} \,,   
\label{likelihood_NS}
\end{eqnarray}
where $s_k$, $b_k$ and $n_k$ denote respectively the number of ALP signal, SM background and observed data events in a given   bin $k$, and the signal strength modifier $\mu$ involves the ALP signal couplings $(c_i/f_a)$ and is the only variable parameter in the likelihood function, with no systematic uncertainties considered for simplicity (for details see Ref.~\cite{Brivio:2017ije}). $L(\mu)$ is maximised for no ALP signal events and corresponds to the background-only hypothesis. It is tested against the combined background and signal hypothesis. No significant excess was observed by the experimental data with respect to the SM expectations. ALP couplings $c_{i}/f_a$ are considered excluded at 95\% C.L. when the negative log-likelihood (NLL) $(-\log L)$ of the combined signal and background hypothesis exceeds 3.84/2 units the NLL of the background-only hypothesis.

For the $WW$ and $WW\gamma$ channels, we perform a $\chi^2$ fit to the data  including systematic errors but no correlations between the bins. The $\chi^2$  function of the Wilson coefficients is minimised to find the best fit value of $c_i/f_a$ and the 95\% C.L. intervals are obtained by requiring $\Delta \chi^2=\chi^2-\chi^2_{\rm min} \le 3.84$.


The bounds extracted from these four process analyses constrain the products, $g_{agg}g_{aV_1V_2}$ and  $g_{agg}g_{aZh}$. For the $Zh$  process, we obtain $g_{agg} a_{2D}<0.075$~TeV$^{-2}$ at 95\% C.L. The limits on the coupling product $g_{agg} g_{aWW}$ at 95\% C.L.  are determined to be :  $g_{agg} g_{aWW}<0.59$~TeV$^{-2}$  from $WW$ analysis and $g_{agg} g_{aWW}<0.27$~TeV$^{-2}$  from $WW\gamma$ analysis. In addition, the $WW\gamma$ process induces a four point interaction of $aWW\gamma$ and the analysis puts a constraint on it of $g_{agg} g_{aWW\gamma}<0.18$ TeV$^{-2}$. The $Z\gamma$ process analysis yields a 95\% C.L. exclusion limit of $g_{agg} g_{aZ\gamma}<0.24$ TeV$^{-2}$.  These limits can be interpreted as constraints on $g_{aV_1V_2}$, assuming a constant $g_{agg}$ value of 1 TeV$^{-1}$. A smaller $g_{agg}$ would result in more stringent limits on $g_{aV_1V_2}$. It is noteworthy that these operator coefficient bounds are more significantly constrained by the higher energy data bins.
\vspace{-0.5cm}
\subsection{Validity of EFT}
\label{eft_validity}
\begin{figure}[b]
	\centering 
	\includegraphics[width=8.8cm]{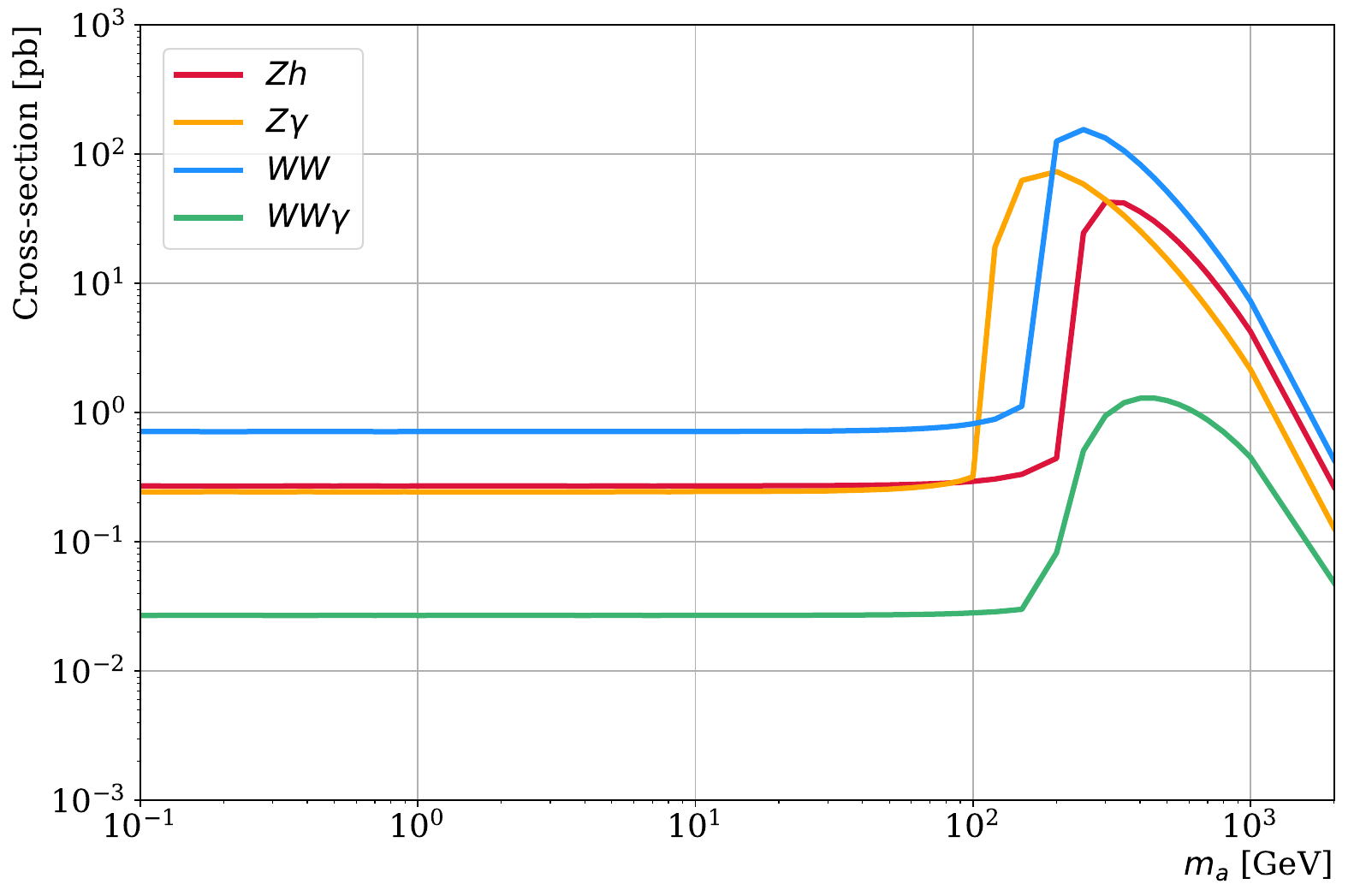}
	\includegraphics[width=8.8cm]{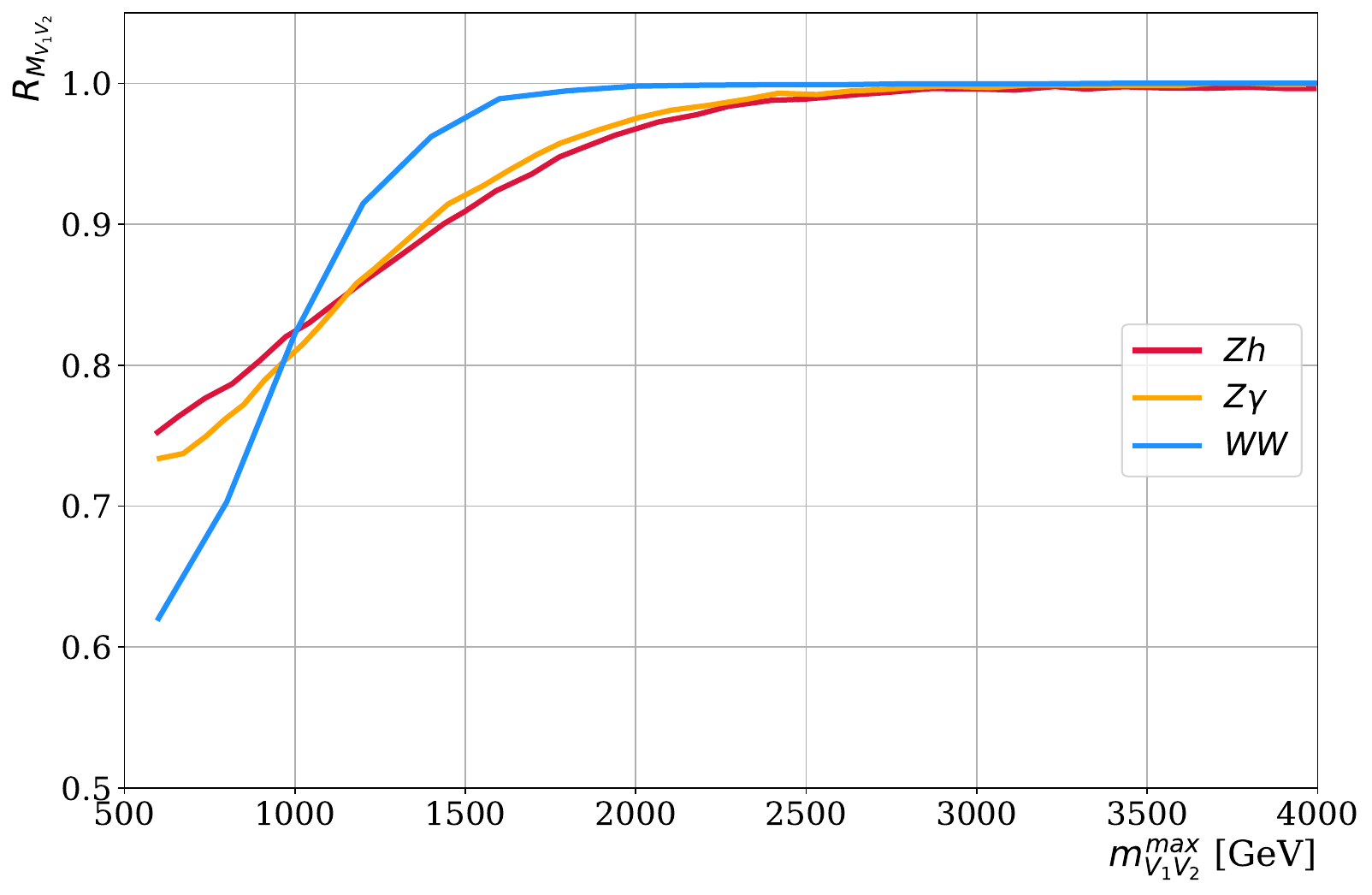}
	\caption{{\it Left: }Total cross-sections at $\sqrt{s} = 13~\rm{TeV}$ for the ALP contributions to the different scattering processes as a function of the ALP mass. The value of $f_a$ in each of these processes is taken to be $4$ TeV and $c_{\tilde G}=1$. The `$Zh$' curve is evaluated at $a_{2D}=1$ and `$WW$' curve at $c_{\tilde{W}}=1$. For the `$Z\gamma$' and `$WW\gamma$' cases, they are evaluated at $c_{\tilde{W}}=1,~c_{\tilde{B}}=-0.305$. At each point in the plot, the ALP decay width was re-estimated as a function of $m_a$ and the Wilson coefficients. {\it Right: } The ratio $R_{M_{V_1V_2}}$ variation as a function of maximum invariant mass of the final state system in $Zh$ (red), $Z\gamma$ (yellow) and $WW$ (blue) production processes.}
	\label{xsec_mass_validity}
\end{figure}

\vspace{-3mm}
In this subsection, we discuss the validity of our theoretical expectations discussed. As we explore the non-resonant $s$-channel ALP signatures, they have several interesting characteristics that could potentially benefit the detection , sensitivity information on its couplings and calls for further study.  When the momentum transfer through the ALP propagator ($p_a$) obeys $\sqrt{|\vec{p}_a|^2}\gg m_a, \Gamma_a$ where $\Gamma_a$ is its decay width, the cross-section and differential distribution of the ALP signal remain largely independent of actual value of $m_a$. This implies that our search strategy retains its validity over a wide range of ALP masses, particularly those significantly below the energy scale of the experiment. For the LHC searches we investigated, this translates into a consistent detection capability for ALP masses below 100 GeV. Fig.~\ref{xsec_mass_validity} (left panel) verifies the off-shell approximation for the processes. It shows the ALP signal cross-section at $\sqrt{s} = 13~\rm{TeV}$, applying the cuts defined in Eqn.~\eqref{preselection}, plotted against $m_a$ for fixed values of $a_{2D}$, $c_{\tilde{W}}$, $c_{\tilde{B}}$ and $f_a$. Here, $\Gamma_a$ is implicitly computed at each point which is dependent on $m_a$ and the ALP couplings, following the relation $\Gamma_a\propto (c_i/f_a)^2 m_a^3$. The lines running almost parallel with the mass of the ALP in Fig.~\ref{xsec_mass_validity} (left panel) confirm that our simulations are relevant even for small  values of $m_a$ and till about $100$ GeV. We perform the analyses on the assumption that the ALP contributes only off-shell in all the processes we considered, setting the ALP mass and decay width in our simulations at $m_a=1~{\rm {MeV}}, \Gamma_a=0$.

As the mass $m_a$ increases, the cross-sections for processes of $Zh, Z\gamma, W^\pm W^\mp$ and $W^\pm W^\mp\gamma$  show a resonance effect when the propagator becomes predominantly influenced by the ALP mass. This is particularly noticeable for all the processes. The chosen point values for $c_i,  f_a$ facilitate resonant ALP exchange in the $Z\gamma, W^\pm W^\mp, W^\pm W^\mp\gamma$ at masses above 150 GeV, close to 250 GeV and around 400 GeV, respectively. The slight shifts in the $WW\gamma$ and $Z\gamma$ processes can be attributed to the photon $p_T$ preselection cut. We evaluated the $Z\gamma$ channel at a point $(c_{\tilde{W}}=1, c_{\tilde{B}}=-0.305)$ to ensure a ``photophobic'' interaction (where $g_{a\gamma\gamma}=0$) and to explore the resonant effect induced by $g_{aZ\gamma}$ coupling. In the $Zh$ process, the resonance effect is apparent near 300 GeV. These observations serve as a validation  that our results hold for ALP masses up to approximately $100~\rm{GeV}$. At this mass, the cross-sections for all four mentioned processes deviate by less than 5\% from their asymptotic values when $m_a$ approaches zero.

Furthermore,  an important feature of the non-resonant process is its lack of dependence on specific assumptions about extra couplings not directly contributing to the process  and any other model specific parameters. This is in contrast to on-shell analyses, which are usually limited to particular mass and width ranges and where the impact of extra ALP couplings becomes evident in their partial decay widths. Conventionally, studies on ALP limits from resonant processes have focused on a single independent $g_{a V_1 V_2}$ coupling, as outlined in Eqn.~\eqref{eq:lagrangian_gs}~\cite{Mimasu:2014nea,Jaeckel:2015jla,Knapen:2016moh,Dobrich:2015jyk,Izaguirre:2016dfi}. However, recent studies have started to explore scenarios incorporating two or three independent couplings simultaneously~\cite{Mariotti:2017vtv,Alonso-Alvarez:2018irt,Gavela:2019wzg}. Thus, the model-independence of non-resonant searches is evident, making them more effective in detecting new physics phenomena.

Estimating the validity of the EFT expansion is crucial for collider bounds, especially given the broad range of energies encountered at hadron collider experiments. We now consider the range of Wilson coefficients constrained and check whether they allow for a valid EFT interpretation of cross-sections. Theoretically, the $g_{a V_1 V_2}$ couplings depend only on the ratio $c_i/f_a$ (as detailed in Eqns.\eqref{eq:gagluon}-\eqref{eq:gaZgamma}). However, the value of $f_a$ is important in assessing the validity of EFT, which in turn restricts the energy range feasible for LHC searches, such as energy bins where $\sqrt{\hat{s}} < f_a$. If the underlying BSM theory operates in a weak coupling regime, leading to the operators in Eqn.~(\ref{operators}) at one-loop, the coefficients might be attenuated by an additional $ 16\pi^2$ factor. This would considerably restrict the valid energy bins for LHC searches.

For illustration, in the $Z\gamma$ production process where $Z$ decays into two bottom quarks and is detected as a fat-jet, the energy scale of the collision is determined by the invariant mass of the jet-photon system, $m_{J \gamma}$. The $Z\gamma$ EFT expansion validity can be maintained by ensuring $m_{J \gamma}$ stays below the cut-off scale, $f_a$. However, precisely defining the EFT cut-off scale in a model-independent manner is difficult without taking into account the specific details of the underlying UV-complete theory. To have an idea of the cut-off scale, we adopt a methodology based on Refs.~\cite{Busoni:2013lha,Bhattacharya:2015vja}. If $m_{J \gamma}$ is consistently smaller than $f_a$ in most collisions, the ratio $R_{M_{V_1V_2}}$ (where `$V_1V_2$'  refers to `final state bosons'), defined below would tend to unity.

\begin{equation}
R_{M_{V_1V_2}} \equiv \frac { \int^{m_{J \gamma} < {m ^{max} _{J \gamma}}} \frac{d \sigma} {d m_{J \gamma}}\; d m_{J \gamma} }  { \int \frac{d \sigma} {d m_{J \gamma}}\; d m_{J \gamma} }
\label{validity_equation}
\end{equation}

In Fig.~\ref{xsec_mass_validity} (right panel), we see this effect in the process at $\sqrt{s}=$13 TeV, involving non-zero EFT couplings, whose values are set at the limits obtained at 95\% C.L. As for the $Z\gamma$ process,  ratio of $R_{M_{V_1V_2}}$ close to 1 suggests that the energy exchange in the process remains considerably below ${m ^{max}_{J \gamma}}$, the maximum value allowed for  $m_{J \gamma}$. Identifying such peak $m_{J \gamma}$ values provides a practical reference for the EFT cutoff scale, $f_a$. Additionally, we examine similar variations for the $Zh$ and $WW$ processes, using their respective invariant mass measurements. When $m_{V_1V_2}^{max}$ ($V_1V_2$ denoting `final state bosons') is 1 TeV, for example, $20\%$ of signal events are lost and this implies that final limits are weakened. In cases involving both the Higgs chiral operator and linear bosonic operators, the ratio $R_{M_{V_1V_2}}$ is approaching unity when $m_{V_1V_2}^{max}> 2.0$ TeV and thus, more than $95\%$ of the collision events respect the EFT validity considerations.
\vspace{-5mm}
\subsection{Collider Analysis with HL-LHC probes}
\label{hllhc_probe}
\vspace{-3mm}
We will discuss the results of our cut-based analysis for a few benchmark points (BPs) to accentuate the distinguishability of the ALP signal from the backgrounds. The BPs are so chosen such that they obey  the experimental constraints obtained from the 13 TeV data. The selected benchmark points are listed in Table~\ref{BM_points}. As some of the operator coefficients probe more than one process at a time, we choose these points to highlight specific regions of parameter space so that they probe one effective coupling at time for a specific process as detailed below.
\begin{table}[b]
	\centering
	\renewcommand{\arraystretch}{1.5}
	\begin{tabular}{|c|c|c|}
		\hline
		Signal & Coupling parameter  & Process \\ \hline \hline
		BP1 & \rm $a_{2D}=0.2$ TeV$^{-1}$, $c_{\tilde{G}}=1.0$ TeV$^{-1}$, $f_a=5$ TeV & $pp \to Zh$ \\ \hline
		BP2 & \rm $c_{\tilde{W}}=0.5$ TeV$^{-1}$, $c_{\tilde{B}}=-0.5$ TeV$^{-1}$, $c_{\tilde{G}}=1.0$ TeV$^{-1},~f_a=5$ TeV& $pp \to Z\gamma$\\ \hline
		BP3 & \rm $c_{\tilde{W}}=0.5$ TeV$^{-1}$, $c_{\tilde{B}}=-1.639$ TeV$^{-1}, ~c_{\tilde{G}}=1.0$ TeV$^{-1},~ f_a=5$ TeV& $pp \to Z\gamma$\\ \hline
		BP4 & \rm $c_{\tilde{W}}=0.5$ TeV$^{-1}$,$c_{\tilde{G}}=1.0$ TeV$^{-1},~ f_a=5$ TeV & $pp \to W^\pm W^\mp$\\ \hline
		BP5 & \rm $c_{\tilde{W}}=0.5$ TeV$^{-1}$, $c_{\tilde{B}}=0.5$ TeV$^{-1}, c_{\tilde{G}}=1.0$ TeV$^{-1},~  f_a=5$ TeV& $pp \to W^\pm W^\mp\gamma$\\ \hline
		BP6 & \rm $c_{\tilde{W}}=0.5$ TeV$^{-1}$, $c_{\tilde{B}}=-0.152$ TeV$^{-1}, c_{\tilde{G}}=1.0$ TeV$^{-1},~   f_a=5$ TeV& $pp \to W^\pm W^\mp\gamma$\\ \hline
	\end{tabular}
	\caption{Summary of selected benchmark points for the study}
	\label{BM_points}
\end{table}

 It is to be noted that all of these four processes depend on ALP-gluon coupling $g_{agg}$ and the relevant ALP-bosonic or -$Zh$ coupling. In the simulation for all the BPs, we choose $c_{\tilde G} = 1$ and $f_{a}=5$ TeV. 
For $Zh$ production, we have an ALP-Higgs operator that contributes at LO and we choose the corresponding operator coefficient value for $a_{2D}=0.2$. The ALP mediated  $Z\gamma$ production is induced from $g_{aZ\gamma}$ coupling which in turn receives contributions from $c_{\tilde{W}}$ and $c_{\tilde{B}}$.  We choose BP2 such that $c_{\tilde{W}}=-c_{\tilde{B}}~(g_{aZ\gamma} \neq 0)$ and BP3 such that $c_{\tilde{W}}=-c_{\tilde{B}}t_{\theta}^2,~i.e.,~g_{aZZ}=0$.  The $W^\pm W^\mp$ production receives bosonic contribution from $g_{aWW}$ coupling only and thus, depends on $c_{\tilde{W}}$. The $W^\pm W^\mp\gamma$ production receives contributions from $g_{a\gamma\gamma},g_{aZ\gamma} $ and $g_{aWW},g_{aWW\gamma}$ couplings. BP5 corresponds to $g_{aZ\gamma}=0$ while BP6 corresponds to $c_{\tilde{W}}=-c_{\tilde{B}}/t_{\theta}^2,~i.e.,~g_{a\gamma\gamma}=0$. Couplings $g_{aWW},g_{aWW\gamma}$  are proportional to $c_{\tilde{W}}$ only. Equipped with these benchmark points, we now discuss some kinematic differences between the ALP signal and the SM backgrounds for each of the above mentioned processes.

\begin{figure*}[t]
	\centering
	\includegraphics[width=8.8cm,height=6.0cm]{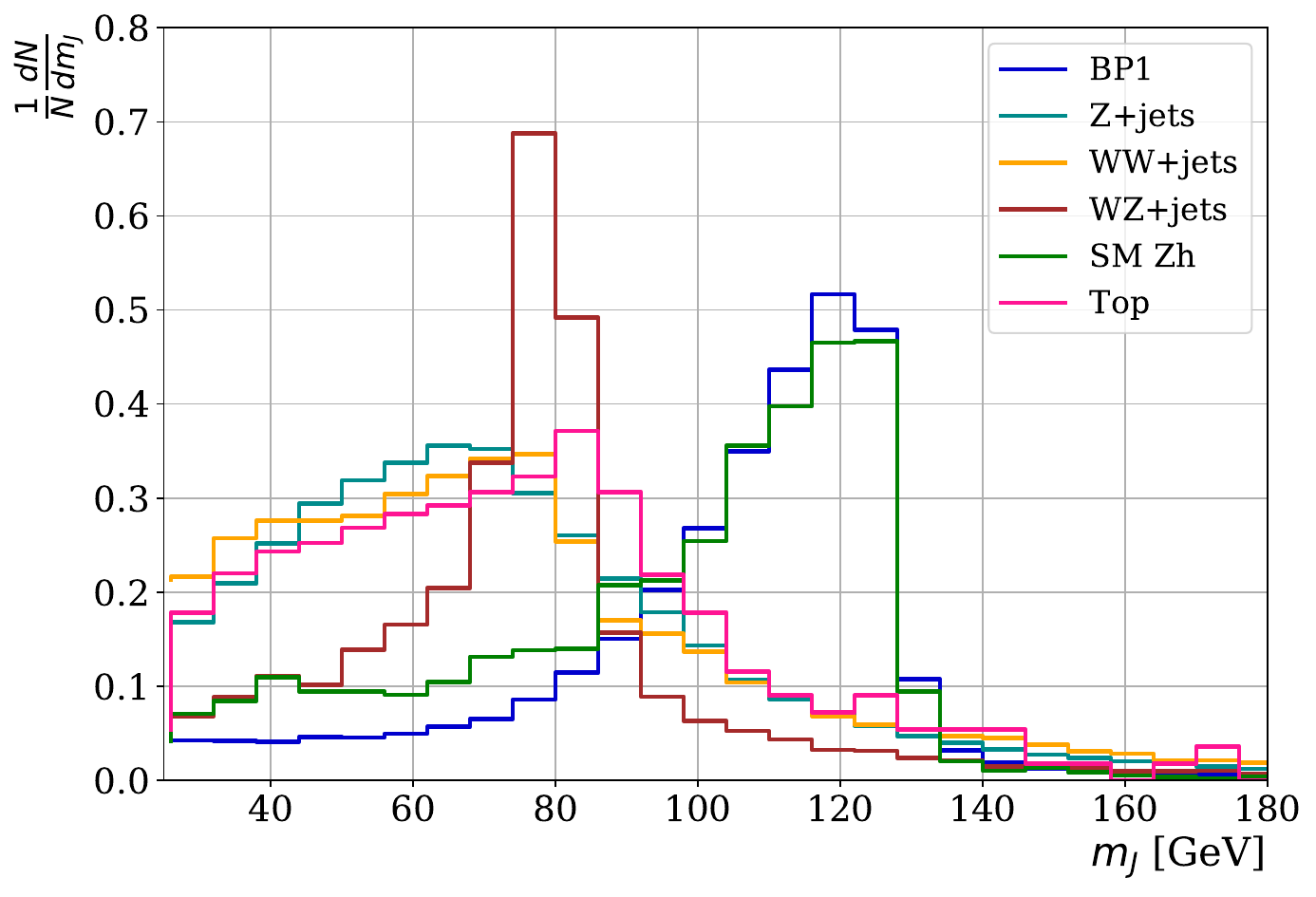}
	\includegraphics[width=8.8cm,height=6.0cm]{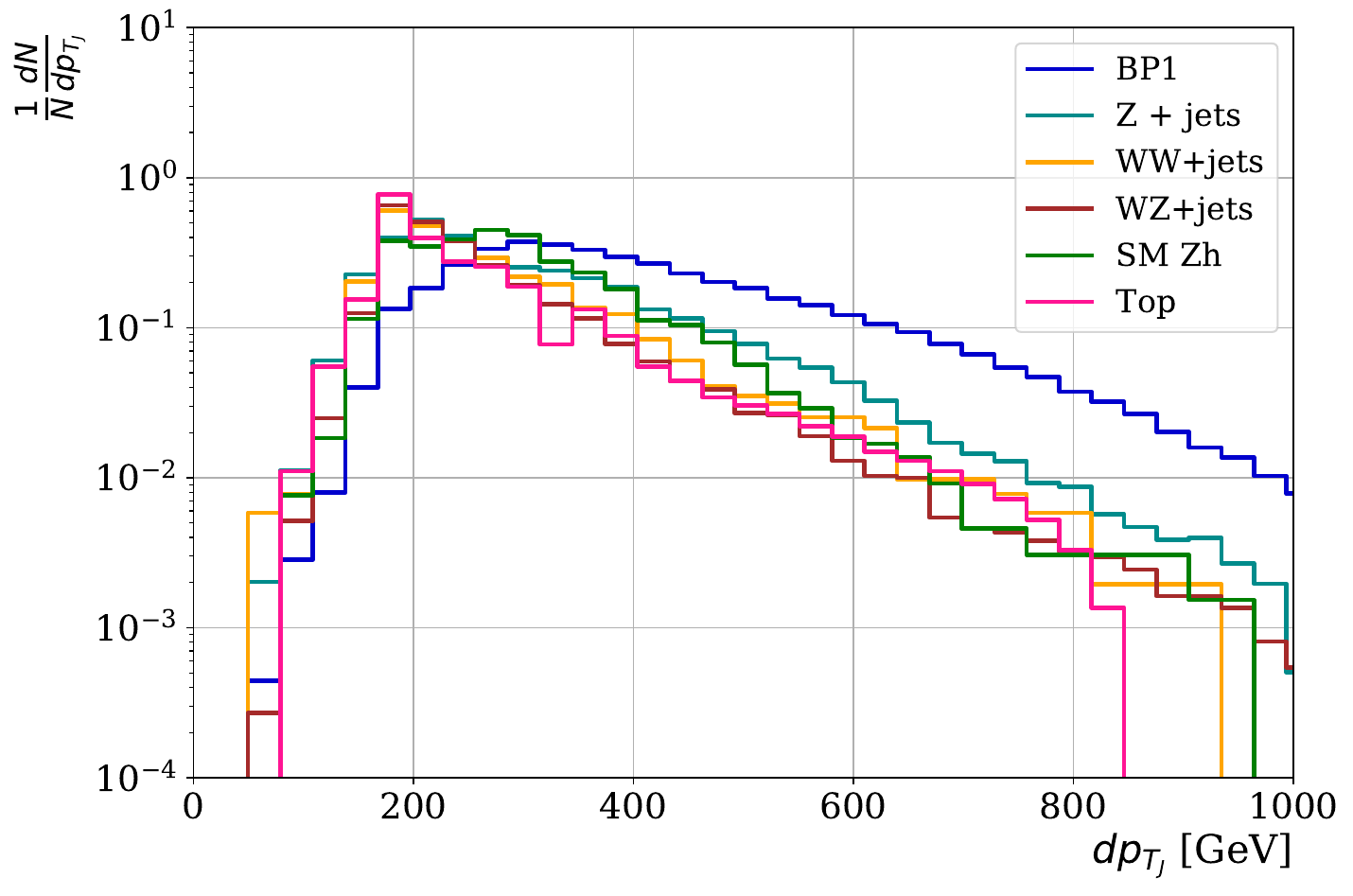}  \\
	\hspace{4mm}(a)\hspace{68mm}(b)
	\caption{Normalized distributions of (a) the mass $m_J$ and (b) the transverse momentum $p_{T_{J}}$ of the jet, both for the $pp\to hZ$ ALP mediated signal and SM backgrounds at $\sqrt{s}=14$ TeV. For the ALP mediated signal, we have chosen BP1 with $a_{2D}=0.2, c_{\tilde G}=1.0$ and $f_a = 5$ TeV (blue).}
	\label{zh_kin_distribtn1}
\end{figure*}

We first consider the Higgs-strahlung process, which has a radius $R=1$ fat-jet and two leptons in the final state. Fig.~\ref{zh_kin_distribtn1} (a) shows the  mass of the leading fat-jet for the signal BP1 and the dominant backgrounds. It is evident from the distributions that the peak around $115-140~\rm GeV$ reflects the Higgs peak for the signal process whereas for most of the backgrounds, the peaks are below 50 GeV reflecting that the fat-jet mimicing  either single prong hard QCD jet or a peak around 90 GeV reflects $Z$ boson or peak about $165-185$ GeV from a top. Numerically, the $m_J \in [115,140]$ GeV selection suppresses the $Z+$jets backgrounds by a factor of $20\%$ at the price of keeping $\sim60\%$ of the signal events.

The variable $p_{T_{J}}$ (Fig.~\ref{zh_kin_distribtn1}(b)) is quite efficient in distinguishing the new interactions from most of the SM backgrounds. The availability of larger parton center-of-mass energy in these derivative interactions pushes the transverse momentum of fat-jet ($p_{T_{J}}$) to higher values. We thus, put slightly tighter cuts on these variables compared to the 13 TeV analysis, namely, $p_{T_{J}}>250$ GeV and $115<m_{{J}}<140$ GeV. We also select events satisfying $m_{Zh}>500$ GeV.

\begin{figure*}[t]
	\centering
	\includegraphics[width=8.8cm,height=6.0cm]{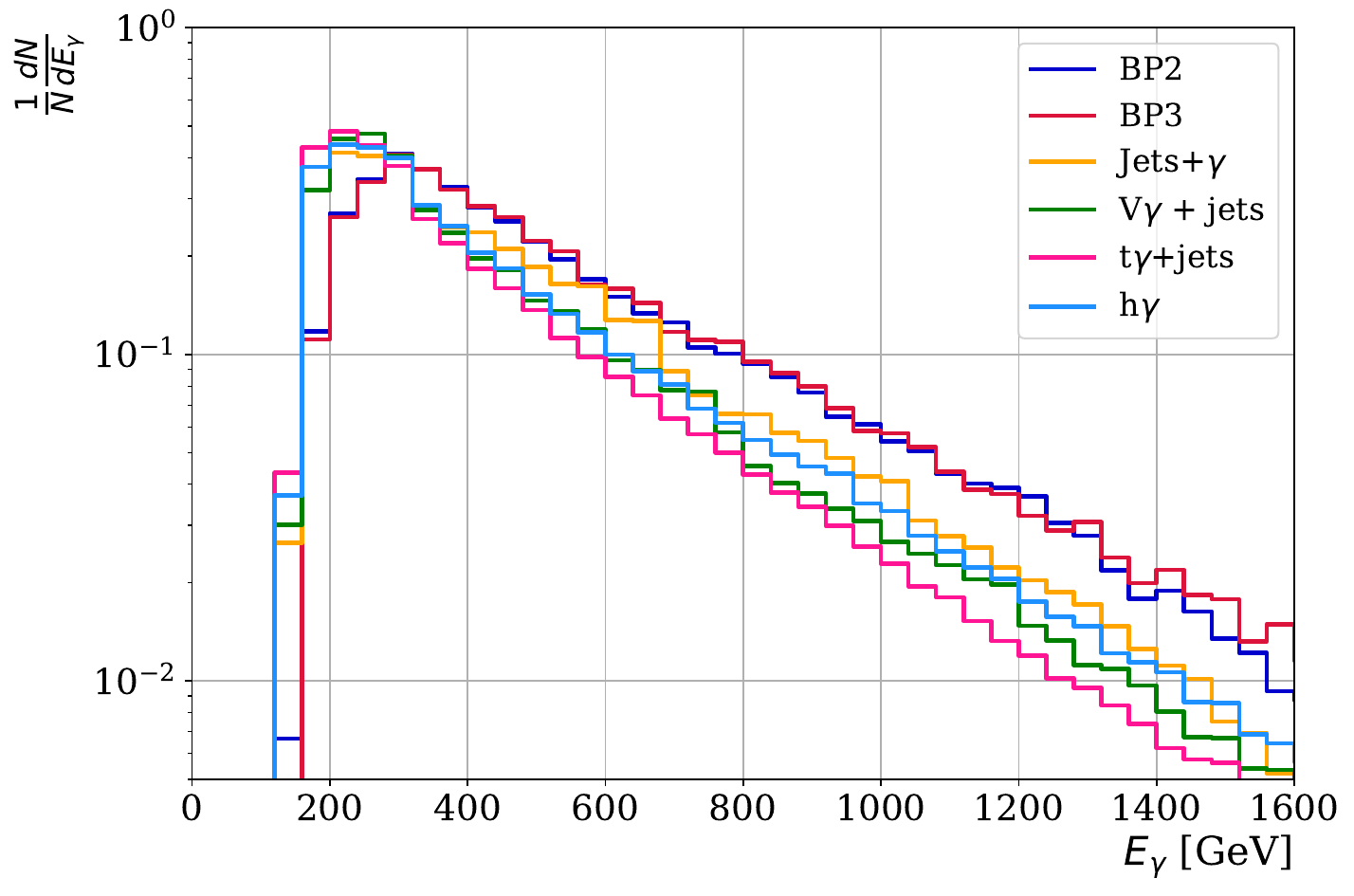}
	\includegraphics[width=8.8cm,height=6.0cm]{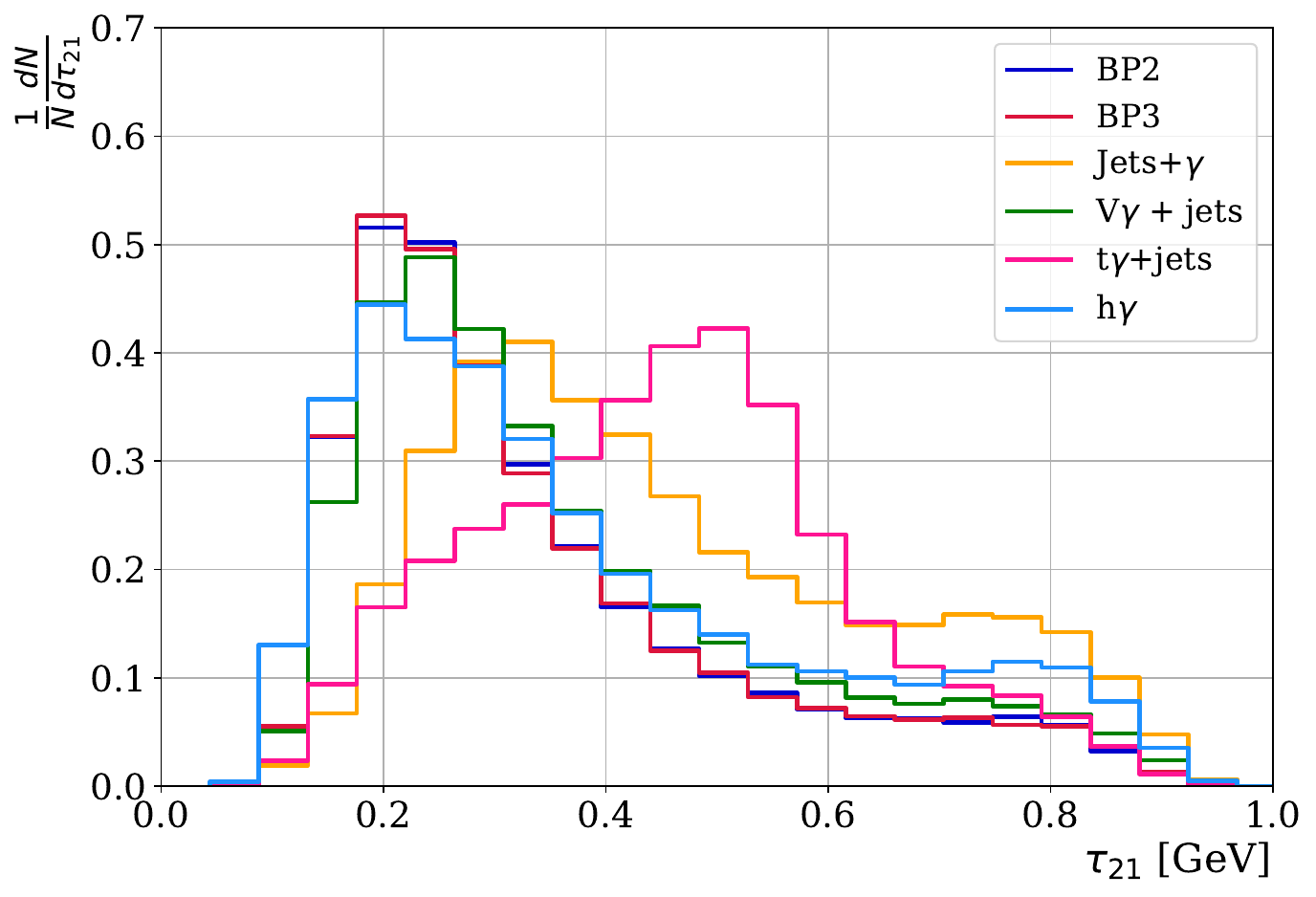}  \\
	\hspace{4mm}(a) \hspace{85mm}(b)
	\caption{Normalized distributions of (a) the transverse energy $p_{T_{J}}$ of the photon and (b) N-subjettiness of the jet, both for the $pp\to Z\gamma$ ALP mediated signal and SM backgrounds at $\sqrt{s}=14$ TeV. For the ALP mediated signal, we have chosen BP2 with with $c_{\tilde{W}}=0.5, c_{\tilde{B}}=-0.5, c_{\tilde G}=1.0,  f_a = 5$ TeV (blue) and BP3 with with $c_{\tilde{W}}=0.5, c_{\tilde{B}}=-1.639, c_{\tilde G}=1.0,  f_a = 5$ TeV (red).}
	\label{zgamma_kin_distribtn1}
\end{figure*}

For $Z\gamma$ production, the photon $\pt$ is a strong discriminator.  The distributions for the ALP signals corresponding to BP2 and BP3 and various SM backgrounds are shown in Fig.~\ref{zgamma_kin_distribtn1} (a). The photons in the signal events exhibit a hard $p_T$. Requiring an energetic photon puts a high $p_T$ threshold on the jet in recoil, above which the $Z$ boson becomes sufficiently boosted. The $E_\gamma$ distributions extend to 1 TeV. In the signal process, which despite being an s-channel process, we see significant enhancement from the SM backgrounds at high energy tails of the distribution due to the contribution of bosonic type of dimension-5 operators. In presence of the effective operators, the cross-section grows faster at higher energies compared to the SM backgrounds whose effect diminish with increasing energy.

The fat-jet resulting from the $Z \to b \bar{b}$ decay can potentially retain information about its two-pronged structure. This characteristic feature is captured by the jet-shape variable known as N-subjettiness~\cite{Thaler:2010tr,Thaler:2011gf}, which is computed as follows:
\begin{eqnarray}
\tau_N^{(\beta)} = \frac{1}{\mathcal{N}_0} \sum\limits_i p_{i,T} \, \min \left\lbrace \Delta R _{i1}^\beta, \Delta R _{i2}^\beta, \cdots, \Delta R _{iN}^\beta \right\rbrace
\label{eq:nsub_N}
\end{eqnarray}

where $N$ refers to the number of subjet axes taken within the fat-jet. The index $i$ runs over the individual jet constituents and $p_{i,T}$ represents their transverse momenta. $\Delta R_{ij}=\sqrt{(\Delta \eta)^{2} + (\Delta \phi)^{2}}$ measures the separation in the $\eta-\phi$ plane between a possible subjet $j$ candidate and a constituent particle $i$. The normalization factor, $\mathcal{N}_{0}$, is computed as $\sum_{i}p_{i,T} R_{0}$, where $R_0$ denotes the fat-jet  radius. The $\beta$ represents the angular exponent and is taken to be $1$ here. Essentially, the ratio $\tau_N/\tau_{N-1}$ serves to differentiate between jets that likely contain $N$ internal energy clusters versus those with $N-1$ clusters. Specifically in our analysis, the jet coming from the $Z$ boson is observed to exhibit smaller values for $\tau_{21}$ in comparison to typical QCD jets, a pattern evident in Fig~\ref{zgamma_kin_distribtn1}(b).  Thus, a cut of $\tau_{21}<0.45$ can reduce a significant amount of background while translating into a signal selection efficiency of $\sim 12\%$.

\begin{figure*}[h]
	\centering
	\subfloat{
		\begin{tabular}{cc}
			\includegraphics[width=8.8cm,height=6.0cm]{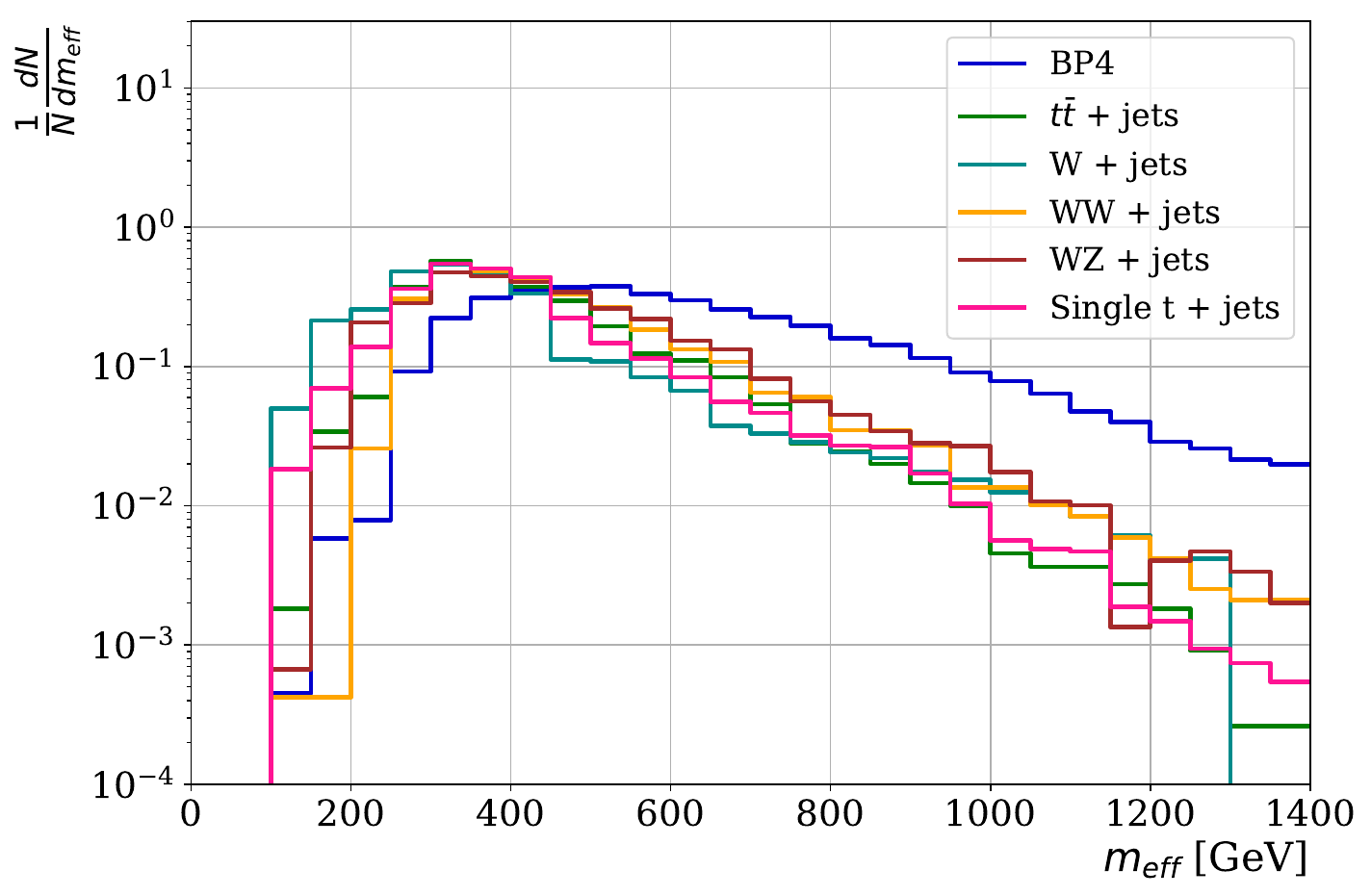}&
			\includegraphics[width=8.8cm,height=6.0cm]{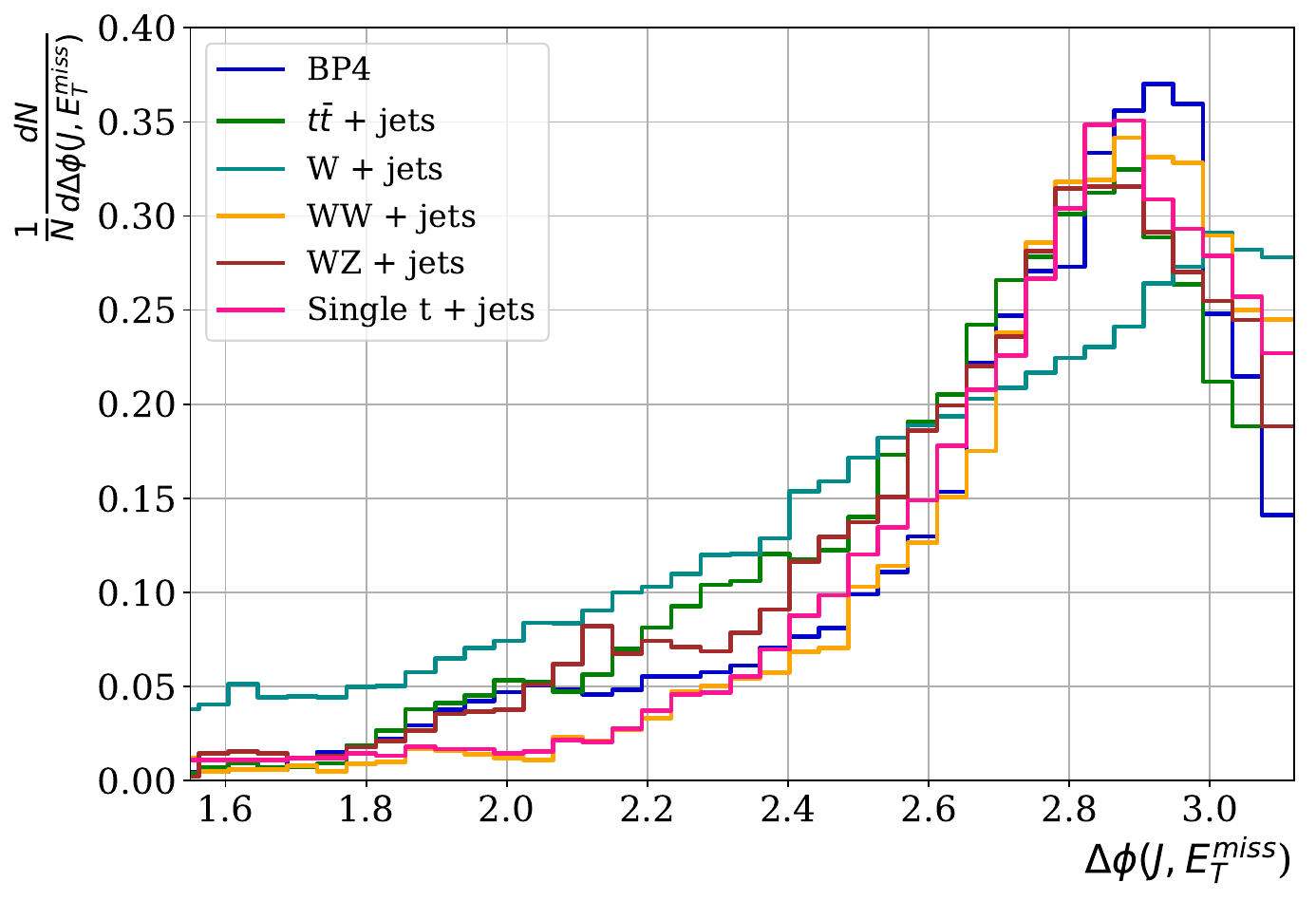} \\
			(a)&(b)
	\end{tabular}}
	\caption{Normalized distributions of (a) the effective mass  $m_{eff}$ variable and (b) $\Delta \phi(jet,\slashed{E}_T)$,  for the $pp\to WW$ ALP mediated signal and SM backgrounds at $\sqrt{s}=14$ TeV. For the ALP mediated signal, we have chosen BP4 with $c_{\tilde{W}}=0.5, c_{\tilde G}=1.0, f_a = 5$ TeV (blue).}
	\label{ww_kin_distribtn1}
\end{figure*}

We analyze the $pp \to W^\pm W^\mp$ process, in which one $W$ boson undergoes a leptonic decay and the other a hadronic decay. Here, we consider the $m_{eff}$ or the effective mass variable which is an important variable for BSM searches. It defined as follows : 
\begin{equation}
m_{eff} = \sum_{i}|\vec{p}_{T_{i}}| + \slashed{E}_T
\end{equation}
Here, $i$ encapsulates all entities in the event, including the reconstructed jets and $p_T$ refers to their transverse momenta and $\slashed{E}_T$ is the total transverse missing energy in the event. This global variable, which does not rely on specific event topology, proves to be highly useful, especially given that signal events receive a high parton-level center-of-mass energy compared to most  SM background processes. In Fig.~\ref{ww_kin_distribtn1} (a), we present the effective mass of the ALP process distribution for this channel for BP4.   It is evident that for most SM backgrounds, the distributions tend to peak at lower values than in the ALP scenario. It is to be noted that these are normalized distributions, providing qualitative insights into potential additional cuts on these variables, rather than quantitative ones. 

In Fig.~\ref{ww_kin_distribtn1} (b), we plot the $\Delta \phi(jet,\slashed{E}_T)$ distribution for both the signal and background processes. In case of the ALP signal, the $\slashed{E}_T$ is most likely to recoil against the leading jet in the azimuthal plane. Therefore, the distribution  peaks around $\sim \pi$ for the signal, and similarly for the SM $WW$ and $W+$jets backgrounds. Moreover, the veto on additional hard jets largely reduce the $WZ$, single top and $t\bar{t}$ (semileptonic) backgrounds.

\begin{figure*}[h]
	\centering
	\includegraphics[width=8.8cm,height=6.0cm]{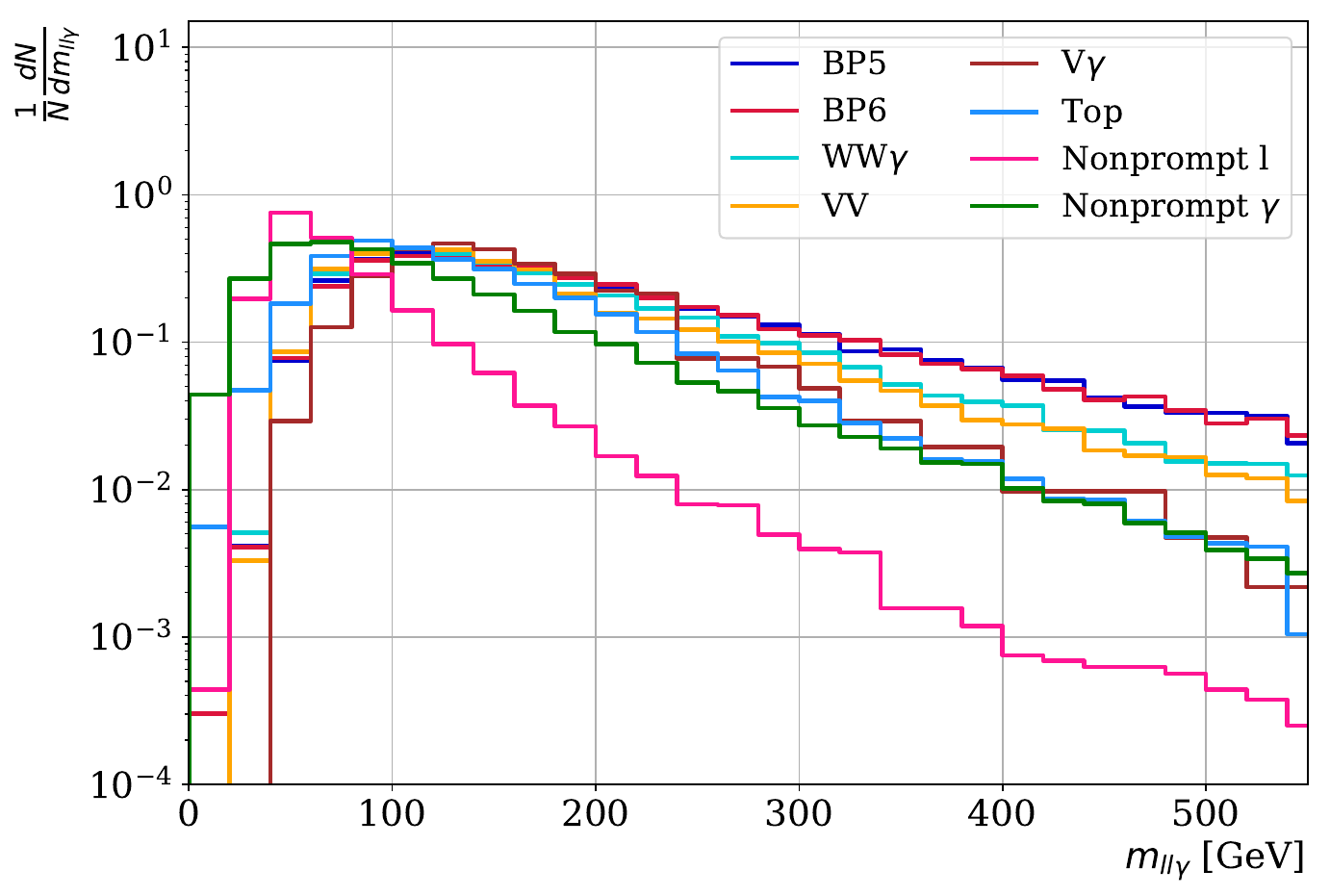}
	\includegraphics[width=8.8cm,height=6.0cm]{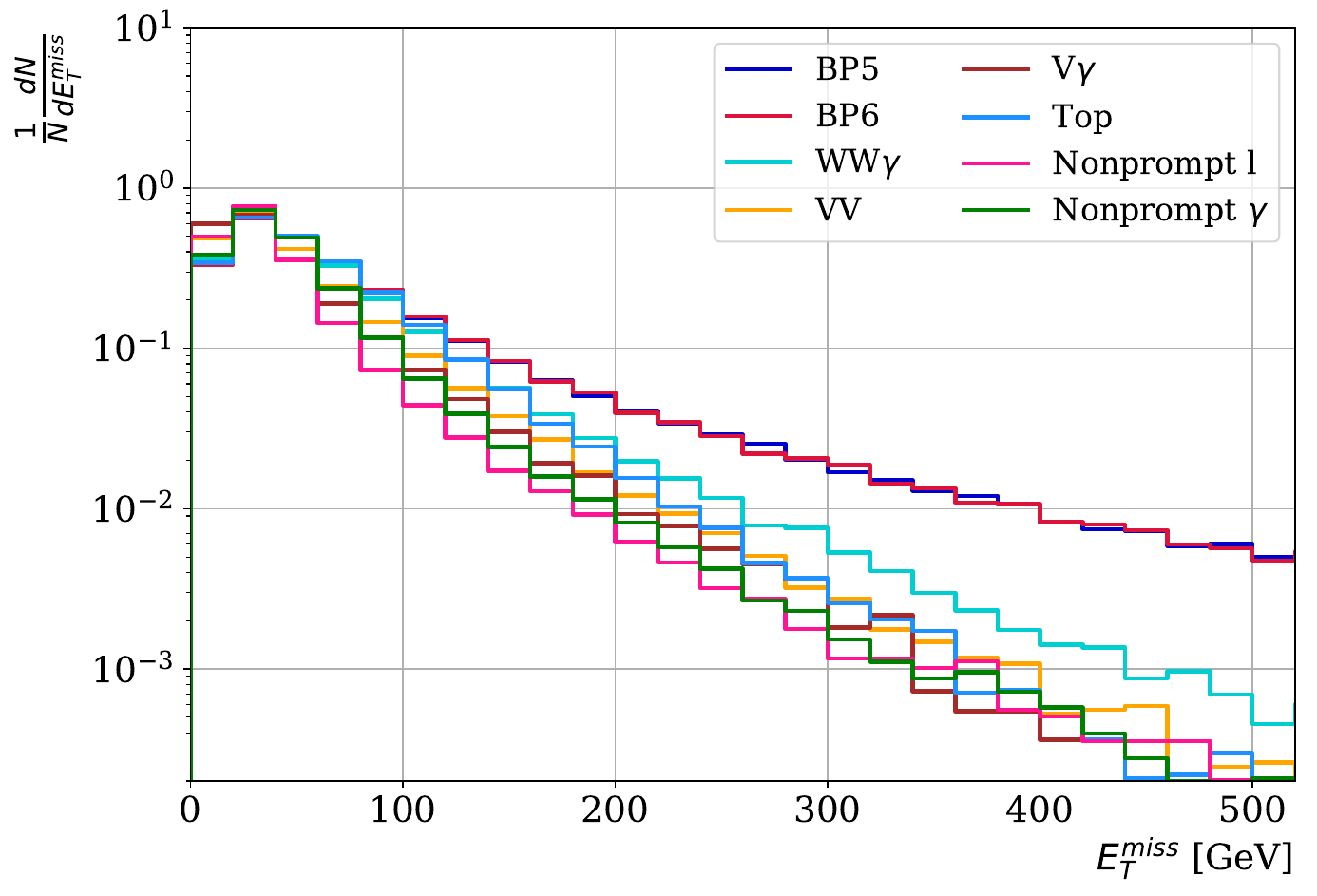}  \\
	\hspace{4mm}(a)\hspace{82mm}(b)
	\caption{Normalized distributions of (a) the invariant mass of the dilepton and photon system, $m_{ll\gamma}$ and (b) transverse missing energy $\slashed{E}_{T}$, for the $pp\to WW\gamma$ ALP mediated signal and SM backgrounds in the fully leptonic decay channel at $\sqrt{s}=14$ TeV. For the ALP mediated signal, we have chosen BP5 with with $c_{\tilde{W}}=0.5, c_{\tilde{B}}=-0.5, c_{\tilde G}=1.0,  f_a = 5$ TeV (blue) and BP6 with with $c_{\tilde{W}}=0.5, c_{\tilde{B}}=-0.152, c_{\tilde G}=1.0,  f_a = 5$ TeV (red).}
	\label{wwa_kin_distribtn4}
\end{figure*}

Next, we examine the $pp \to W^\pm W^\mp \gamma$ process in a fully leptonic channel, characterized by two DFOS leptons, a photon and missing energy in the final state. In Fig.~\ref{wwa_kin_distribtn4} (a), the invariant mass distribution of the dilepton-photon system is shown. In  case of the signal, the leptons originating from the $W$ bosons are boosted due to the influence of the ALP coupling. As a result, the distribution for the signal shows a prominent enhancement towards higher values of the invariant mass system, in contrast to the SM background processes. We also show the distribution of $\slashed{E}_T$ in Fig.~\ref{wwa_kin_distribtn4} (b) for both the SM backgrounds and the ALP signal events. It is evident that for all the benchmark points considered, the event distribution in the presence of an ALP is shifted towards increased missing transverse energy, distinguishing it from the typical SM scenarios. Thus, these variables play a significant role in isolating the ALP interaction effects in the events.
It is relevant to mention here that the kinematic distributions for BP5 and BP6 look quite similar. As the mass of the ALPs are the same and it indicates that the process receives dominant contribution from $g_{aWW}$  as we re-iterate that the benchmark points have been chosen such that BP5 leads to $g_{aZ\gamma}=0$ and BP6 leads to $g_{a\gamma\gamma}=0$.

\begin{figure*}[h!]
	\centering
	\includegraphics[width=8.9cm,height=6.0cm]{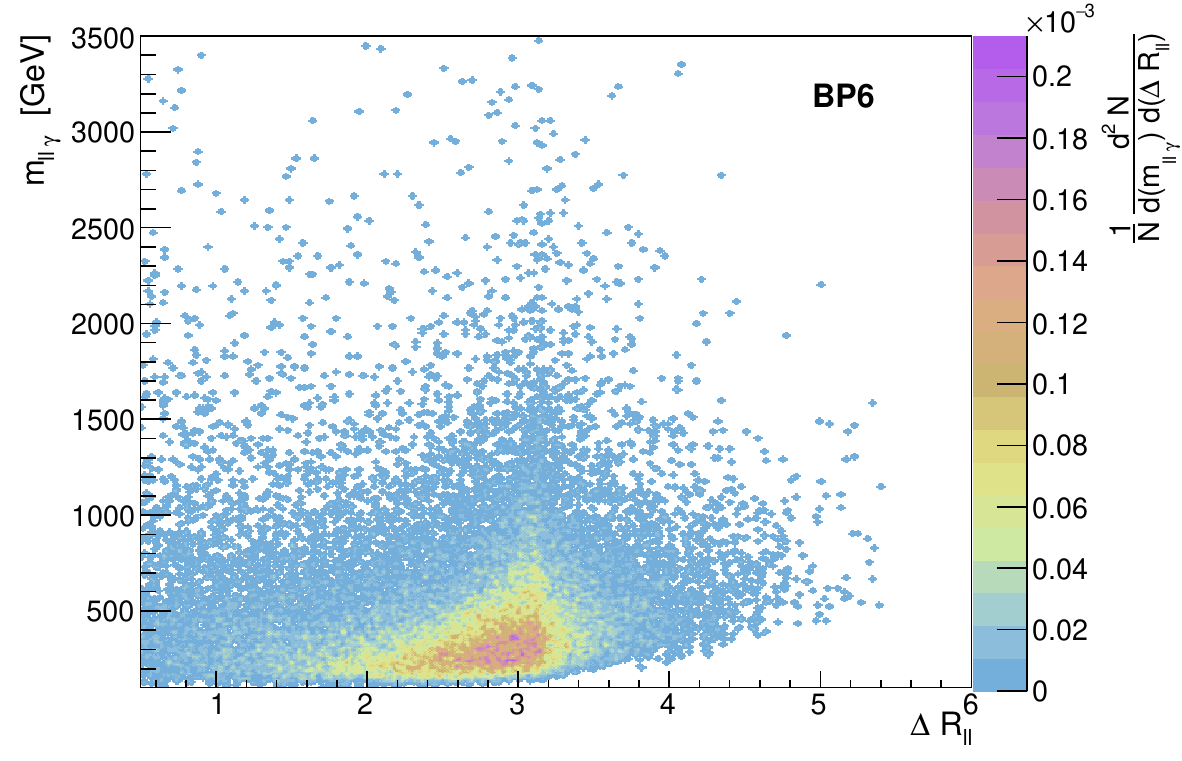}
	\includegraphics[width=8.9cm,height=6.0cm]{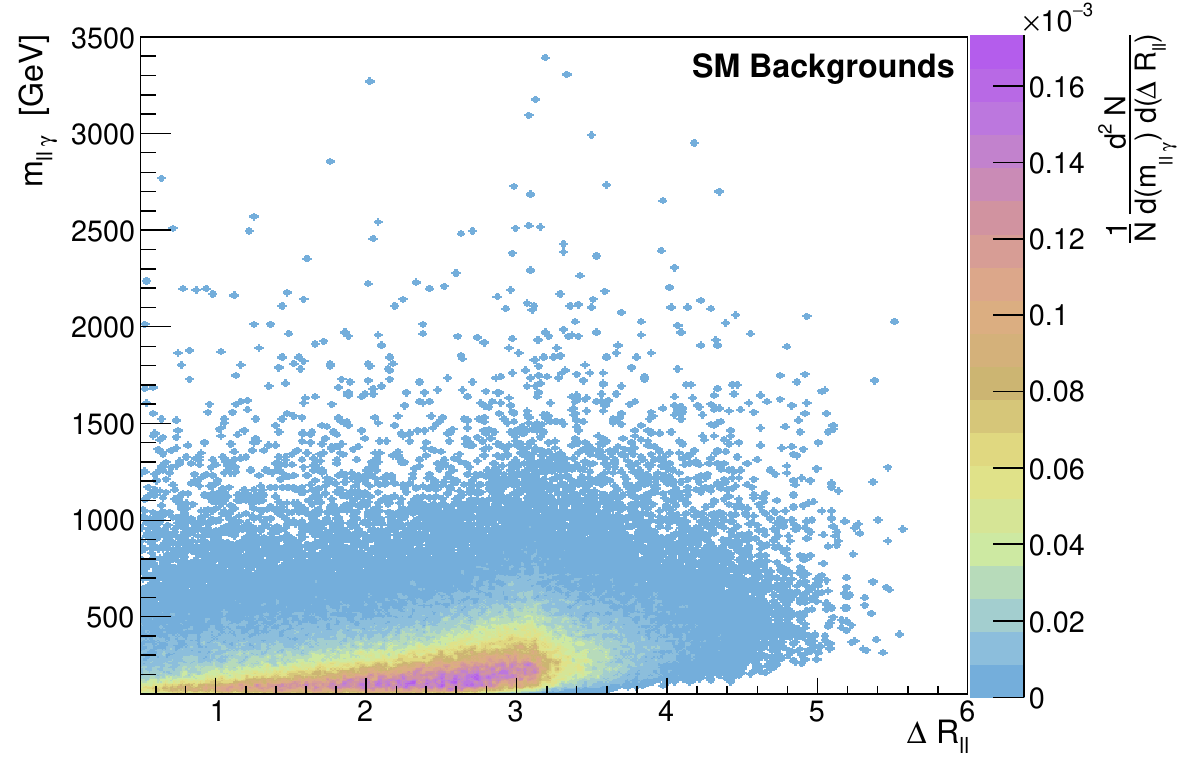}  \\
	\hspace{4mm}(a)\hspace{82mm}(b)
	\caption{Two dimensional histograms showing the correlation between invariant mass of the dilepton and photon system, $m_{ll \gamma}$ and the separation between the two leptons, $\Delta R_{ll}$. The $z$-axis indicates the normalized frequency of events, in arbitrary units. Fig~(a) represents BP6 ALP scenario with $c_{\tilde{W}}=0.5, c_{\tilde{B}}=-0.152, c_{\tilde G}=1.0,  f_a = 5$ TeV and (b) represents the SM backgrounds, comprising, $SM~WW\gamma$, $WZ\gamma$, $V \gamma$, $t \bar{t}\gamma$ and  backgrounds from non-prompt leptons and non-prompt photon at $\sqrt{s}=14$ TeV.}
	\label{mllgamma_deltaRll_correlation}
\end{figure*}

We now delve into an interesting feature in this $2\to3$ process. We will explore the relationship between two variables in the $WW\gamma$ final state: the invariant mass of the dilepton-photon system, $m_{ll\gamma}$ and the $\Delta R$ separation between the two leptons. Fig.~\ref{mllgamma_deltaRll_correlation} highlights how the populated regions in the phase-space shift with the inclusion of new physics effects from higher dimensional operators. The following observations emerge from this figure:
\begin{itemize}
	\item In the background scenarios, such as $WZ\gamma$ and events with non-prompt leptons or photon, the $m_{ll\gamma}$ distribution typically decreases smoothly and rapidly. However, in scenarios involving new physics, this distribution fall more slowly. We observe that background dileptons are more likely to appear in the same hemisphere, contrasting with the signal events. In case of the ALP signal, most of the events are produced with all the three bosons being equally energetic. There is a notable increase in event density as $\Delta R_{ll}$ approaches $\pi$, indicating that the leptons from the W bosons have greater separation (indicated by red color for a higher number of events). Implementing a cut on the invariant mass of the dilepton and photon at $200$ GeV would distinctly highlight these new phase space regions. Additionally, an angular separation cut of $\Delta R_{ll} \ge 2.5$ could effectively filter out a significant portion of the background events, which tend to cluster at lower angular separations.
	
	\item In case of the backgrounds such as $Z (\to \tau^+\tau^-)\gamma$ (and tau leptons decaying leptonically) and non-prompt photons, the photon is significantly energetic and is in recoil to the heavy boson. Thus, the decay leptons appear boosted with less separation between them. Overall, this implies that ALP interactions which result in both the dilepton and the photon gaining higher energy, also results in the angular separation between the leptons tending to be larger compared to that in the SM backgrounds. This correlation is especially evident in Fig.~\ref{mllgamma_deltaRll_correlation}(a), where the most populated event regions are around $\Delta R_{ll}\sim \pi$, especially in high $m_{ll\gamma}$ regions (around 200 GeV).
\end{itemize}

We assess the sensitivity reach  for the various benchmark points at the 14 TeV LHC.  To quantify the signal significance, we use the following definition:
\begin{equation}
{\mathcal S} = \sqrt{2[(S+B)ln(1+S/B)-S]}
\label{eqnsignificance}
\end{equation}
Here, $S$ and $B$ are the numbers of signal and background events, respectively, corresponding to the residual signal and background cross-sections after applying the selection criteria that isolates the signal events from the backgrounds.

The calculated signal significance for each benchmark point across the four different processes is presented in Table~\ref{significance_CBA}. This includes for different possible choices of integrated luminosities, specifically at ${\mathcal L}=300, 1000$ and $3000$ fb$^{-1}$.
\begin{table}[h!]
	\small
	\begin{center}
		\begin{tabular}{|c||c|c|c||c|c|}
			\cline{2-6}
			\multicolumn{1}{c||} {}&\multicolumn{3}{c||}{{Statistical significance}} & \multicolumn{2}{c|}{Required Luminosity} \\
			\multicolumn{1}{c||}{}&\multicolumn{3}{c||}{{(${\mathcal S}$)}} & \multicolumn{2}{c|}{(in $fb^{-1}$)} \\
			\hline
			Signal &   
			${\mathcal L}=300$ fb$^{-1}$  & 
			${\mathcal L}=1000$ fb$^{-1}$  & ${\mathcal L}=3000 $ fb$^{-1}$ & ${\mathcal S}=3{\sigma}$ & ${\mathcal S}=5{\sigma}$  \\ 
			\cline{1-6}
			BP1 & 1.733 & 3.164 & 5.479 & 850.69 &  2363.02\\ 
			BP2 & 1.282 & 2.341 & 4.054 & 1631.76 & 4532.66 \\
			BP3 & 1.902 & 3.473 & 6.015 & 739.15 & 2053.19 \\
			BP4 & 1.816 & 3.317 & 5.744 & 817.01 & 2269.48 \\ 
			BP5 & 1.336 & 2.439 & 4.225 & 1473.89 & 4094.13 \\ 
			BP6 & 1.523 & 2.781 & 4.817 & 1129.66 & 3137.95 \\ 
			\hline
		\end{tabular}
		\caption{Signal statistical significance at various benchmark points for the distinct four processes of our study at the 14 TeV LHC. The significance levels are evaluated for integrated luminosities of ${\mathcal L}=300, 1000,$ and $3000$ fb$^{-1}$. We also estimate the integrated luminosity required to attain a $3\sigma$ and $5\sigma$ excess over the background for each benchmark point at the LHC running at $\sqrt{s}=14$ TeV.}
		\label{significance_CBA}
	\end{center}
\end{table}
We can see from Table~\ref{significance_CBA} that BP1 for the ALP mediated Higgs-strahlung signal will have substantial significance at $3000$ fb$^{-1}$ luminosity. The main reason is large production cross-section of the ALP signal.  Detecting signatures of the $aZh$ interaction in this process, a phenomenon not expected in linear expansions up to NNLO,  would essentially serve as the smoking gun evidence for non-linearity. The $Z\gamma$ process via BP2 and BP3 shows to have the most prominent separation between the signal and the background. The benchmark point BP4 uniquely probes the $g_{aWW}$ coupling, reaching a $3\sigma$ level sensitivity at $1000$ fb$^{-1}$.  The $WW\gamma$ BP5 and BP6 benchmark points are only slightly less sensitive in probing $g_{aWW}$ coupling.
\vspace{-5mm}
\subsection{Direct probes of ALP coupling}
\label{direct_probe}
\vspace{-3mm}
In this subsection, we focus on another ALP production mechanism which involves the ALPs produced in association with Higgs or vector bosons or the `ALP-strahlung' process and study the constraints it puts on ALP-Higgs and ALP-vector boson interactions.   We assume the ALP to be stable within the collider, meaning that it has a sufficiently long lifetime to leave the detector without decaying. This assumption depends on the decay modes available to the ALP, which in turn depend on its mass and couplings. For an ALP with a mass around 1 MeV, decaying to fermions or heavier particles is not kinematically possible. The possible decay channels include $a\to\nu\bar{\nu}\nu\bar{\nu}$ (indistinguishable from a pure missing energy signature), $a\to \gamma\gamma$ and $a\to \gamma \nu \bar{\nu}$. Both of the latter decays would typically allow the ALP to traverse distances much greater than the detector's dimensions before decaying. When the ALP mass exceeds 1 MeV, new decay channels to fermions become feasible once the ALP mass becomes greater than twice the mass of the fermions in the final state. Also, $m_a \ge 3m_\pi$ ($\sim 0.5$ GeV) would enable hadronic decay channels. However, this introduces a dependence on complex model-specific factors, which we do not delve into in this study. One motive in this subsection is to compare the constraints derived from direct ALP searches with those obtained from non-resonant ALP-mediated processes. Direct ALP probes involve additional model-based assumptions, limiting  the generality of fit results. Since we ignore the ALP couplings to SM fermions, the associated production at colliders is dominated by the s-channel diagram through a vector boson propagator. Here, the production rates drop faster as $m_a$ increases, due to the power suppression of energy from s-channel propagator. For our simulations in the {\tt{MG5aMC\_@NLO}} framework, we assume an ALP mass of 1 MeV, consistent with our non-resonant ALP analysis and treat the ALPs as stable within the collider for the purposes of detector simulation.

\vspace{3mm}
\paragraph{\bf{ATLAS measurement of Higgs boson production in association with missing energy and h decays to b-quarks:}} We will study the ALP signal $pp  \rightarrow h(\to b\bar{b}) + a$ and reinterpret the ATLAS search for dark matter via missing energy in association with a SM Higgs boson  channel~\cite{ATLAS:2021shl} in the context of the ALP signal. ATLAS has recently provided measurements of the missing transverse energy ($E_{T}^{miss}$) distribution in events with a large-radius jet with two b-tags and missing energy, using the Run II data from the LHC at $\sqrt{s}=13$ TeV with an integrated luminosity of 139 fb$^{-1}$, along with an estimation of the SM background.  The analysis is confined to a fiducial region, which is closely replicated by the phase space cuts outlined in Table \ref{tab:monox_cuts}.

\begin{figure*}[h!]
	\centering
	\includegraphics[width=8.9cm,height=6.1cm]{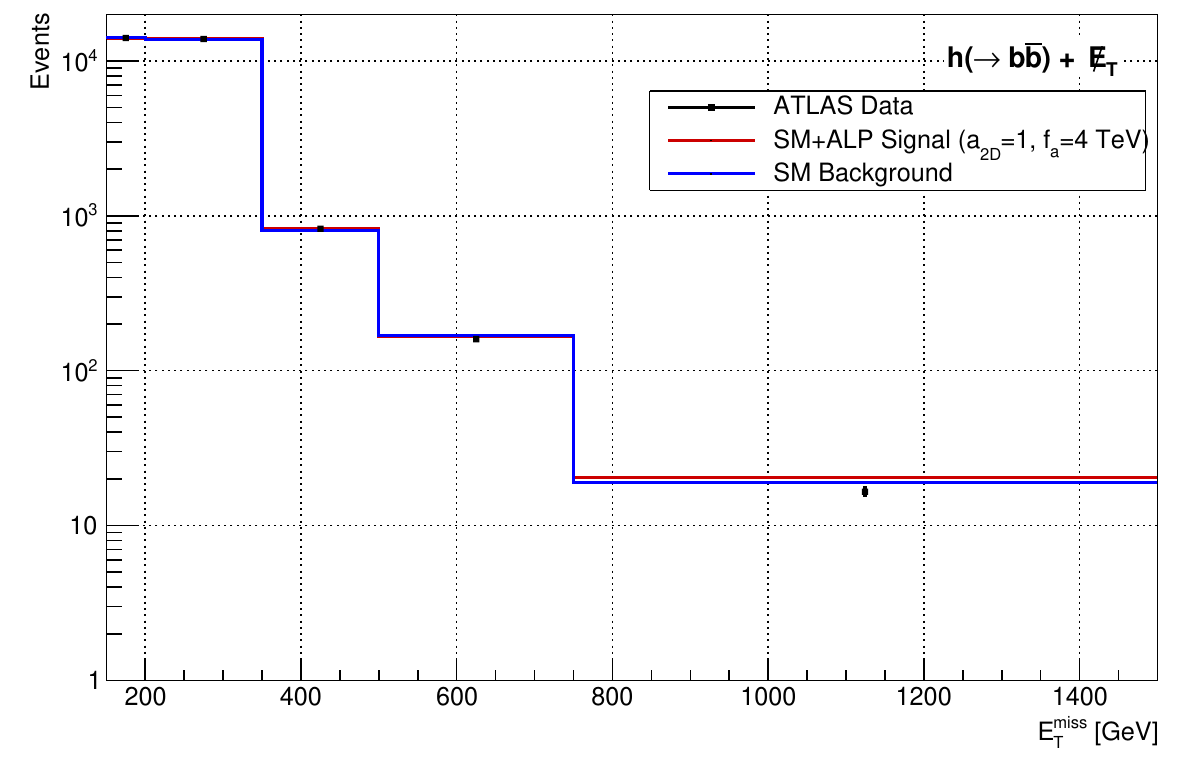}
	\includegraphics[width=8.9cm,height=6.1cm]{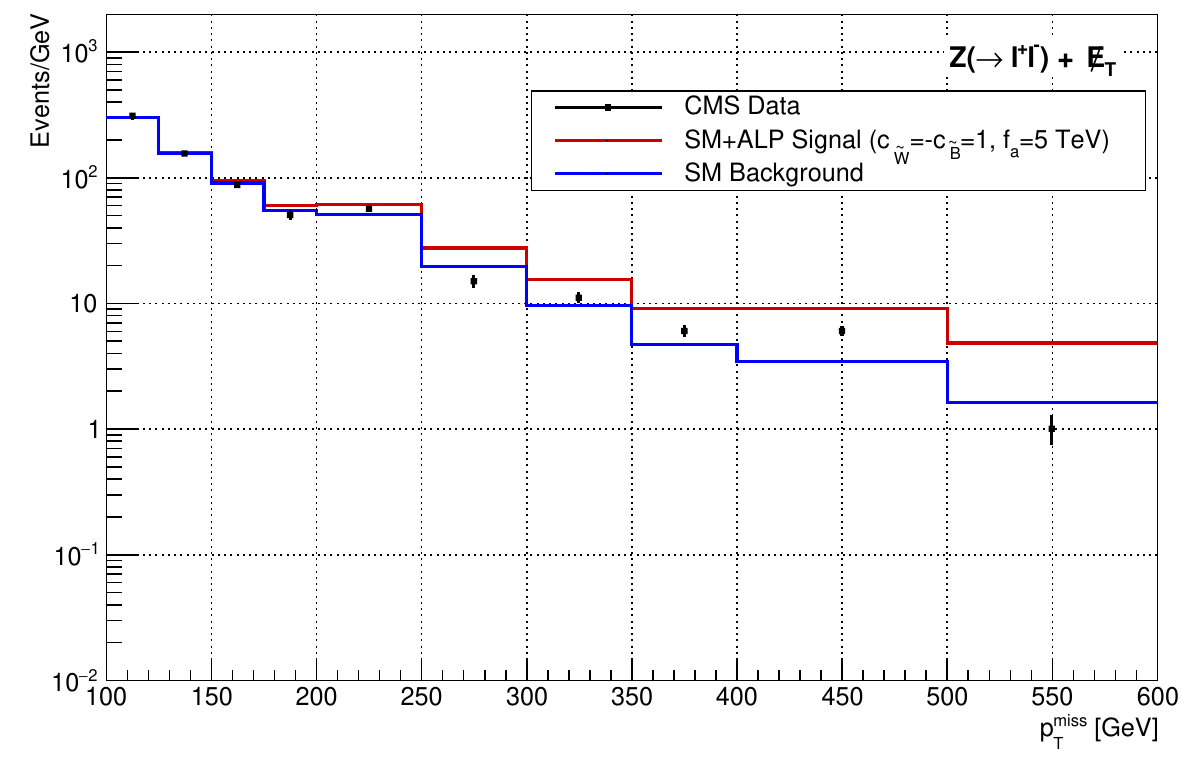}  \\
	\hspace{4mm}(a)\hspace{82mm}(b)
	\includegraphics[width=8.9cm,height=6.1cm]{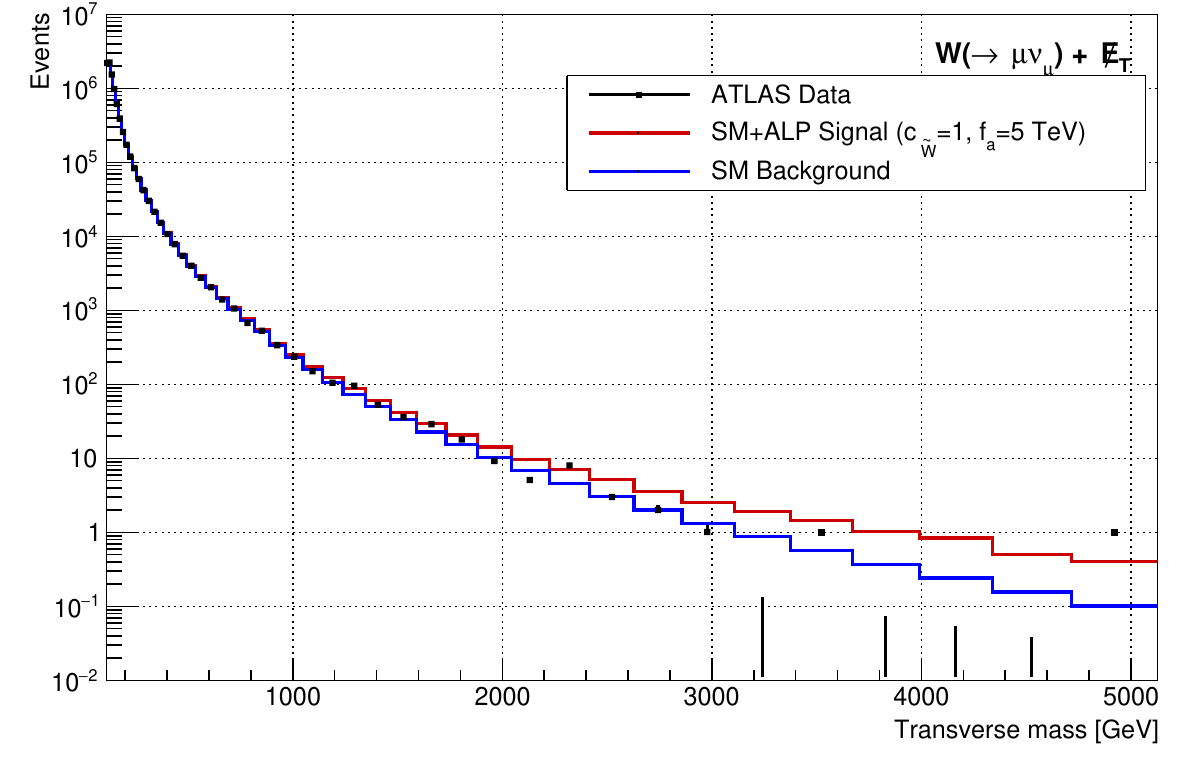}  \\
	\hspace{4mm}(c)
	\caption{(a) The differential distribution of $E_T^{\rm miss}$ for $h (\to b\bar{b}) + E_T^{\rm miss}$ signal and background for $\sqrt{s}=13$ TeV and $139~{\rm {fb}}^{-1}$ of integrated luminosity, following the selection cuts from Table ~\ref{tab:monox_cuts} (a).  The total SM $E_T^{\rm miss}$ background  (blue) distribution is obtained from~\cite{ATLAS:2021shl} and for the signal $pp \to a h$ ($h \to b\bar{b}$) $, E_T^{\rm miss}$ distribution involves contribution from coefficients  $a_{2D}=1$ and $f_a=4$ TeV along with SM contribution (red). (b) Distribution of $p_T^{\rm miss}$  for $a\,Z$ ($Z \to \ell^{+} \ell^{-}$) production, with coefficients $c_{\tilde{W}}=1, c_{\tilde{B}}=-1,  f_a = 5$ TeV (red) and the experimental data and SM backgrounds  from the  CMS analysis~\cite{CMS:2017nxf} at 13 TeV and $35.9\,\mathrm{fb}^{-1}$. (c) Distribution of transverse mass $m_T$ for $a\,W^{\pm}$ ($W^{\pm} \to \ell^{\pm} \nu_{\ell}$) production  in the $\mu + \slashed{E}_T$ final state, obtained with $c_{\tilde{W}}=1, f_a = 5$ TeV (red), compared with experimental data and SM backgrounds from the  ATLAS analysis~\cite{ATLAS:2019lsy} at 13 TeV and $139\,\mathrm{fb}^{-1}$ integrated luminosity.}
	\label{monox_kin_distribtn}
\end{figure*}

We consider a 5 bin-data set, with the bin widths increasing with higher values of $E_{T}^{miss}$. The boundaries of these bins are set at (150, 200, 350, 500, 750) GeV. For the ALP signal simulation, we consider the process
\begin{equation}
p p \rightarrow h + a, \left(h \rightarrow b \bar{b} \right)  
\end{equation}
with the reconstructed Higgs jet having a radius parameter $R=1$. The comparative results between the ALP signal and the ATLAS data are depicted in Fig.~\ref{monox_kin_distribtn} (a), showcasing a slight increase in energy across the $E_{T}^{miss}$ bins.  We perform a $\chi^2$ fit to obtain a limit of :
\begin{equation}
\label{monoh}
\left|\frac{f_a}{a_{2D}}\right| >1.75\, {\rm TeV} \qquad {\rm at}\,\, 95\% \,~{\rm C.L.}
\end{equation}
By studying the indirect probe in the non-resonant  ALP mediated $m_{Zh}$ bins, we have an enhanced sensitivity to the ALP-Higgs coupling.
\begin{table}[t]
	\centering
	\scalebox{0.8}{
		\begin{tabular}{|c|c||c|c||c|c|}
			\hline
			\multicolumn{2}{|c||}{(a) Mono-Higgs}&\multicolumn{2}{|c||}{(b) Mono-Z} &\multicolumn{2}{|c|}{(c) Mono-W} \\
			\hline
			Observable & Selection &Observable & Selection &Observable & Selection \\
			\hline \hline
			$p_T$ leading jet & $> 100$ GeV &$N_l$ & $2$&$p_T$ of leading lepton & $> 55$ GeV ($\mu^-$),$>60$ GeV ($e^-$) \\
			$E_T^{miss}$  & $> 150$ GeV &	$p_T$ of lepton & $> 20$ GeV&$|\eta|$ of leading lepton & $<2.5$ ($\mu^-$),$<2.47$ ($e^-$) \\
			Lepton \& extended $\tau$-lepton vetos& &Reconstructed $Z$ boson mass   & $|m_{ll}-m_Z|<15$ GeV && excluding $1.37 < |\eta| < 1.52$\\
			$\Delta \phi$(jet,$E_T^{miss}$) & $> 20$ rad &$p_T^{ll}$  & $> 60$ GeV&Exactly 1 lepton & \\
			$E_T^{miss}$  & $> 500$ GeV &$p_T^{miss}$  & $> 100$ GeV& Jet veto&\\ 
			large-R jet & $\ge 1$ &b-jet \& extended $\tau$-lepton vetos& &&\\
			b-tagged subjets within the R-jet & $ > 2$ &$\Delta \phi({\vec{p}_T^{j},\vec{p}_T^{miss}})$  & $> 0.5$ &$E_T^{miss}$  & $> 55$ GeV ($\mu^-$),$>65$ GeV ($e^-$) \\ 
			Reconstructed Higgs jet mass & $\in [50,270]$ GeV &$\Delta \phi({\vec{p}_T^{ll},\vec{p}_T^{miss}})$ & $>2.6$ &$m_T$ & $>110$ GeV ($\mu^-$),$>130$ GeV ($e^-$)\\
			&&$|p_{T}^{miss}-p_{T}^{ll}|/p_{T}^{ll}$ & $ <0.4$&&\\
			&&$\Delta R_{ll}$ & $<1.8$&&\\
			\hline
	\end{tabular}}
	\caption{Phase space cuts defining the selection criteria for signal region in (a) ATLAS search for $h (\to b \bar{b})$ + MET~\cite{ATLAS:2021shl}, (b) CMS search for $Z (\to l^+ l^-)$ + MET~\cite{CMS:2017nxf} and (c) ATLAS search for $W (\to l \nu_l)$ + MET~\cite{ATLAS:2019lsy}.}
	\label{tab:monox_cuts}
\end{table}

\vspace{3mm}
\paragraph{\bf{CMS search for new physics events with $Z$ production and large missing energy:}}
We consider now  ALP production in association with a $Z$ boson, in hadronic collisions. We will study the impact of the ALP signal on the CMS measurement of $Z + \ETmiss$ search~\cite{CMS:2017nxf} with $\sqrt{s} = 13$ TeV  and  integrated luminosity $35.9\,\mathrm{fb}^{-1}$. This time, we will be considering a measurement in the leptonic channel  to assess the sensitivity to the effective ALP interaction.

We will use the $p_T^{miss}$ distribution as a key kinematic discriminator between signal and background. Data within the fiducial region, as opposed to the full phase space, will be used to refine the search. The selection cuts from the second and third column of Table~\ref{tab:monox_cuts} are employed. The comparison of signal and background $p_T^{miss}$ distributions for $\ell = \mu$ can be seen in Fig.~\ref{monox_kin_distribtn} (b), with the maximum $p_T^{miss}$ bin being set at 600 GeV. To ensure the EFT applicability, we  remove events in each bin where $\sqrt{\hat{s}}$ exceeds $2 p_T^{miss}{^\text{,max}}$.

The ALP-photon-$Z$ and ALP-$Z$-$Z$ couplings could potentially lead to a mono-$Z$ final state. This indicates contributions from both Wilson coefficients $c_{\tilde{W}}$ and $c_{\tilde B}$ to this process. We establish constraints on $c_{\tilde W}$, assuming $c_{\tilde B} = -t_\theta^2 c_{\tilde W}$. Similar to previous processes, a $\chi^2$ fit, as outlined in Eqn.~\eqref{chisq}, will be used to derive constraints on $c_{\tilde W}$, giving : 
\begin{equation}
\label{monoz}
\left|\frac{f_a}{c_{\tilde W}}\right| > 7.75 ~{\rm TeV} \qquad {\rm at}~~95\%~{\rm C.L.}
\end{equation}
The mono-$Z$ search proves to be useful in constraining the effect of $c_{\Wt}/f_a$. Notably, in the higher-energy regime of $p_{T}^{\rm miss}>250$ GeV, the ALP contribution becomes considerable, especially in the tail of the $p_{T}^{\rm miss}$ distribution, which is where the most significant constraints originate. Nonetheless, this results in a constraint on $c_{\tilde W}$ that is less stringent than what is derived from the CMS $m_{WW}$ and $m_T^{WW}$ distributions.
\vspace{3mm}
\paragraph{\bf{ATLAS measurement of charged lepton  with missing energy:}} 
Let us now concentrate on the ALP production in association with a $W$ boson. We reinterpret the ATLAS search for $W'$ decaying to $\ell+\slashed{E}_T$ final states with $139$ fb$^{-1}$ integrated luminosity~\cite{ATLAS:2019lsy}.  We employ the transverse mass  distribution of the leptonically decaying $W$ for our analysis, as depicted in Fig.\ref{monox_kin_distribtn}(c). To study the influence of the ALP signal on the $m_{T}$ distribution, we apply  the selection criteria outlined in the final two columns of Table\ref{tab:monox_cuts}. The figure also includes the $m_T$ spectrum of the SM background. The ALP coupling involved in this signal is $c_{\tilde{W}}$, with the high-$m_{T}$ bins playing a significant role in shaping the constraints on $c_{\tilde W}$.

Background data for the electron and muon samples are taken from Ref.\cite{ATLAS:2019lsy} and the depicted bins correspond to those with available experimental background information, following  $m_T < m_T^\text{max} = 2.6$ TeV for electrons and $m_T < m_T^\text{max} =3$ TeV for muons. From this analysis, we derive a constraint of:
\begin{equation}
\label{monow}
\left|\frac{f_a}{c_{\tilde W}}\right| > 9.11 ~{\rm TeV} \qquad {\rm at}~~95\%~{\rm C.L.}
\end{equation}
from the ATLAS $m_{T}$ data. Thus, the mono-$W$ analysis yields stronger constraints than those obtained from the non-resonant $pp \to WW$ process. On the other hand, a dedicated search in the channel $W\gamma+$MET has not yet been performed at the LHC. This channel has several advantages over the $W+MET$ channel search. First, the high efficiency of reconstruction of high energy photons will lead to better sensitivity to the new physics effect and the SM background is also expected to be lower. Second, this channel like the non-resonant ALP mediated $WW\gamma$ production will be able to probe couplings such as the four-point interaction of $aWW\gamma$ and also can help disentangle more than one direction in non-linear ALP EFT parameter space. Thus, a combination of such probes will lead to better refining of the observables.

It is important to note that most direct bounds usually depend on specific model assumptions, which often involve setting all other coefficients to zero, unlike indirect bounds. As such, the indirect limits presented in this study act as a good complementary probe, proving useful even in instances where direct probes might provide  more stringent constraints.
\vspace{-5mm}
\section{Projected Sensitivities on ALP EFT couplings}
\label{sensitivity_reach}
\paragraph{\bf Sensitivity to ALP-Higgs coupling:}

The results presented in Table~\ref{significance_CBA} provide the sensitivities for different benchmark points.  This section outlines the sensitivity projections  within the  parameter space of ALP couplings using the relevant ALP-mediated non-resonant $Zh, Z\gamma, WW, WW\gamma$ production processes. These processes are sensitive to the product of the ALP coupling to gluons with the respective ALP coupling to bosons. The ALP-gluon coupling, in principle, is an independent free parameter. We present the results for the ALP-boson couplings in this section  assuming $g_{agg}=1$ TeV$^{-1}$.

\begin{figure}[b!]
	\begin{center}
		\includegraphics[width=8.9cm,height=7.0cm]{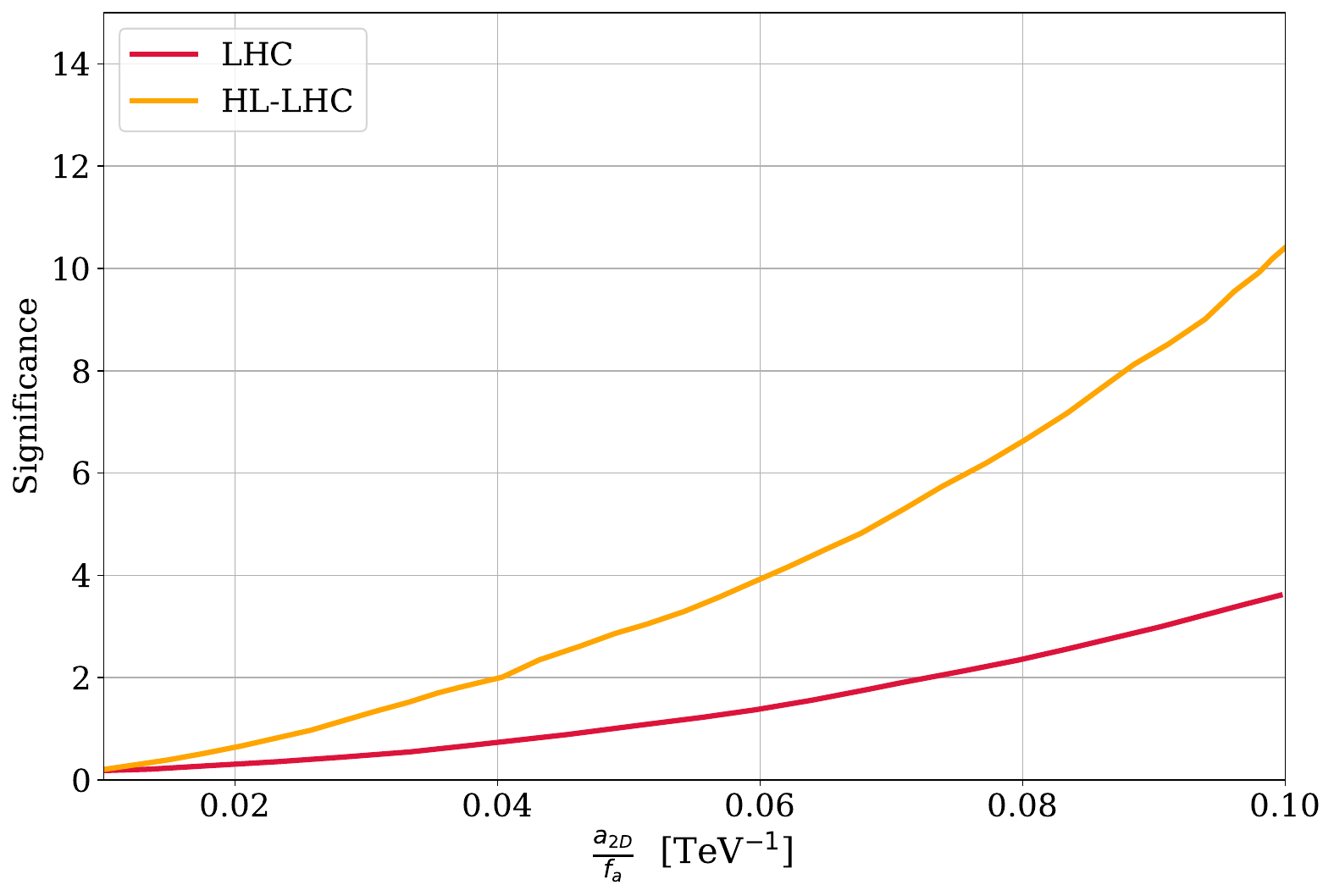}
		\includegraphics[width=8.9cm,height=7.0cm]{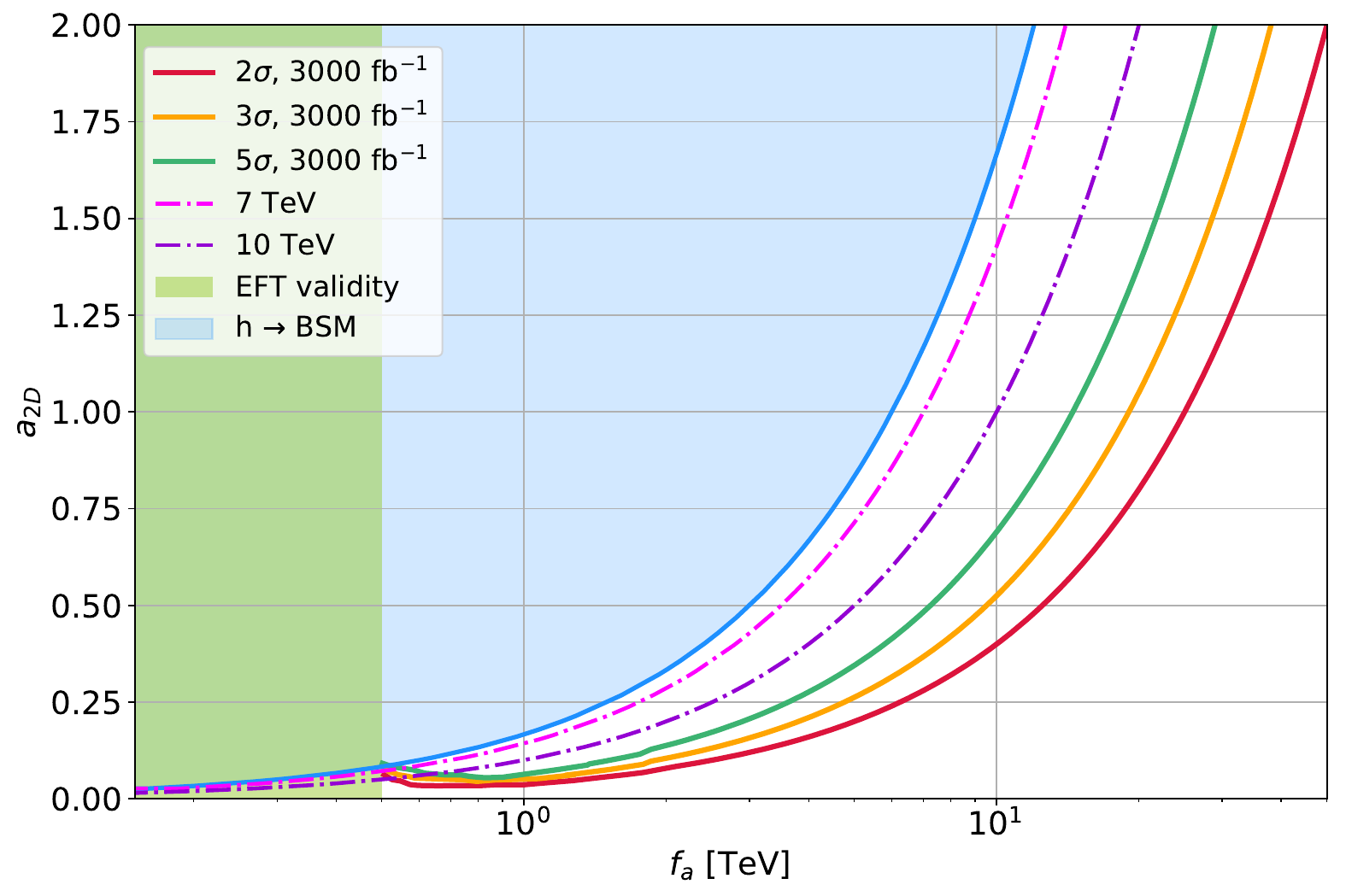}
		\caption{\textit{Left}: Significance, $\mathcal{S}$ (calculated using total cross-sections) as function of $\frac{a_{2D}}{f_a}~(\rm TeV^{-1})$  at $\sqrt{s} = 13~$TeV,  $139 ~$fb$^{-1}$ (red) and $\sqrt{s} = 14~$TeV,  $3000 ~$fb$^{-1}$ (yellow) for $p p \to Z h$  for $g_{agg}=1$ TeV$^{-1}$. \textit{Right}: Sensitivity contours at $2\sigma$ (red), $3\sigma$ (yellow) and $5\sigma$ (green) levels  for the ALP mediated $p p \to Z h$  signal at $\sqrt{s}=14$ TeV LHC and for an integrated luminosity of $3000$ fb$^{-1}$, in $a_{2D}$-$f_a$ plane  assuming $g_{agg}=1$ TeV$^{-1}$. The green shaded region depicting $f_a < \sqrt{s}^\mathrm{min}$, is  excluded by the criterion of EFT validity. The blue region is excluded by the limits from $\Br(h\to {\rm BSM})$~\cite{CMS:2022dwd}. The {\em dash}-{\em dotted} lines represent constant values of $f_a/a_{2D}$.}
		\label{zh_sensitivity}
	\end{center}
\end{figure}

In Fig.~\ref{zh_sensitivity} (a), we present the variation of significance of ALP mediated $hZ$ signal with operator coefficient $\frac{{a}_{2D}}{f_a}$ for an integrated luminosity of $139$ fb$^{-1}$ at $\sqrt{s}=13$ TeV (red curve) and $3000$ fb$^{-1}$ at $\sqrt{s}=14$ TeV (yellow curve). Signal stands over the background  a $3\sigma$ level for $\frac{{a}_{2D}}{f_a}\simeq$ 0.095 TeV$^{-1}$ (0.058 TeV$^{-1}$) at 13 TeV (14 TeV).

Fig.~\ref{zh_sensitivity} (b) shows the sensitivity levels at $2\sigma$ (red), $3\sigma$ (yellow) and $5\sigma$ (green) for the $pp \to Zh$ signal  with $\sqrt{s}=14~\rm{TeV}$ and for an integrated luminosity of $3000~\rm{fb^{-1}}$ in the parameter space of $f_a$-${a}_{2D}$. The green shaded region represents $f_a \leq \sqrt{\hat{s}^{\rm min}}$ and is thus, excluded since all signal events will break the validity criterion of EFT. The $5\sigma$ sensitivity level is achieved for $f_a / a_{2D} \simeq 15{\,\rm TeV}$   for an integrated luminosity of $3000{\,\rm fb^{-1}}$ of data.  Thus, the region is allowed to observation at the HL-LHC.

The dash-dotted reference  lines correspond to constant values of $f_a/{a}_{2D}$. In the region with higher values of $f_a$, we find that the sensitivity curves run almost parallel to the lines of constant  $f_a/{a}_{2D}$ indicating a stable detection range for $f_a/{a}_{2D}$ here, despite a loose constraint on $a_{2D}$. When $f_a$ decreases below $1~\rm TeV$, the sensitivity curves fall slowly compared to the reference lines. This indicates that the analysis in this lower $f_a$ region is limited to smaller $f_a/{a}_{2D}$ ratios compared to the higher $f_a$ regions. This change in sensitivity is attributed to the reason that as $f_a$ decreases, more and more events from the higher energy bins are excluded to ensure the applicability of the EFT in the region. This leads to  loss in discerning power of the signal. 

The interaction between ALP and the Higgs boson also induces non-standard decays of Higgs. These decay modes of the Higgs boson will put constraints on ALP interactions through the unobserved Higgs branching fraction (h$\to$ BSM). Considering that these exotic decays are the only LO modifications to Higgs properties, the global signal strength measurements can be used to constrain $a_{2D}/f_a$. This is because the invisible Higgs branching fraction will be  proportional to $1-\rm BR(h\to SM)$. The latest combined CMS global signal strength measurement restricts $\rm BR(h\to BSM)<$ 0.11~\cite{CMS:2022dwd}. Assuming  $\Gamma_{BSM}\simeq\Gamma_{h\to aZ}$, we obtain a limit of $\Gamma_{h\to aZ} < 0.5$ MeV at 95\% C.L. This limit  translates into a constraint of $f_a/{a}_{2D} \geq 5.95$ TeV for $m_a \le 34$ GeV. However, this expected sensitivity is less stringent than the current limit derived from the $pp\to a^* \to Zh$ process, as depicted in the blue shaded region  of Fig.~\ref{zh_sensitivity} (b).

\begin{figure*}[t]
\begin{center}
	\includegraphics[width=7.8cm]{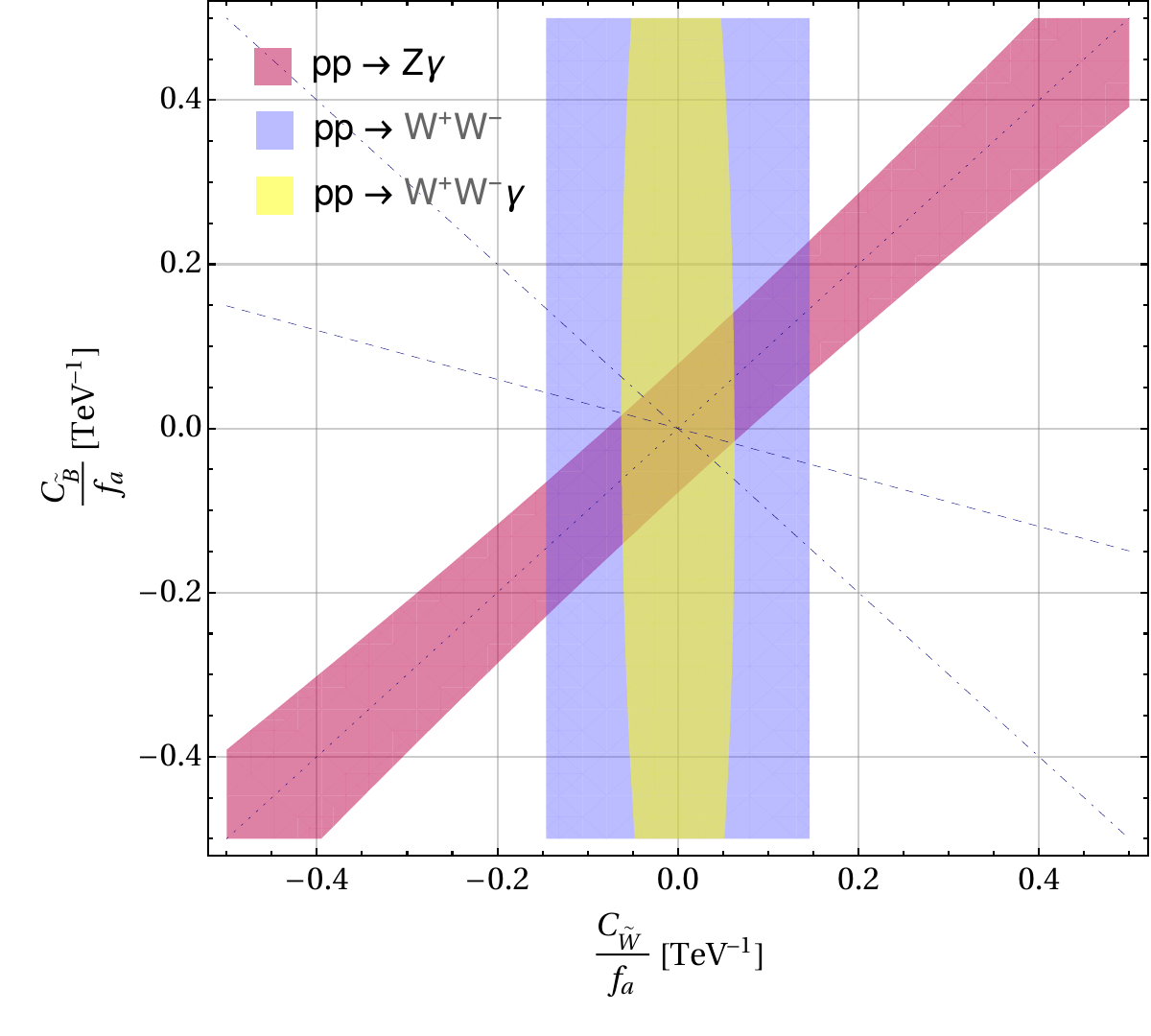}
	\includegraphics[width=7.8cm]{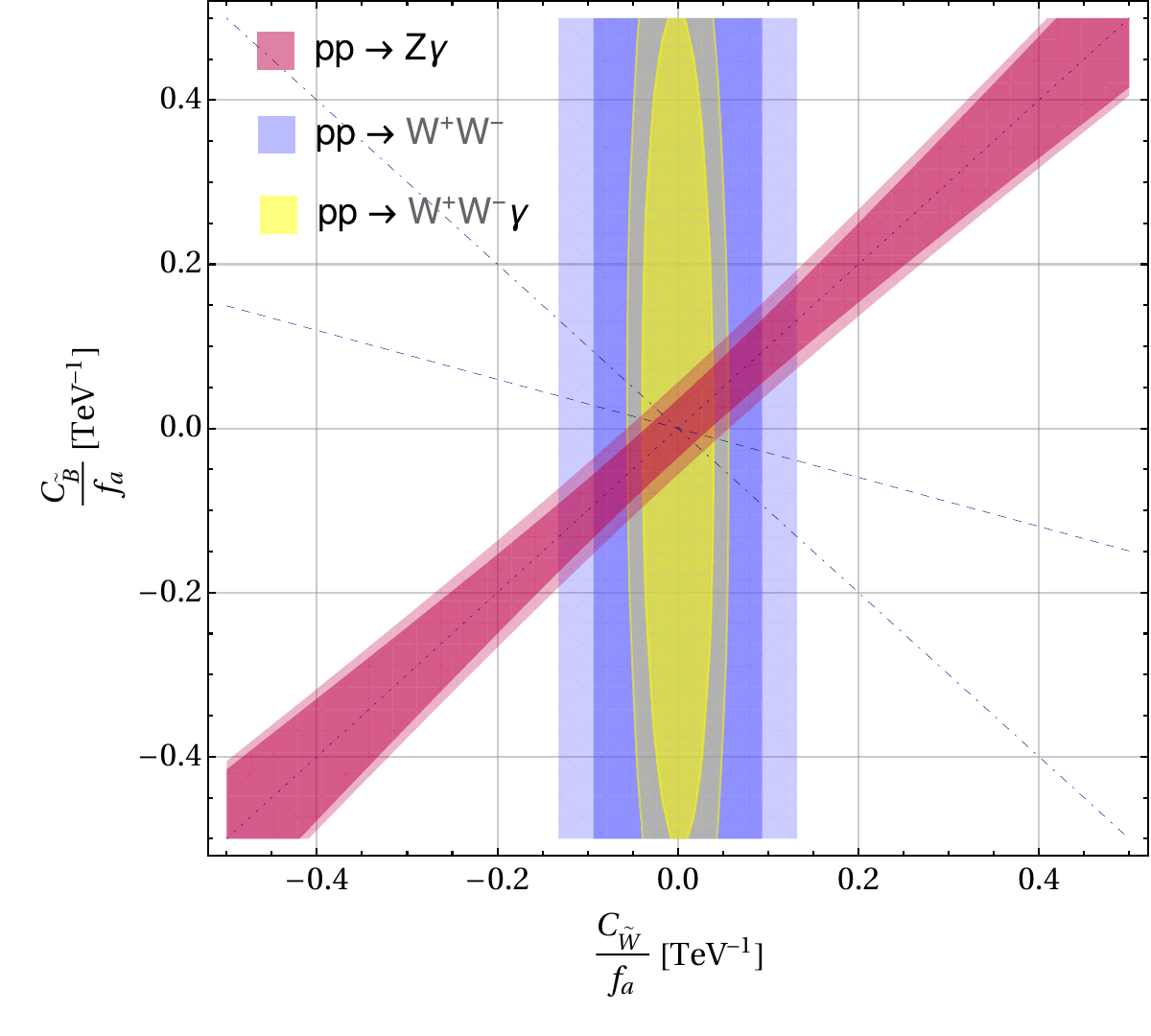}
	\caption{\textit{Left}: 95\% C.L. allowed region in $\frac{c_{\tilde{W}}}{f_a}$-$\frac{c_{\tilde{B}}}{f_a}$ plane from 13 TeV analysis of differential distribution measurements for $Z\gamma$ (dark pink), $WW$ (blue) and $WW\gamma$ (yellow) production processes with $g_{agg}=1$ TeV$^{-1}$. \textit{Right}: Projected $2\sigma$ limit (darker-shaded) and $5\sigma$ discovery level (light-shaded) regions  in $\frac{c_{\tilde{W}}}{f_a}$-$\frac{c_{\tilde{B}}}{f_a}$ plane  for the respective ALP mediated signals at $\sqrt{s} = 14~\rm{TeV}$ and integrated luminosity of $3000$ fb$^{-1}$ with $g_{agg}=1$ TeV$^{-1}$. The thin dashed, dotted and dot-dashed lines represent the directions of vanishing couplings for $g_{a\gamma\gamma}$, $g_{aZ\gamma}$ and $g_{aZZ}$ respectively. The vertical axis at $c_{\tilde{W}}=0$ respresents $g_{aWW}=0$.}
\label{fig:bounds_projected}
\end{center}
\end{figure*}

\paragraph{\bf Sensitivity to ALP-electroweak gauge bosons coupling : }
Fig.~\ref{fig:bounds_projected} presents the upper bounds on the coefficients $\frac{c_{\tilde{W}}}{f_a}$ and $\frac{c_{\tilde{B}}}{f_a}$ (in TeV$^{-1}$), derived from $Z\gamma, WW$ and $WW\gamma$ analyses. These limits  can also be interpreted as products of ALP couplings in the plane of $\{\frac{c_{\tilde{G}}c_{\tilde{W}}}{f_a^2}-\frac{c_{\tilde{G}}c_{\tilde{B}}}{f_a^2}\}$ as all of these processes involve ALP-gluon coupling. They are calculated for each individual experimental channel and based on the differential measurements of relevant energy-dependent variables (refer to Sec.~\ref{13tev_collider_probe}).  We will present these limits  assuming $g_{agg}=1$ TeV$^{-1}$. The $Z\gamma$ process which gets modified by both $c_{\tilde{W}}$ and $c_{\tilde{B}}$ coefficients, constrains the difference $|c_{\tilde B}-c_{\tilde W}|<$ 0.074 TeV$^{-1}$, as derived from the 13 TeV $m_{Z\gamma}$ differential measurement. The $WW\gamma$ analysis imposes a stricter limit on $c_{\tilde W}$. The expected limit for $WW\gamma$, based on $m_{WW}$, is $|c_{\tilde{W}}|<0.147$ TeV$^{-1}$, which is twice as stringent as that from the $WW$ analysis based on $m_{T}^{WW}$. The $WW$ process is not affected by $c_{\tilde{B}}$, whereas $WW\gamma$ has a slight dependence on it, as seen in Fig.~\ref{fig:bounds_projected}. Combining the results from $WW$ and $WW\gamma$, along with other diboson channels like $ZZ$, $W\gamma$ and triboson channels such as $ZZ\gamma$, could potentially yield improved sensitive limits and is a prospect for global analysis in future work.  The non-linear framework of ALP EFT  generates other operators that could modify the interactions of the charged weak bosons with the ALP. Exploring the $WW\gamma$ process further could help disentangle more than one directions in the ALP parameter space, an endeavor to be taken up in the follow-up. When all constraints are considered together, only a narrow overlapping region near zero remains viable, with $|c_{\tilde{W}}|<0.06$ and $|c_{\tilde{B}}|<0.072$.  The limits from $Z\gamma$ measurement provide the most stringent constraints along the $c_{\tilde{B}}$ axis. These constraints can also be interpreted in the plane  of effective couplings like $g_{a\gamma\gamma}, g_{aZ\gamma}$ and $g_{aZZ}$ (using Eqn.~(\ref{eq:alp_coupling})), which are depicted in the dashed, dotted, dot-dashed lines in the Fig.~\ref{fig:bounds_projected}.
				
The $Z$ boson can decay into a light ALP and a photon. The upper limit on the width of $Z$ boson to exotic channels is $\Gamma(Z\to\text{BSM}) \lesssim 2~{\rm MeV}$ at a 95\% C.L.~\cite{Workman:2022ynf}. This  puts a strong limit on the tree-level  decay of $Z\to a\gamma$. This contribution $\Gamma(Z\to a \gamma)$ is given by:
				
\begin{equation}
\Gamma(Z\to a \gamma)=\frac{m_Z^3}{384\pi} g_{aZ\gamma}^2\left(1-\frac{m_a^2}{m_Z^2}\right)^3
\end{equation}
				
Using the $Z$ boson width data, the coefficient $g_{aZ\gamma}$ can be constrained which is largely independent of $m_a$ for values of $m_a\lesssim m_Z$ GeV:
				
\begin{equation}\label{aZg_constraint}
|g_{aZ\g}|< 1.8~{\rm TeV^{-1}} ~~{\rm at}~~95\% ~{\rm C.L.}
\end{equation}
				
Constraints from LEP experiments on the $Z \to 3 \g$ decay process~\cite{Jaeckel:2015jla} cosntrains a combination of $g_{a\gamma\gamma}$ and $ g_{aZ\gamma}$. However, based on the already strong limits of $g_{a\gamma\gamma}$, the resulting bound on $g_{aZ\gamma}$ turns out to be less stringent than the one derived in Eqn.~(\ref{aZg_constraint}).
				
We will discuss the projected constraints of ALP couplings to EW gauge bosons via these ALP mediated processes  at the HL-LHC. Fig.~\ref{fig:bounds_projected} (b) shows the sensitivity regions at $2\sigma$ (darker shaded region)  and $5\sigma$ (lighter shaded region) significance levels on the $\frac{c_{\tilde{W}}}{f_a}- \frac{c_{\tilde{B}}}{f_a}$ plane for a 14 TeV LHC and  $3000~\rm{fb^{-1}}$ of data. At $2\sigma$ level, a more stringent region for each channel is seen, with the $Z\gamma$ channel exhibiting the most significant individual improvement ($|c_{\tilde B}-c_{\tilde W}| = 0.05$ TeV$^{-1}$). The combined limits are mostly constrained by the $WW\gamma$ and $Z\gamma$ channels. Additionally, Fig.~\ref{fig:bounds_projected} (b) highlights the expected discovery threshold (lighter shaded region) at 14 TeV, where the SM point would be excluded by 5 standard deviations if the measurements align with the predicted ALP signal. This region is within the exclusion limits of the current 13 TeV LHC data and suggests that the absence of results from the current LHC data does not necessarily rule out the possibility of a discovery at the HL-LHC.
\vspace{-5mm}
\section{Multivariate analysis}
\label{multivariate_analysis}
\vspace{-3mm}
After performing a cut-based analysis for each of the signals in the four distinct non-resonant ALP processes at the LHC in Sec.~\ref{collider_probe}, we now delve into investigating for potential improvement in the analysis with some advanced techniques like {\em Gradient Boosted Decision Trees}~\cite{Coadou:2022nsh}. The usefulness of these methods have been extensively studied in recent studies~\cite{Baldi:2014kfa,Oyulmaz:2019jqr}, particularly in the Higgs sector~\cite{Bakhet:2015uca,Lasocha:2020ctd} and have demonstrated better  efficacy in differentiating between signal and background characteristics compared to conventional rectangular cut-based analyses. Their application in ALP scenario searches at colliders is yet to be thoroughly explored.  In our study, we assess the possibility for maximizing signal significance in the specific signal processes under our consideration. To achieve this, we utilized the AdaBoost classifier from the {\tt{scikit-learn}} library in Python.

At first, we discuss the details of our analysis for the Higgs-strahlung process, considering the BP1 benchmark scenario for the ALP mediated signal. We take into account  all relevant SM backgrounds in the process $-$ here, $Z$+jets which includes $Z$+$b\bar{b}$ is the most dominant background for $Zh$ production. To optimize the classifier's performance in identifying the signal region, we impose slightly looser cuts compared to the cut-based analysis, thereby ensuring better training. The selection criteria we employed are as follows: $75~{\rm{GeV}}<m_{ll}<105~{\rm{GeV}},~p_{T_{ll}} > 160~{\rm{GeV}},~\Delta R_{ll} > 0.2,~p_{T_{J}} > 60~{\rm{GeV}},~95~{\rm{GeV}}<m_{J}<155~{\rm{GeV}},~\Delta R_{b_i,b_j} > 0.4$ and $\slashed{E}_T<70~{\rm{GeV}}$. After these pre-selections, we trained the classifier on the signal and background samples with the following set of variables:

\begin{itemize}
	\item Transverse momenta ($p_T$) of the two isolated leptons
	\item Reconstructed $Z$ boson and its $p_T$
	\item $\Delta R$ separation between the two b-tagged subjets ($\Delta R_{b_i,b_j}$), subjet $i$ and lepton $j$ ($\Delta R_{b_i,l_j}$) and two leptons ($\Delta R_{l_i,l_j}$)
	\item Scattering angle of reconstructed $Z$ boson.
	\item N-subjettiness of the leading fat-jet ($\tau_{21}$)
	\item $\Delta \phi$ separation between the leading fat-jet and the reconstructed $Z$ boson
	\item Mass of the reconstructed Higgs jet and its $p_T$
\end{itemize}

\begin{figure}[t]
	\includegraphics[width=0.48\textwidth]{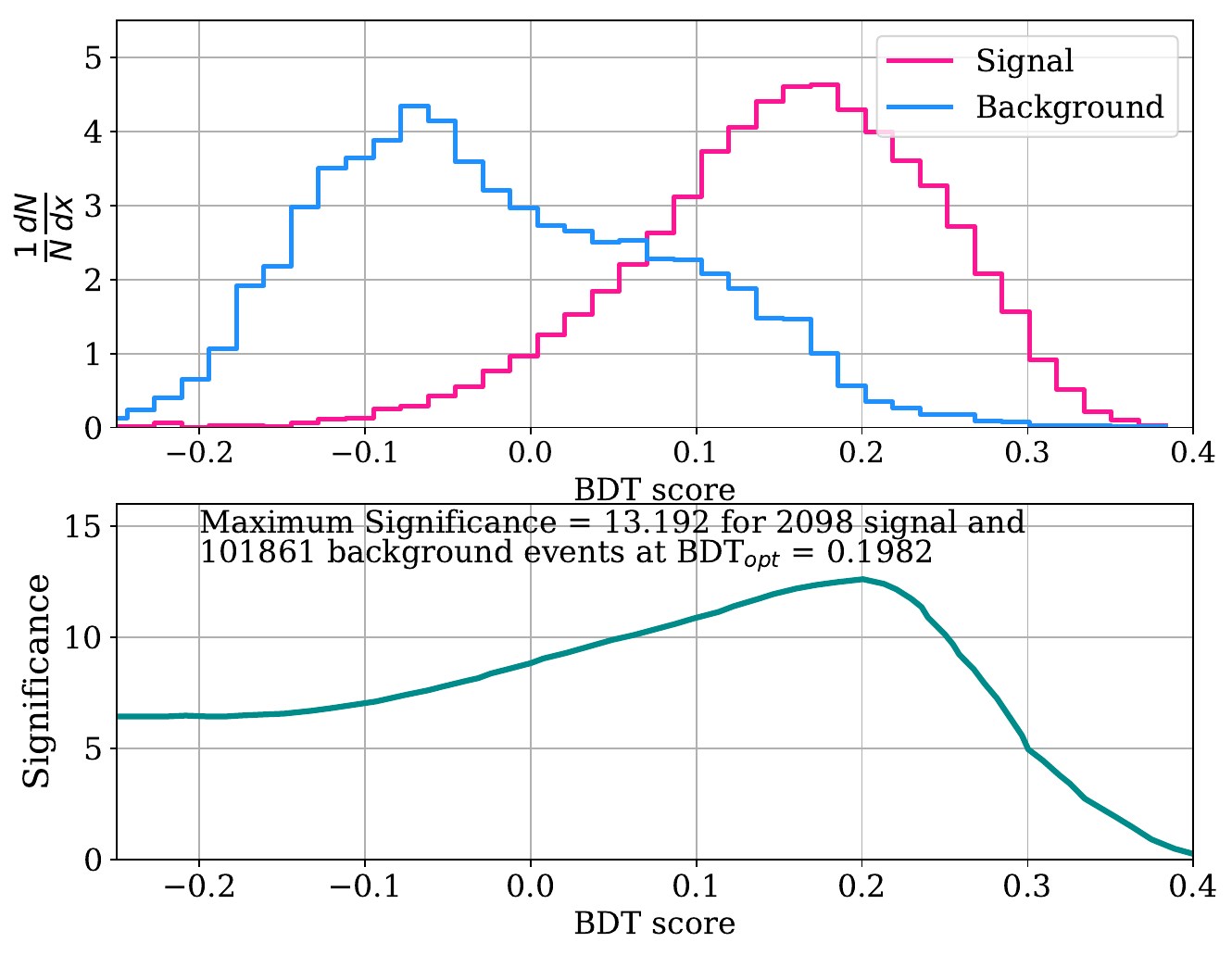}
	\includegraphics[width=0.48\textwidth]{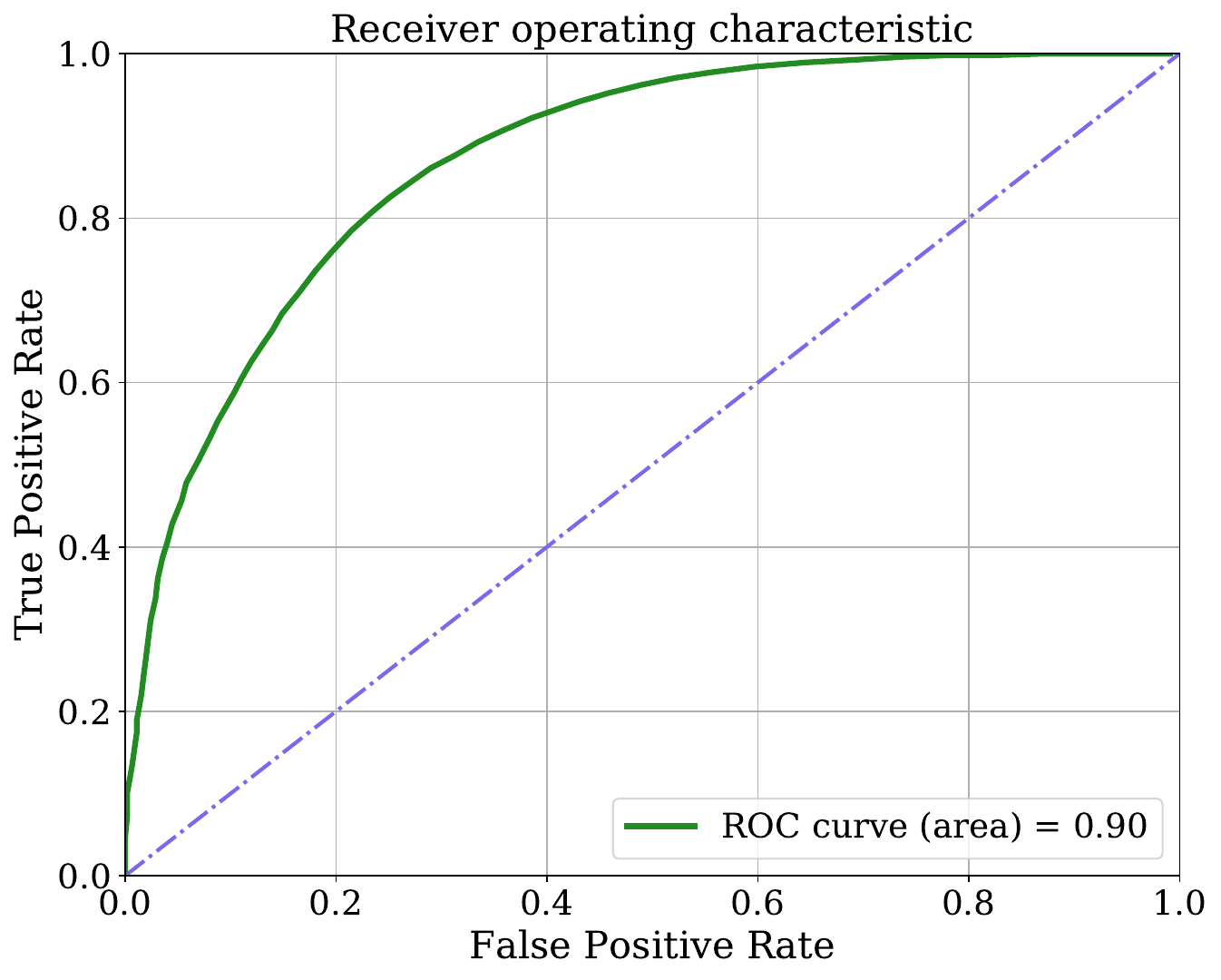}
	\caption{$pp\to Z(\to \ell^+\ell^-)h(\to b\bar{b})$. \textit{Left}: The normalized BDT score distributions for the signal and the background. Significance as a function of the BDT cut value for BP1 at $\sqrt{s} = 14~$TeV,  $\mathcal{L}_{int}=3000\,\mathrm{fb}^{-1}$. \textit{Right}: ROC curve for BP1.}
	\label{BDT1}
\end{figure}

For the gradient boosted decision tree method of separation, we have taken $1000$ estimators and maximum depth of 4 with learning rate 0.1. We have used 75\% of the total dataset for training purpose and 25\% for validation.  After implementing the BDT algorithm, we obtain the distribution of the response of the BDT classifier for the signal and total background events for Higgs-strahlung process as shown in Fig.~\ref{BDT1} (top-left panel). We can see a clear distinction between the signal and the background distributions. We have checked that in this process, $p_T$ distribution of the leading lepton plays the role of the most important input variable. The $\Delta R$  separation between the two b-tagged subjets and $p_T$ of the reconstructed Higgs jet are the second and third best discriminators, respectively. Thus, stronger transverse momenta of the leading fatjet and the lepton are favourable to retain the correct classification of these variables. We have plotted the Receiver Operating Characteristic (ROC) curve (that estimates the degree of rejecting the backgrounds with respect to the signal) for the benchmark signal process BP1 in Fig.~\ref{BDT1} (right panel). One of the possible demerits of these techniques is over-training of the data sample. In case of over-training, the training sample gives extremely good accuracy but the test sample fails to achieve that. We have explicitly checked that with our choice of parameters, the algorithm does not over-train. The ROC curve remains almost same for training and testing samples. The area under the ROC curve is 0.90 for BP1. At the $\sqrt{s}=14$\,TeV LHC with 3000\,fb$^{-1}$ of integrated luminosity, we expect to observe 833 signal events and 3542 background events for an optimal cut of 0.1982 on the \texttt{BDT} output. The signal significance  computed using the formula in Eqn. \eqref{eqnsignificance}, is 13.192. Upon assuming a systematic uncertainty, $\sigma_{sys\_un}$, the signal significance formula is modified in the following form :
\begin{equation}
\mathcal{S}_{\rm sys}=\sqrt{2\left((S+B){\rm log}\left(\frac{(S+B)(B+\sigma_B^2)}{B^2+(S+B)\sigma_B^2}\right)-\frac{B^2}{\sigma_B^2}{\rm log}\left(1+\frac{\sigma_B^2 S}{B(B+\sigma_B^2)}\right)\right)}
\label{syst_significance}
\end{equation}
where $\sigma_B=\sigma_{sys\_un}\times B$. The performance of the multivariate analyses was optimized to maximize the signal significane while also maintaining a reasonably good value of $S/B$.
Adding a 5\% systematic uncertainty translates to a significance  of 4.164. We present our results for $\sqrt{s}=14$\,TeV to make it easier to translate to the case of Run-3 ($\sqrt{s}=13.6$\,TeV) and HL-LHC ($\sqrt{s}=14$\,TeV) as the cross-sections are not expected to change much.

For the $Z\gamma$ process, we study the BP2 and BP3 ALP mediated signals. We considered all the backgrounds listed in Table~\ref{tab:zgamma_cutflow} in the background class. For the MVA, we have adopted cuts that are slightly less stringent than those used in the cut-based approach (detailed in Table~\ref{tab:zgamma_cutflow}). Along with the preliminary selection cuts, we have applied a requirement for the leading fat-jet and the photon to have a minimum transverse momentum of 175 GeV. We have set a minimum threshold for the reconstructed fat-jet mass at 60 GeV. These criteria effectively minimize the dominant background while retaining most of the signal events. This approach is important because the MVA tends to be less effective with only pre-selection cuts, given the small signal size relative to the large background. It is also worth noting that the more stringent cuts from Table~\ref{tab:zgamma_cutflow} do not necessarily lead to better results in the MVA context. Therefore, the cuts chosen for MVA are carefully calibrated to be neither too strict nor too relaxed compared to the cut-based analysis.

The BDT classifier is configured with the following hyperparameters:  {\tt `n\_estimators':800, `learning\_rate':0.1, `trees':10, `max\_depth':4}

For the training, we have selected a range of observables that are effective at distinguishing between the signal and background. These observables are chosen as input variables for the BDT to optimize its discerning potency.

\vspace{-2mm}
\begin{equation}
\begin{split}
p_{T}^{\gamma},~m_{J},~\tau_{21},~p_{T}^{J},~cos \theta_{\gamma}^*,
~\Delta R_{J\gamma},~E^{\gamma},~\eta^{J},~p_T^{b_i},~\Delta R_{b_ib_j}
\end{split}
\end{equation}
where the symbols have their usual meaning. $p_T^{b_i}$ denotes the transverse momentum of $i^{\rm th}$ b-tagged sub-jet and cos $\theta_\gamma^*$ is the scattering angle of photon in the $Z\gamma$ rest frame. Among these variables used, the four most important variables to distinguish the ALP signal from the backgrounds are : $m_{J}, ~\tau_{21},~p_T^{J}, ~E^{\gamma}$.

\begin{figure}[t]
	\includegraphics[width=0.48\textwidth]{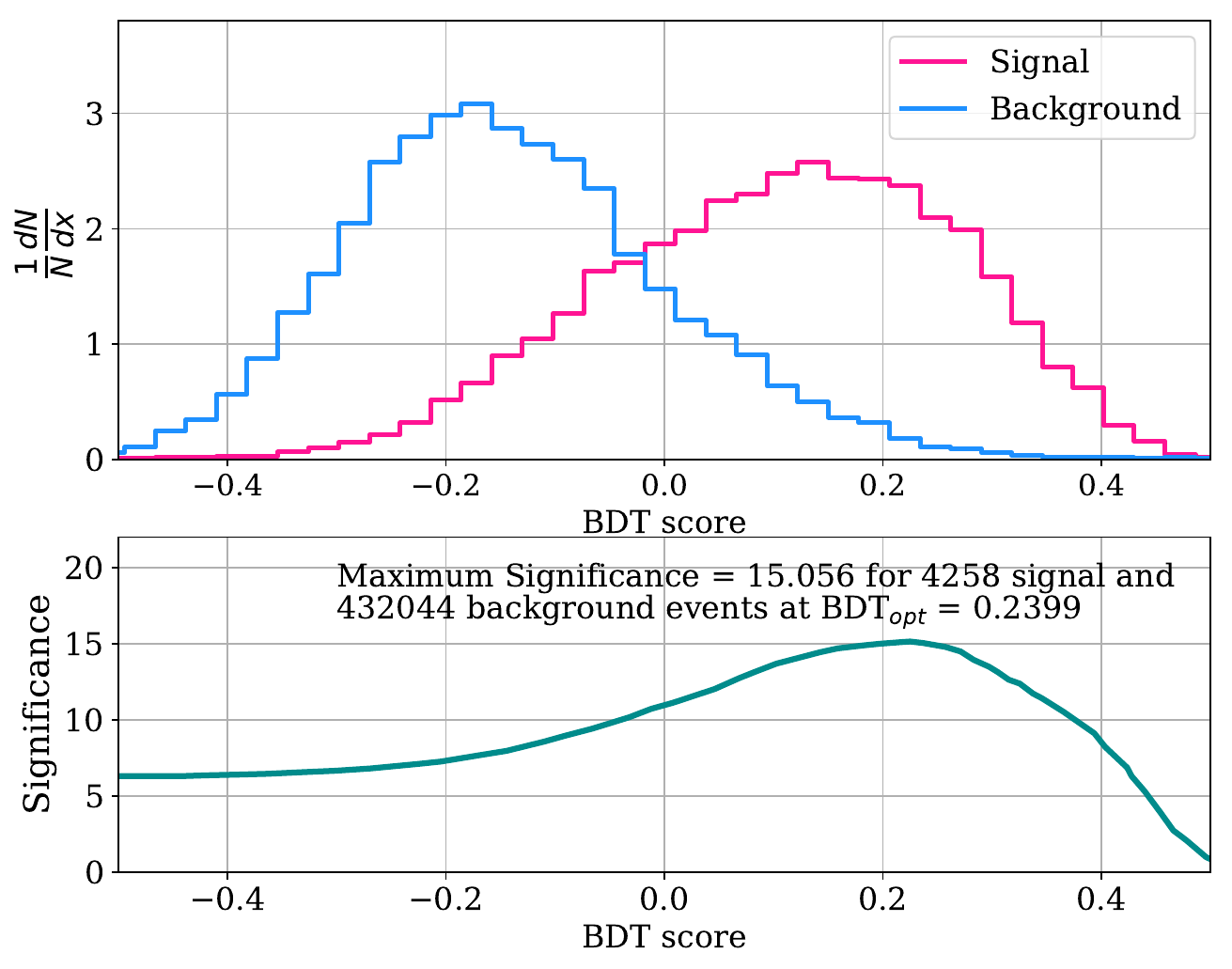}
	\includegraphics[width=0.48\textwidth]{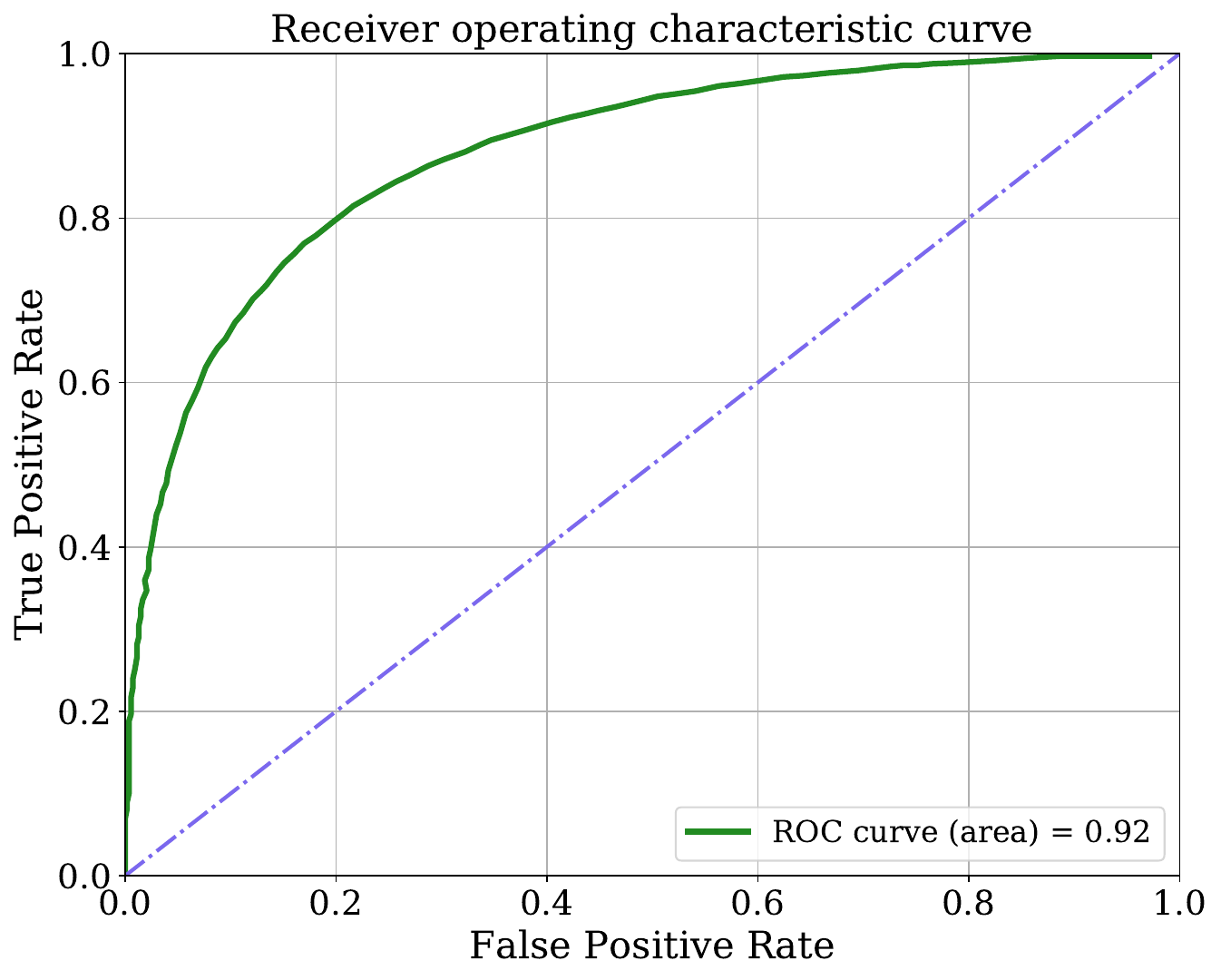}
	\caption{$pp\to Z(\to b\bar{b})\gamma$. \textit{Left}: The normalized BDT score distributions for the signal and the background. Significance as a function of the BDT cut value for BP2 at $\sqrt{s} = 14~$TeV,  $\mathcal{L}_{int}=3000\,\mathrm{fb}^{-1}$. \textit{Right}: ROC curve for BP2.}
	\label{BDT2}
\end{figure}

The classifier, after being trained with these kinematic variables, is used to discriminate the signal benchmark from the background class by computing the significance of observing the signal over the background events. We find that the signal significance over background for the benchmark scenarios BP2 (shown in Fig.~\ref{BDT2} (bottom-left)) and BP3 are $15.056~(4.653)$ and $19.89~(6.102)$, respectively, assuming zero $(5\%)$ systematic uncertainty at 14 TeV HL-LHC.  It is to be noted that there is no significant difference in the spread of the background BDT score for the two BPs with the change in effective coupling $g_{aZ\gamma}$ but the signal distribution spreads away from the background as $g_{aZ\gamma}$ increases. This is also reflected in the signal significance since the signal and background discrimination becomes more obvious with the increase in $g_{aZ\gamma}$. The ROC curve is shown in Fig.~\ref{BDT2} (right).

 For the MVA of the semileptonic channel of the $WW$ production, we consider BP4 ALP mediated signal category and all the relevant background processes which the mimic $1J+1\ell+\slashed{E}_T$ final state in the background class.  The different backgrounds are mixed according to their proper weights to obtain the kinematical distributions for the combined background class. In order to be quantitative, we have applied some weak kinematical cuts than discussed in Sec.~\ref{13tev_ww}, eg. $\slashed{E}_T > 100$ GeV, $m_{W}^{lep} > 65$ GeV, $p_{T,W^{lep}} > 120$ GeV and $p_{T}^J > 100$ GeV on signal and background events in addition to the	pre-selection criteria mentioned in Sec.~\ref{collider_probe}. Upon inspecting various kinematic distributions, we choose the following 12 variables for our multivariate analysis:
\begin{equation}
\tau_{21},\slashed{E}_T,~m_{T}^{W},~m_{eff},~\Delta R_{jj},~m_{J},~p_{T}^{J},~p_{T}^{l},~\Delta \phi (J,p_{T}^{\rm {miss}}),~\Delta R(J,l),~m_{W}^{\rm{lep}}, p_{T,W^{lep}} \nonumber
\end{equation}
where $\Delta \phi (J,p_{T}^{\rm {miss}})$   implies the azimuthal angle separation between the directions of leading fat-jet and missing $p_T$. The final number of signal and background events along with the significance are listed in Table~\ref{BDT_signi}. The four best discriminatory variables are $\tau_{21}, \; ~m_{T}^{W}, \; m_{eff}$ and $\Delta R_{jj}$.

\begin{figure*}[t]
	\begin{center}
		\includegraphics[width=0.48\textwidth]{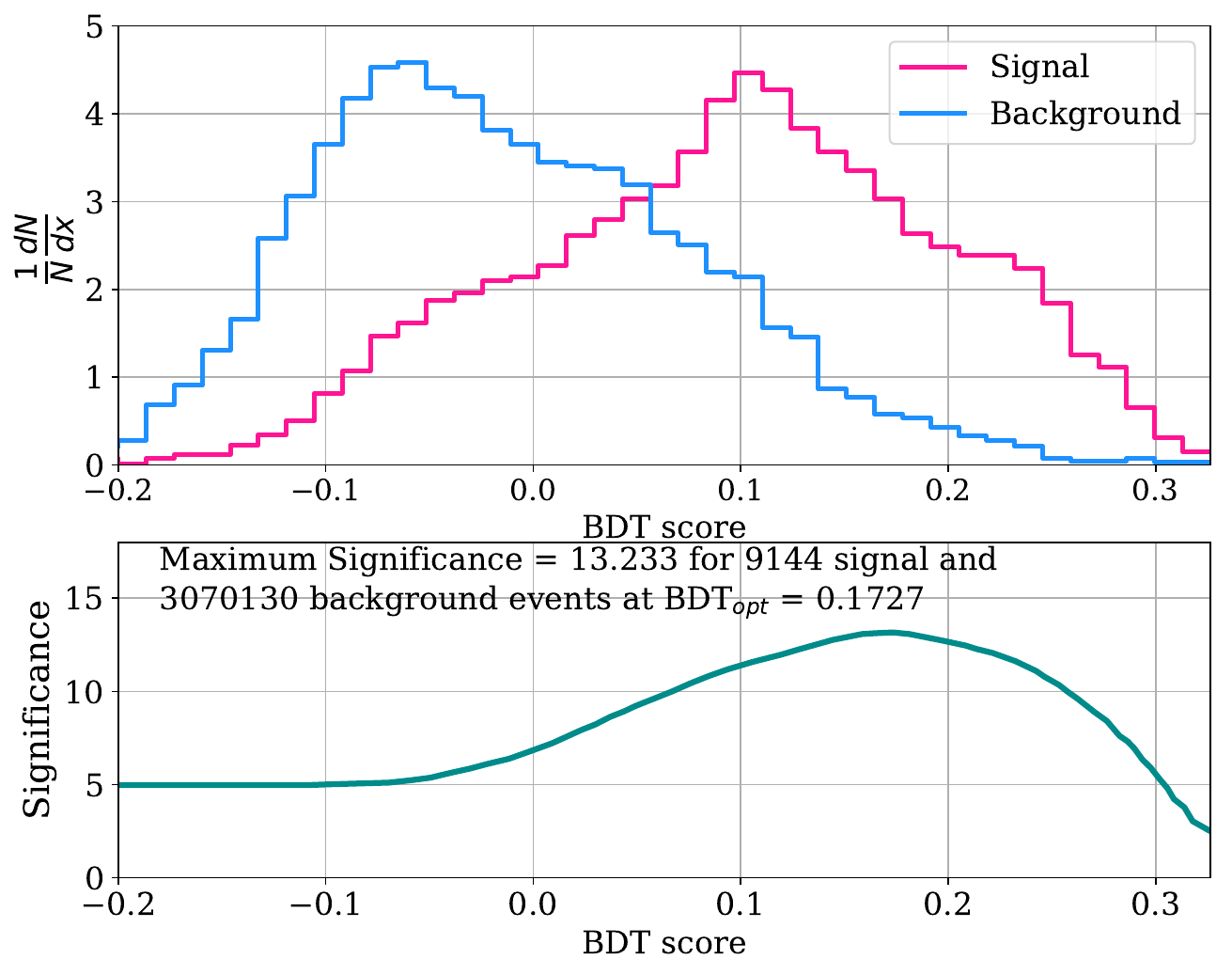}
		\includegraphics[width=0.48\textwidth]{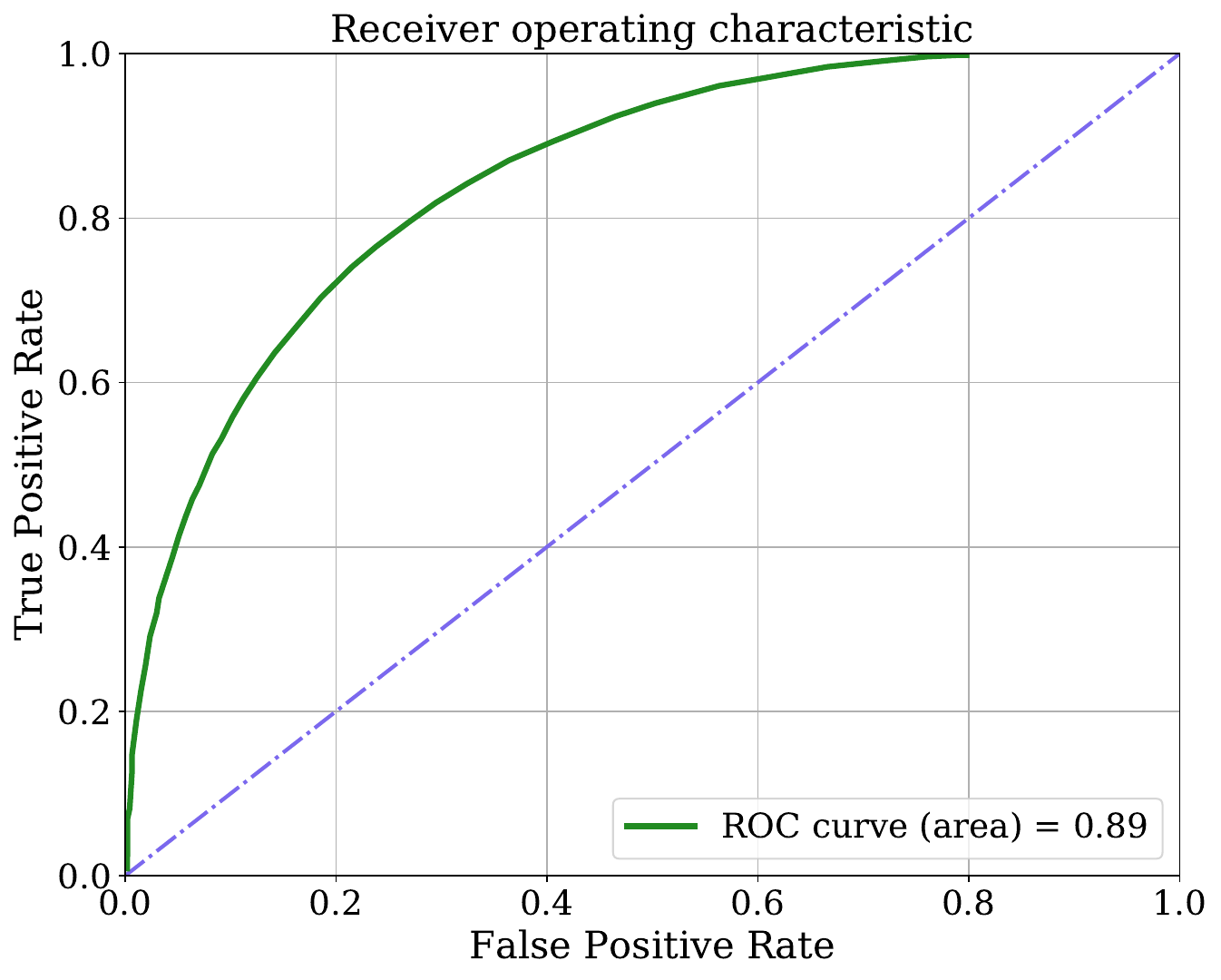}
		\caption{$pp\to WW$ in semileptonic channel. \textit{Left}: The normalized BDT score distributions for the signal and the background. Significance as a function of the BDT cut value for BP4 at $\sqrt{s} = 14~$TeV,  $\mathcal{L}_{int}=3000\,\mathrm{fb}^{-1}$. \textit{Right}: ROC curve for BP4.}
		\label{BDT3}
	\end{center}
\end{figure*}

Finally, with a judicious cut on the BDT score, we find $3052$ signal and $ 49699$ background events, yielding a significance of $13.233$ upon neglecting systematic uncertainties and a significance of $1.985$ taking into account 5\% systematic uncertainties.  In Fig.~\ref{BDT3}, the ROC curve for the benchmark BP4 is shown and an area of $\sim89\%$(BP4) is obtained under ROC curve. 

Before concluding this subsection, we make an attempt to decipher the potential of the leptonic final state for the $WW\gamma$ channel. We study the benchmark scenarios BP5 and BP6 separately for the signal. We consider the same set of cuts as for this channel before performing the multivariate analysis as the cuts are neither too strong nor too loose. For this case, we find the following variables to have the best discriminatory properties.
\begin{equation}
m_{T}^{WW},\slashed{E}_T,~m_{ll},~\Delta R_{ll},~p_{T}^{l_{1}},~p_{T}^{l_{2}},~p_{T}^{\gamma},~\Delta \phi_{ll},~p_{T,ll},~E_{\gamma},~\Delta \eta_{ll},
~m_{ll\gamma},~\eta^{\gamma} \nonumber
\end{equation}
where $p_{T,ll}$ and $\Delta \eta_{ll}$ refer to the $p_T$ of the dilepton system and the rapidity separation between the leptons respectively. The best four variables among these are $\Delta R_{ll},~m_{T}^{WW},~\Delta \eta_{ll}$ and $p_{T}^{\gamma}$.

\begin{figure}[t]
	\includegraphics[width=0.48\textwidth]{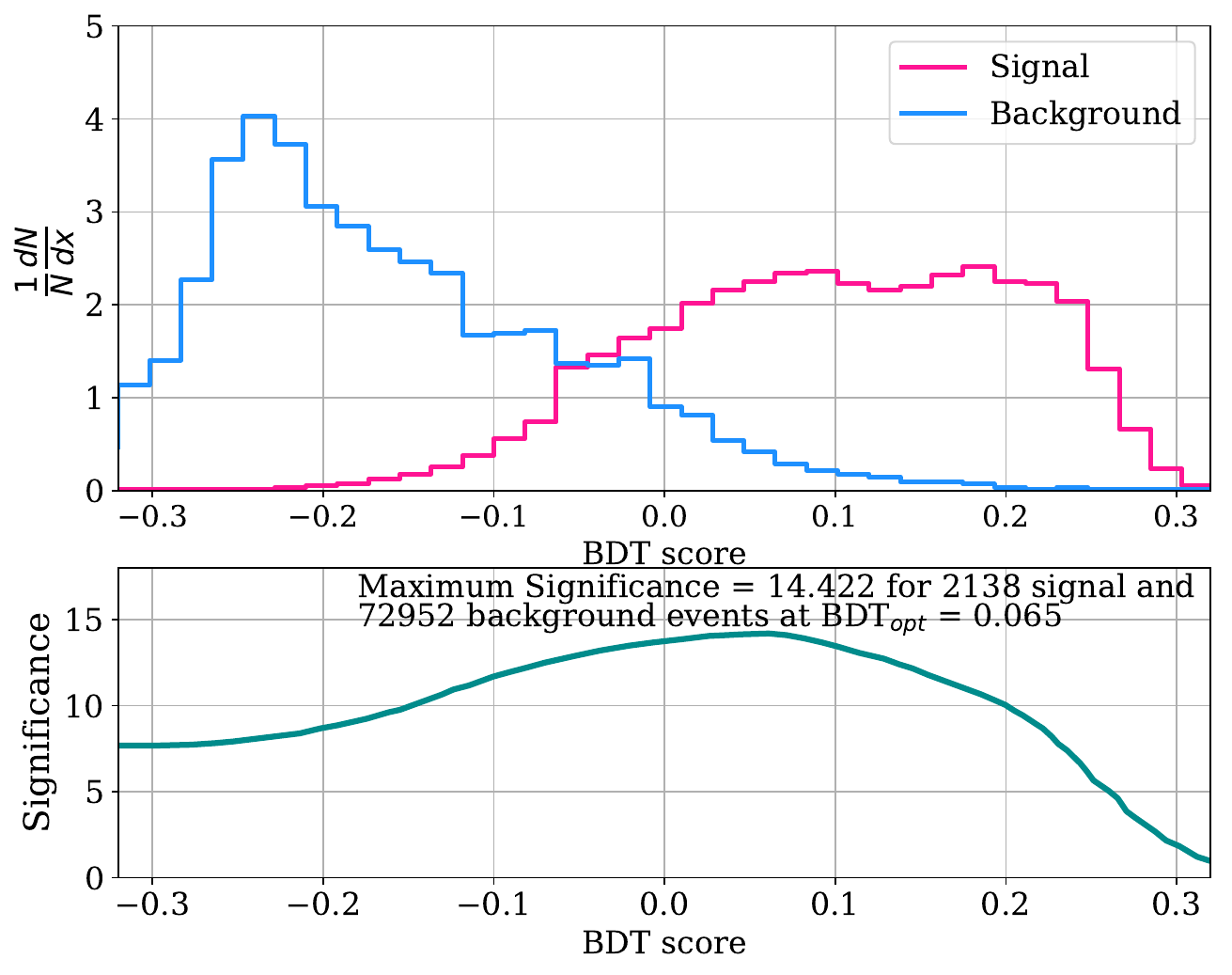}
	\includegraphics[width=0.48\textwidth]{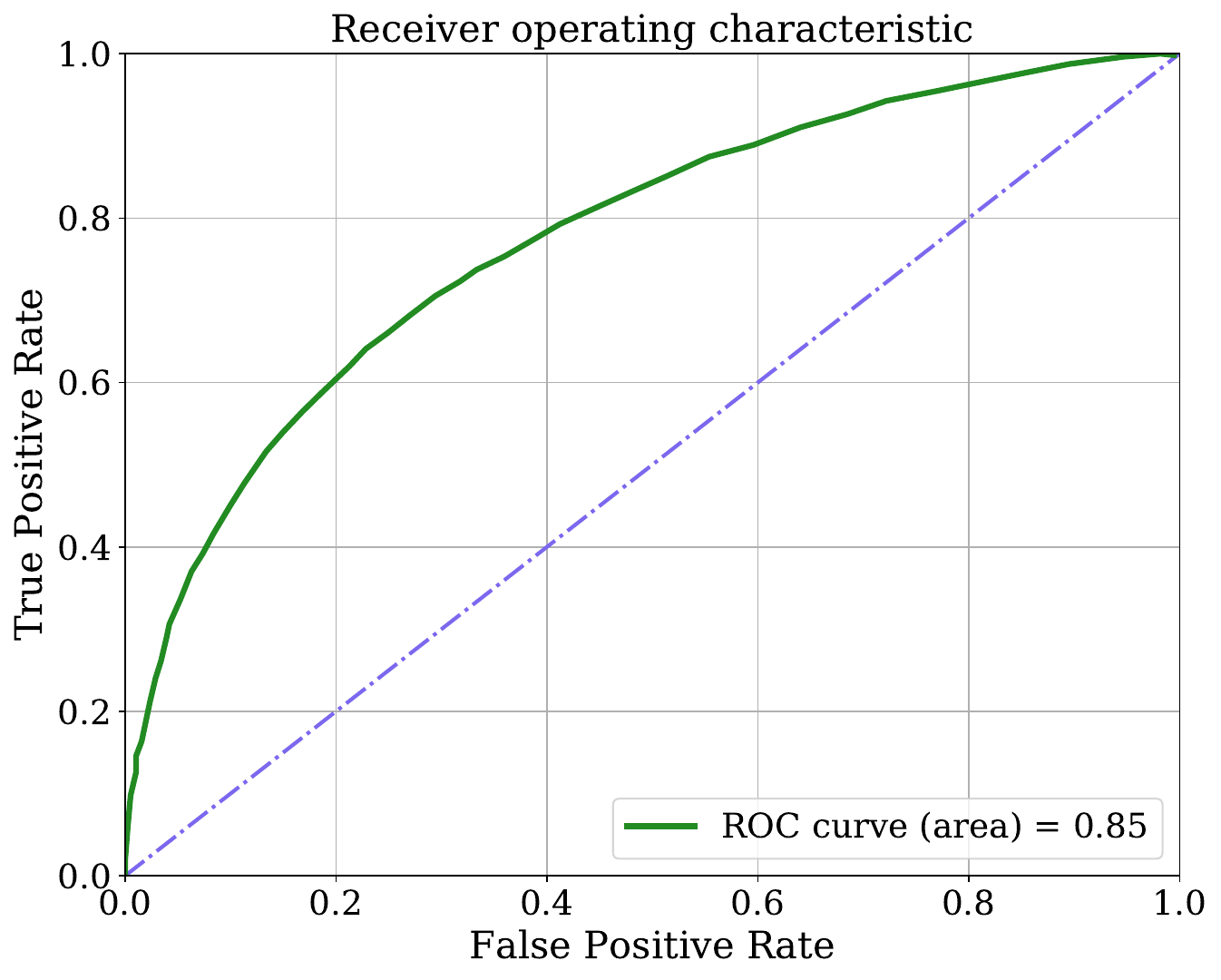}
	\caption{$pp\to WW\gamma$ in fully leptonic channel. \textit{Left}: The normalized BDT score distributions for the signal and the background. Significance as a function of the BDT cut value for BP5 at $\sqrt{s} = 14~$TeV,  $\mathcal{L}_{int}=3000\,\mathrm{fb}^{-1}$. \textit{Right}: ROC curve for BP5.}
	\label{BDT4}
\end{figure}

Hence, in an analogous way to the $WW$ case, we train the classifier with the signal and the background samples, albeit with proper weight factors for the backgrounds. We find a similar significance of $14.422$ and $14.614$  for the benchmark scenarios BP5 and BP6 respectively, assuming zero systematic uncertainties. The results are summarised in Table~\ref{BDT_signi}.  Assuming $5\%$  systematic uncertainties, we obtain a significance of	$3.678$ and $3.726$  for BP5 and BP6 respectively. The response of the classifier and the ROC curve for BP5 are shown in Fig.~\ref{BDT4}.

The signal significance computed for all the benchmark points with Adaptive BDT algorithm is presented in Table~\ref{BDT_signi}. One can compare these results with the ones presented in Table~\ref{significance_CBA}. It is clear that in all cases there is significant improvement from rectangular cut-based analysis. We particularly point out the BP2 and BP3 in case of $Z\gamma$ production. Here, we observe a considerable improvement from the cut-based results. The BDT algorithm finds the best possible combination of feature variables to separate the signal and background by choosing the best possible set of cuts on the most relevant observables. We remark here that the data sample used for training purpose may in principle be subjected to some pre-assigned additional cuts, such as demanding specific invariant masses for opposite-sign dileptons in $WW\gamma$ or using variables that directly are proportional to the energy scale of the process, for instance, the invariant mass of the final state system in the $2\to2$ scattering processes is one of the most important distinguishing features between signal and background. However, to minimize the bias, we do not use it as an input variable to the BDT. Thus, the analysis always has the scope of improvement, by choosing a better set of variables and cuts. However, the variables that we have used are good discriminators as demonstrated in the following.
\begin{table}[h!]
	\centering
	\begin{tabular}{|c||c|c|c|c|c|c|c|}
		\hline \hline
		BPs & $\mathcal{N}_S^{bc}$ &$\mathcal{N}_{\textrm{Bkg}}^{bc}$ & $\mathrm{BDT}_{opt}$ & $\mathcal{N}_{S}(\epsilon_S)$ & $\mathcal{N}_{B}(\epsilon_B)$ &  $\mathcal{S}$  & $\mathcal{S} (5\%$ sys)  \\ 
		\hline 
		BP1 & 2098 &101861 & 0.1982 & 835(0.397) & 9843(0.03477) &13.192 &4.164\\ 
		\hline
		BP2 & 4258  &432044& 0.2399 & 1040(0.2443) & 3652(0.008452)  &15.056&4.653\\ 
		\hline
		BP3 &  5304  &432044& 0.2399 & 1296(0.2443)  &3652(0.008452)   & 19.890 &6.102  \\ 
		\hline
		BP4 & 9144  &3070130& 0.1727 & 3052(0.3338) &49699(0.01619) & 13.233&1.985\\ 
		\hline
		BP5 & 2138   & 72952 &  0.063 &  1914(0.8956)&  16178(0.2218)  &14.422 &3.678  \\ 
		\hline
		BP6 & 2166  &72952& 0.065  & 1940(0.8956) & 16178(0.2218)  & 14.614 &3.726 \\ 
		\hline \hline 
	\end{tabular} 
	\caption{Evaluation of signal and background events at 14 TeV LHC for an integrated luminosity of 3000~fb$^{-1}$. The table includes the number of signal ($\mathcal{N}_S^{bc}$) and background ($\mathcal{N}_{\text{Bkg}}^{bc}$) events before and after applying the optimal BDT cut ($\mathrm{BDT}_{\text{opt}}$), along with the signal ($\epsilon_S$) and background ($\epsilon_B$) acceptance efficiencies at the $\mathrm{BDT}_{\text{opt}}$ cut value are given. The statistical significance ($\mathcal{S}$ with no systematic uncertainty) for each benchmark point is presented. The last column presents the signal significance for $5\%$ systematic uncertainty.}
	\label{BDT_signi}
\end{table}
\vspace{-5mm}
\section{ALP couplings and masses}
\label{existing_constraints}
\vspace{-3mm}

\begin{figure*}[b]
	\centering
	\includegraphics[width=.495\textwidth]{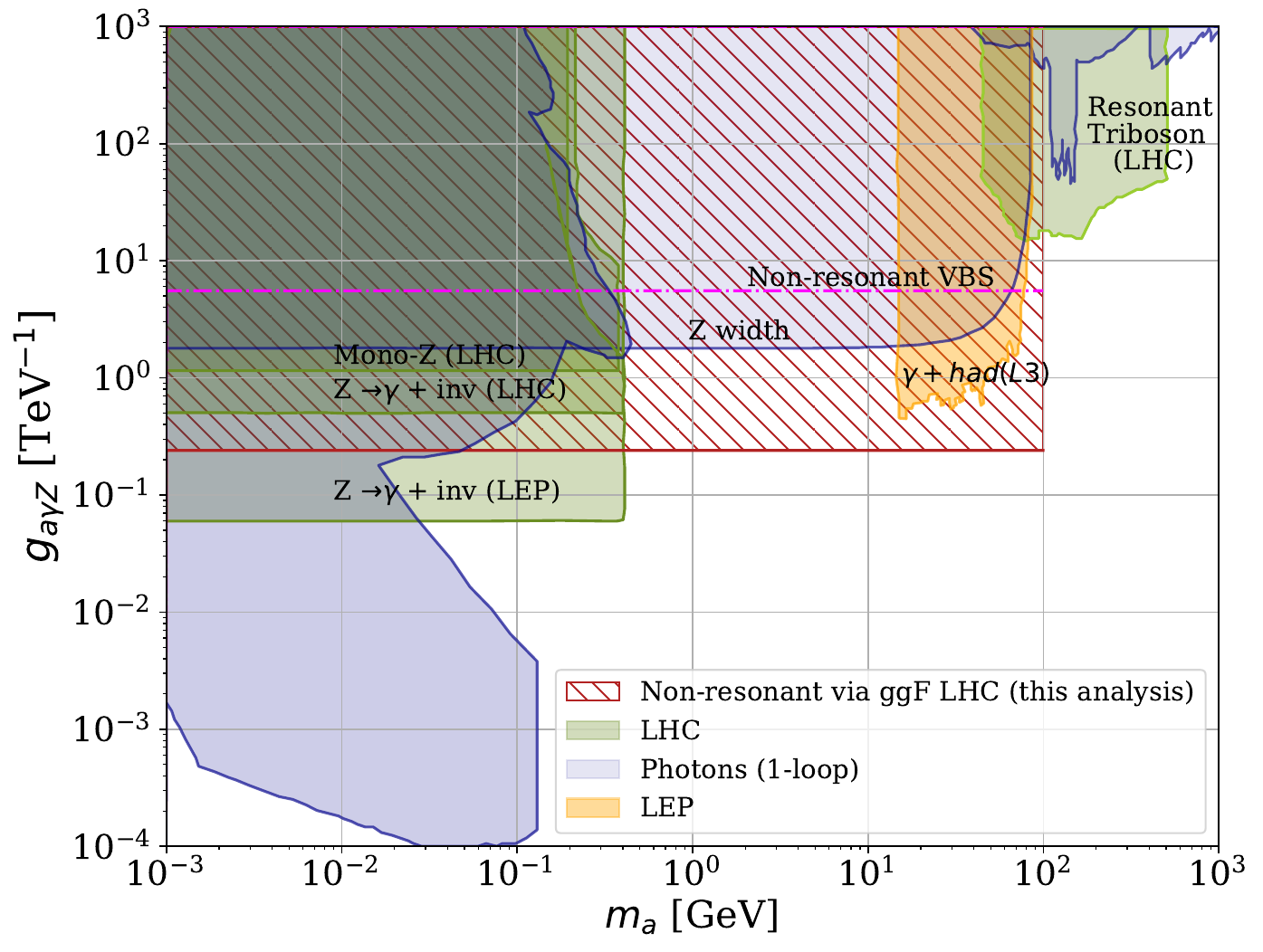}
	\includegraphics[width=.495\textwidth]{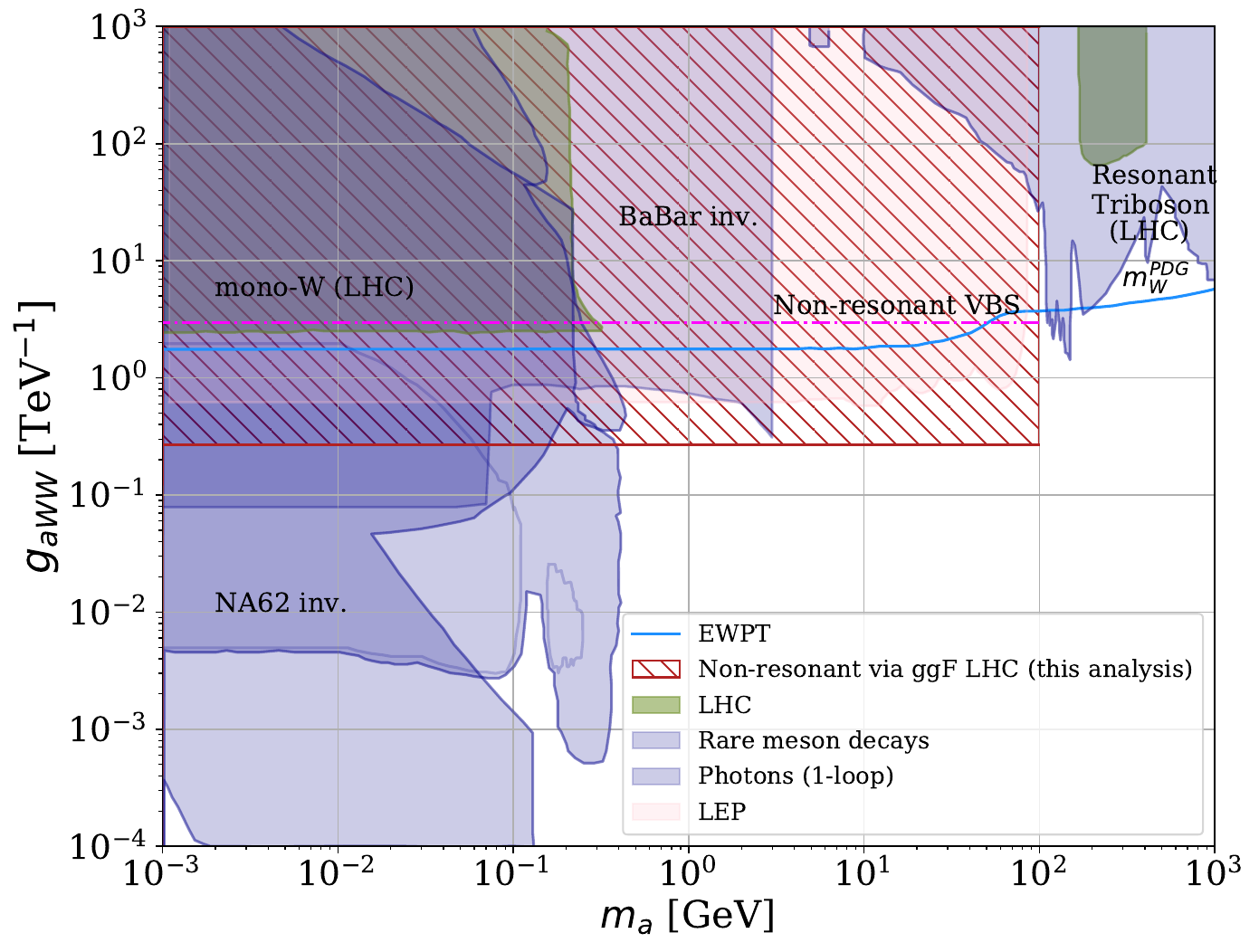}
	\caption{Summary of current constraints as a function of the ALP mass and couplings $g_{aZ\gamma}$ (left) and $g_{aWW}$ (right).  Limits derived in this work are labeled ``Non-resonant via ggF at LHC'' and shown in brown. These constraints are normalised with $g_{agg}=1$ TeV$^{-1}$. Bounds from ``Non-resonant VBS'' are shown in magenta. Orange region refers to an assumed gluon dominance $g_{agg} \gg g_{aV_1V_2}$ for constraints from $\gamma+$hadrons search. Green region (constraints from LHC searches such as mono-$W,Z$, resonant triboson production) indicates more complex assumptions on the ALP EW couplings. Bounds with minimal assumptions on the ALP model  are in blue. See the main text for more details.}
	\label{fig.comparison}
\end{figure*}

Fig.~\ref{fig.comparison} illustrates the constraints obtained in our study at 13 TeV, plotting them in the subspace of the EW $g_{aV_1V_2}$ couplings as defined in Eqns.~\eqref{eq:gaZZ} and \eqref{eq:gaZgamma} and the ALP mass $m_a$. We compare the constraints on $g_{aWW}$ and $g_{aZ\gamma}$ with those from various other experiments (See, for instance, Refs.~\cite{Alonso-Alvarez:2018irt,Aiko:2023trb}). A comment is in order. It is important to note that most measurements often rely on several ALP couplings. To depict these constraints on a two-dimensional plane of ($m_a,g_{aV_1V_2}$), it is necessary to employ a specific underlying rationale or theoretical assumptions, which can differ widely among the various constraints applied. In collider searches, the interplay between specific EW couplings $g_{aXY}$ and gluon couplings $g_{agg}$ is important. This relationship is often modelled as $\frac{g_{agg}}{g_{aV_1V_2}} = \frac{\alpha_s}{\alpha_{V_1V_2}}$, motivated by the pseudo Nambu-Goldstone bosons with anomalous couplings generated by the triangle diagram with $\mathcal{O}(1)$ group theory factors (Ref.~\cite{Alonso-Alvarez:2018irt}). For $m_a > 3m_\pi$, with these assumptions and for LHC searches with resonant processes, it is equivalent to consider $g_{agg} \gg g_{aV_1V_2}$. Also, for loop induced contributions, bounds on fermionic or photonic couplings could be translated to EW gauge boson couplings and they involve a logarithmic dependence on the cut-off scale $f_a$, related to $g_{aV_1V_2}$ by $f_a = \frac{\alpha_{V_1V_2}}{2\pi g_{aV_1V_2}}$. To compare the constraints from other experiments, some of these assumptions for LHC searches and loop-induced couplings are incorporated. The constraints derived from the allowed region in the $\frac{c_{\tilde{W}}}{f_a}-\frac{c_{\tilde{B}}}{f_a}$ plane, inherently incorporate gauge invariance relations. The constraints depicted in brown-hatched region in Fig.~\ref{fig.comparison} (left and right panels) are obtained from the non-resonant $gg\to a^*\to V_1V_2$ processes. They scale with $1/g_{agg}$ and for $c_{\tilde G}\to0$, are lifted completely. For visualization purposes, these figures are normalized to $g_{agg}=1~\rm{TeV^{-1}}$. Bounds derived on $g_{aZ\gamma},g_{aWW}$ from the analysis of non-resonant VBS processes in Ref.~\cite{Bonilla:2022pxu} are shown in magenta.
We will discuss some constraints that involve more complex assumptions about the ALP parameter space. Majority of these constraints, particularly those relating to the interactions of ALP with massive gauge bosons, assume that the ALP is stable and focus on the mass range $m_a<1$ GeV. These constraints are derived from mono-$W$ and mono-$Z$ searches at the LHC and for $g_{a\gamma Z}$, from the hitherto unobserved exotic $Z\to \gamma+\text{inv.}$ decays at LEP~\cite{Craig:2018kne} and LHC~\cite{ATLAS:2020uiq}. It is to be noted that, resonant triboson constraints on $g_{aWW}$ and $g_{a\gamma Z}$ are based on a photophobic ALP model~\cite{Craig:2018kne} and they provide dominant bounds for ALP masses above 100 GeV.
		
All these searches for a stable ALP (including mono-$W$, mono-$Z$, $Z\to \gamma+{\rm inv.}$) implicitly assume a sufficiently small ALP decay width, which, in the relevant mass range, implies certain assumptions about its coupling to photons, electrons and muons. If we move away from the stable ALP assumption, a more conservative constraint arises from the total $Z$ decay width measurements at LEP, applicable up to $m_a\lesssim m_Z$\cite{Brivio:2017ije,Craig:2018kne}. The LEP constraints are predicated on negligible decay rates into leptons. It is also important to note that this bound cuts off at $m_a\leq 3m_\pi\simeq 0.5~\rm{GeV}$, as beyond this point, hadronic decay channels for the ALP become kinematically feasible. This leads to potential $Z\to\gamma+\text{hadrons}$ decays~\cite{L3:1992kcg}, introducing additional dependence on ALP-gluon coupling that would require a detailed analysis~\cite{Alonso-Alvarez:2018irt}, possibly weakening the LEP constraints.
		
Various precise SM measurements would be modified due to the presence of a light state, the ALP, coupled to the SM through the electroweak gauge bosons.   This has been extensively discussed in Ref.~\cite{Aiko:2023trb}, where the impact of the ALP on precision observables is explored. The EWPO set an upper limit on the coupling constant $g_{aWW}$, illustrated by the blue line, at  95\% C.L. It is to be noted that the EWPT results align with the SM expectation of $g_{aWW}=0$ at  95\% C.L.  For ALPs with a mass greater than 500 GeV, the EWPT emerges as the most sensitive method for probing their effects. The model becomes less favored for values of $g_{aWW}\gtrsim4-6$ TeV$^{-1}$.

Precise limits on the rare Kaon and B-meson decays can be used to set bounds on the ALP. In particular, for an invisible axion, the relevant searches are transitions from $K\to \pi$+invisible and $B\to K$+invisible.

The recent NA62 measurement of $K \to \pi \nu \bar \nu$\cite{NA62:2021zjw} has established new constraints on the new $X$ particles in the decay of $K$. Specifically, it has reported limits on the branching ratio $BR(K\to \pi + X)\lesssim (3-6) \times 10^{-11}$ at 90\% C.L. for $m_a < 110$ MeV, and $BR(K\to \pi + X) \lesssim 10^{-11}$ at 90\% C.L. for $m_a \in [160,260]$ MeV. From the searches of B-decays, the most stringent limit currently comes from BaBar\cite{BaBar:2013npw}, setting $BR(B\to K + \textrm{ inv.}) < 3.2 \times 10^{-5}$ at 90\% C.L. for $m_a \lesssim 5$ GeV. However, Belle II has already achieved comparable results with a limit of $4.1\times 10^{-5}$\cite{Kurz:2022rsg}, and is expected to reach approximately $10^{-6}$\cite{Belle-II:2022cgf} with 1 ab$^{-1}$ data. These decays would be mediated by a loop with a W as virtual states, and where the W would radiate an ALP. We can compare with the current NA62 and Babar limits (shown in blue bound regions) to obtain mass-dependent limits on $g_{aWW}$  that uniquely contributes to rare meson decays at the 1-loop level~\cite{Izaguirre:2016dfi,Gavela:2019wzg} (in blue shaded region).  In the case of $g_{aZ\gamma}$ also, much of the mass range addressed by this analysis was already covered by LEP studies. However, our analysis expands the detection scope to lower couplings by nearly an order of magnitude. 
		
While the resonant triboson production yields stringent constraints in the mass range above 100 GeV, the non-resonant $WW\gamma$ process provides constraints valid over a mass window from 1 MeV to 100 GeV. 
		
The constraints labelled as ``Photons (1-loop)'' derived are from a combination of beam dump experiments, observations from supernova SN1987a and LHC studies. For ALP masses below the GeV scale, beam dump searches (in blue region)~\cite{Riordan:1987aw,Bjorken:1988as,Blumlein:2013cua} as compiled in Ref.~\cite{Dobrich:2015jyk} and energy-loss considerations related to supernova SN1987a~\cite{Payez:2014xsa,Jaeckel:2017tud}  set limits on $g_{aZ\gamma}$. These parameters are primarily constrained by the absence of additional cooling and a lack of photon bursts from decaying axions. Due to radiative corrections of axion-boson couplings to axion-photon couplings,  these results after translation can  help establish bounds on $g_{aWW}$ and $g_{aZ\gamma}$, assuming minimal dependence on $f_a$~\cite{Alonso-Alvarez:2018irt}.
		
The use of MVA techniques and improved search strategies are also likely to significantly refine these constraints.

Summarizing, the primary advantage of non-resonant searches lies in their ability to directly probe ALP interactions with EW bosons at the tree level, across a broad range of ALP masses, with minimal dependence on specific model assumptions. This work included processes initiated by gluons, which are influenced by the value of $g_{agg}$. For this analysis, we set $g_{agg}$ at 1 TeV$^{-1}$. 
In Ref.~\cite{Carra:2021ycg}, the ALP-mediated $WW$ and $Z\gamma$ production processes have been studied in the fully leptonic decays of the massive gauge bosons. The 95$\%$ C.L. exclusion limits valid upton $m_a \le 100$ GeV, assuming $g_{agg}=1$ TeV$^{-1}$ are $g_{aWW}<0.62$~TeV$^{-1}$ and $g_{aZ\gamma}<0.37$~TeV$^{-1}$.
In cases where $g_{agg}$ falls below a certain level, non-resonant constraints from EW processes, such as those from vector boson scattering processes, could become more prominent, depending on the specific EW coupling being probed. These constraints have been studied in Ref.~\cite{Bonilla:2022pxu} and the 95$\%$ C.L.  limits derived on the aforementioned two couplings are :  $g_{aWW}<2.98$~TeV$^{-1}$ and $g_{aZ\gamma}<5.54$~TeV$^{-1}$.
		
\vspace{-5mm}
\section{Summary and Conclusions}
\label{summary}
\vspace{-3mm}
Exploring the phenomenology of new, light, propagating particles such as the axion-like particles are pivotal to beyond the SM endeavours as investigated at, for eg. the LHC. The LHC allows for a plethora of processes that are sensitive to the ALPs and in recent times, it has expanded the range of probing the ALP interactions particularly with electroweak bosons and the top quark. Similarly the Higgs particle, whose global understanding  still remains elusive, presents a vital area for potential discoveries in new physics. As experiments at the LHC gain in their sensitivity to rare phenomena, they may  unveil evidence of new physics linked to the Higgs. Our study focuses on the interactions of ALP with the SM Higgs boson and the electroweak gauge bosons through non-resonant searches at the LHC. In particular, we have studied the potential impact of ALP couplings in the effective theory framework, on the production of $Zh, Z\gamma, WW$ and $WW\gamma$ processes at the LHC. Here, the ALP serves as an off-shell mediator in these scattering processes. The key strategy utilises the presence of explicit dependence of derivative interaction of the ALP with the SM bosons. As a consequence, there is a high energy growth of these scattering processes which deviates significantly from the SM. This has been exhibited in the regime where $\sqrt{\hat{s}}\gg v$ and the ALP mass respects $m_a \ll \sqrt{\hat{s}}$. Additionally, we ensure the consistency of the ALP EFT expansion with $\sqrt{\hat{s}}\ll f_a$. With reinterpretation of the public data from the ATLAS and CMS collaborations at 13 TeV for the measurement of the aforementioned SM processes, we obtained constraints on ALP couplings to SM gauge bosons in the set $\{g_{aZh},g_{aZ\gamma},g_{aWW}, g_{aWW\gamma}\}$. We underline the importance of using information from differential distributions in the high energy tails of the final system mass spectrum. The limits we get are rigorous across a broad mass window of ALP from 1 MeV to 100 GeV, assuming an ALP-gluonic coupling exists. For the $Zh$ and $Z\gamma$ production processes, depending on the value of the scale $f_a$ and $g_{agg}=1$ TeV$^{-1}$, upper limits on  the ALP coupling to $Zh$ and $Z\gamma$ of $a_{2D}=0.078$ TeV$^{-1}$ and $|c_{\tilde B}-c_{\tilde W}|=0.073$ TeV$^{-1}$ have been extracted at 95\% C.L. We also carried out the analyses for $WW$ and $WW\gamma$ processes which provide a handle to probe the coupling $c_{\tilde{W}}$. We find that these processes impose a constraint of $c_{\tilde{W}} < $ 0.068 TeV$^{-1}$ and 0.147  TeV$^{-1}$, respectively. Combining these channels yield an additional constraint of $c_{\tilde B} < $ 0.075 TeV$^{-1}$. Among the multi-boson final states, the $Z\gamma$ channel enjoys the highest sensitivity.

We have chosen a few representative benchmark points which give distinct signatures from the SM backgrounds in the boosted regime. The potential of HL-LHC in probing these ALP interactions via non-resonant searches with the chosen BPs are examined and projections for integrated luminosities up to 3000 fb$^{-1}$ at 14 TeV LHC are presented. The upcoming HL-LHC program will allow for improved sensitivity of ALPs through their relevant electroweak boson couplings at discovery level. Detection of statistically significant $Zh$ signal events mediated by the ALP  at LHC would essentially indicate an evidence of {\emph{non-linear}} EWSB.

To explore for potential improvement of the sensitivity for the non-resonant signals at the LHC, we employed a {\em multivariate analysis}. This method differs from the rectangular cut-based analysis by considering all the input kinematic variables at one go and providing  an optimal separation between the signal and the background yields. We utilized a boosted decision tree network algorithm and trained it with  a variety of kinematic variables specific to each of the relevant process to enhance the signal distinction. The results show a clear improvement in the LHC sensitivity to detect new interactions using this method, especially for the benchmark points we considered.

The associated production of ALP is another complementary probe. We also concluded that if the ALP is collider stable and escapes detection, the $W+$ MET (mono-$W$) signature in terms of a direct search for ALP production with a $W$ boson is more sensitive than the off-shell mediated processes involving the ALP-$W$ interactions while processes such as mono-Higgs and mono-$Z$ are less sensitive in the direct probes than the corresponding non-resonant ALP signal analysis carried out. Nevertheless,  a comprehensive global analysis of both the direct and indirect ALP searches would yield more information on the constraints  of the various ALP operators, both in the linear and non-linear frameworks, with emphasis on effects responsible for electroweak symmetry breaking.

The non-resonant searches offer a complementary probe for very light ALP masses. The main advantage lies in the independence of specific assumptions on the ALP characteristics.  Exploring phenomenology of additional processes like di-higgs production, vector boson fusion channels, $WWZ, ZZ\gamma$ processes and other multi-particle productions could further refine our understanding of the ALP parameter space, providing access to disentangle between various operators in both, linear and non-linear mechanisms. While the EFT usually serves as a useful model-independent theoretical framework for experimental searches, expanding the works in the direction of UV completions could predict sensitivity of (model-dependent) degrees of freedom and signals. With the LHC entering a new phase with higher energy and luminosity, it becomes increasingly important to focus on the possible ALP-mediated processes and dedicated designs of observables and analyses which offer significant sensitivity to phenomena beyond the standard paradigm.
\vspace{-0.5cm}
\section*{Acknowledgements}
\vspace{-0.3cm}
The author would like to thank the organisers of 2023 ``ICTP Summer School on Particle Physics'' where this work was initiated and acknowledges the helpful lectures on the relevant topics. TB is thankful to Prof. Anindya Datta for the useful discussions. The author also acknowledges helpful exchanges with Ilaria Brivio during the course of the work and is thankful to Prof. Sabyasachi Chakraborty and Prof. Nirmal Raj for insightful comments on the manuscript. The work of TB is funded by the Council of Scientific and Industrial Research, Government of India through Senior Research Fellowship [File No. 09/028(1107)/2019-EMR-I].


\begin{thebibliography}{99}
\bibitem{Peccei:1977hh} 
R.~D.~Peccei and H.~R.~Quinn,
{\it CP Conservation in the Presence of Instantons}, \href{https://doi.org/10.1103/PhysRevLett.38.1440}{Phys.\ Rev.\ Lett.\  {\bf 38}, 1440 (1977).}
\bibitem{Peccei:1977ur} 
R.~D.~Peccei and H.~R.~Quinn,
{\it Constraints Imposed by CP Conservation in the Presence of Instantons}, \href{https://doi.org/10.1103/PhysRevD.16.1791}{Phys.\ Rev.\ D {\bf 16}, 1791 (1977)}.
\bibitem{Weinberg:1977ma} 
S.~Weinberg,
{\it A New Light Boson?}, \href{https://doi.org/10.1103/PhysRevLett.40.223}{Phys.\ Rev.\ Lett.\  {\bf 40}, 223 (1978).}
\bibitem{Hook:2014cda} 
A.~Hook,{\it Anomalous solutions to the strong CP problem}, \href{https://doi.org/10.1103/PhysRevLett.114.141801}{Phys.\ Rev.\ Lett.\  {\bf 114}, no. 14, 141801 (2015)}, \href{https://arxiv.org/abs/1411.3325}{arXiv:1411.3325 [hep-ph]}.
\bibitem{Wilczek:1977pj} 
F.~Wilczek,
{\it Problem of Strong  $P$  and  $T$  Invariance in the Presence of Instantons}, \href{https://doi.org/10.1103/PhysRevLett.40.279}{Phys.\ Rev.\ Lett.\  {\bf 40}, 279 (1978).}
\bibitem{Chikashige:1980ui}
Y.~Chikashige, R.~N.~Mohapatra and R.~D.~Peccei,
{\it Are There Real Goldstone Bosons Associated with Broken Lepton Number?}, \href{https://doi.org/10.1016/0370-2693(81)90011-3}{Phys. Lett. B \textbf{98} (1981), 265-268.}
\bibitem{Froggatt:1978nt}
C.~D.~Froggatt and H.~B.~Nielsen,
{\it Hierarchy of Quark Masses, Cabibbo Angles and CP Violation}, \href{https://doi.org/10.1016/0550-3213(79)90316-X}{Nucl. Phys. B \textbf{147} (1979), 277-298.}
\bibitem{Izawa:2002qk}
K.~I.~Izawa, T.~Watari and T.~Yanagida,
{\it Higher dimensional QCD without the strong CP problem}, \href{https://doi.org/10.1016/S0370-2693(02)01663-5}{Phys. Lett. B \textbf{534} (2002), 93-96}, \href{https://arxiv.org/abs/hep-ph/0202171}{arXiv:hep-ph/0202171 [hep-ph]}.
\bibitem{Chang:2000ii}
D.~Chang, W.~F.~Chang, C.~H.~Chou and W.~Y.~Keung,
{\it Large two loop contributions to g-2 from a generic pseudoscalar boson}, \href{https://doi.org/10.1103/PhysRevD.63.091301}{Phys. Rev. D \textbf{63} (2001), 091301}, \href{https://arxiv.org/abs/hep-ph/0009292}{arXiv:hep-ph/0009292 [hep-ph]}.
\bibitem{Graham:2015cka}
P.~W.~Graham, D.~E.~Kaplan and S.~Rajendran,
{\it Cosmological Relaxation of the Electroweak Scale}, \href{https://doi.org/10.1103/PhysRevLett.115.221801}{Phys. Rev. Lett. \textbf{115} (2015) no.22, 221801}, \href{https://arxiv.org/abs/1504.07551}{arXiv:1504.07551 [hep-ph]}.
\bibitem{Jeong:2018jqe}
K.~S.~Jeong, T.~H.~Jung and C.~S.~Shin,
{\it Adiabatic electroweak baryogenesis driven by an axionlike particle}, \href{https://doi.org/10.1103/PhysRevD.101.035009}{Phys. Rev. D \textbf{101} (2020) no.3, 035009}, \href{https://arxiv.org/abs/1811.03294}{arXiv:1811.03294 [hep-ph]}.
\bibitem{Abbott:1982af}
L.~F.~Abbott and P.~Sikivie,
{\it A Cosmological Bound on the Invisible Axion}, \href{https://doi.org/10.1016/0370-2693(83)90638-X}{Phys. Lett. B \textbf{120} (1983), 133-136.}
\bibitem{Dine:1982ah}
M.~Dine and W.~Fischler,
{\it The Not So Harmless Axion}, \href{https://doi.org/10.1016/0370-2693(83)90639-1}{Phys. Lett. B \textbf{120} (1983), 137-141.}
\bibitem{Preskill:1982cy}
J.~Preskill, M.~B.~Wise and F.~Wilczek,
{\it Cosmology of the Invisible Axion}, \href{https://doi.org/10.1016/0370-2693(83)90637-8}{Phys. Lett. B \textbf{120} (1983).}
\bibitem{Graham:2015ouw}
P.~W.~Graham, I.~G.~Irastorza, S.~K.~Lamoreaux, A.~Lindner and K.~A.~van Bibber,
{\it Experimental Searches for the Axion and Axion-Like Particles}, \href{https://doi.org/10.1146/annurev-nucl-102014-022120}{Ann. Rev. Nucl. Part. Sci. \textbf{65} (2015), 485-514}, \href{https://arxiv.org/abs/1602.00039}{arXiv:1602.00039 [hep-ex]}.
\bibitem{Bruggisser:2023npd}
S.~Bruggisser, L.~Grabitz and S.~Westhoff,
{\it Global Analysis of the ALP Effective Theory}, \href{https://arxiv.org/abs/2308.11703}{arXiv:2308.11703 [hep-ph]}.
\bibitem{Irastorza:2018dyq}
I.~G.~Irastorza and J.~Redondo,
{\it New experimental approaches in the search for axion-like particles}, \href{https://doi.org/10.1016/j.ppnp.2018.05.003}{Prog. Part. Nucl. Phys. \textbf{102} (2018)}, \href{https://arxiv.org/abs/1801.08127}{arXiv:1801.08127 [hep-ph]}.
\bibitem{Georgi:1986df}
H.~Georgi, D.~B.~Kaplan and L.~Randall,
{\it Manifesting the Invisible Axion at Low-energies}, \href{https://doi.org/10.1016/0370-2693(86)90688-X}{Phys. Lett. B \textbf{169} (1986), 73-78.}
\bibitem{Mimasu:2014nea}
K.~Mimasu and V.~Sanz,
{\it ALPs at Colliders}, \href{https://doi.org/10.1007/JHEP06(2015)173}{JHEP \textbf{06} (2015), 173}, \href{https://arxiv.org/abs/1409.4792}{arXiv:1409.4792 [hep-ph]}.
\bibitem{Brivio:2017ije}
I.~Brivio, M.~B.~Gavela, L.~Merlo, K.~Mimasu, J.~M.~No, R.~del Rey and V.~Sanz,
{\it ALPs Effective Field Theory and Collider Signatures}, \href{https://doi.org/10.1140/epjc/s10052-017-5111-3}{Eur. Phys. J. C \textbf{77} (2017) no.8, 572}, \href{https://arxiv.org/abs/1701.05379}{arXiv:1701.05379 [hep-ph]}.
\bibitem{Bauer:2017ris} 
M.~Bauer, M.~Neubert and A.~Thamm,
{\it Collider Probes of Axion-Like Particles}, \href{https://doi.org/10.1007/JHEP12(2017)044}{JHEP {\bf 1712}, 044 (2017)}, \href{https://arxiv.org/abs/1708.00443}{arXiv:1708.00443 [hep-ph]}.
\bibitem{Esser:2023fdo}
F.~Esser, M.~Madigan, V.~Sanz and M.~Ubiali, {\it On the coupling of axion-like particles to the top quark}, \href{https://doi.org/10.1007/JHEP09(2023)063}{JHEP \textbf{09} (2023), 063}, \href{https://arxiv.org/abs/2303.17634}{arXiv:2303.17634 [hep-ph]}.
\bibitem{Bauer:2016zfj} 
M.~Bauer, M.~Neubert and A.~Thamm,
{\it Analyzing the CP Nature of a New Scalar Particle via S$\to$Zh Decay}, \href{https://doi.org/10.1103/PhysRevLett.117.181801}{Phys. Rev. Lett. \textbf{117} (2016), 181801}, \href{https://arxiv.org/abs/1610.00009}{arXiv:1610.00009 [hep-ph]}.
\bibitem{Biekotter:2022ovp}
A.~Biek\"otter, M.~Chala and M.~Spannowsky,
{\it New Higgs decays to axion-like particles}, \href{https://doi.org/10.1016/j.physletb.2022.137465}{Phys. Lett. B \textbf{834} (2022), 137465}, \href{https://arxiv.org/abs/2203.14984}{arXiv:2203.14984 [hep-ph]}.
\bibitem{Bjorkeroth:2018dzu}
F.~Bj\"orkeroth, E.~J.~Chun and S.~F.~King,
{\it Flavourful Axion Phenomenology}, \href{https://doi.org/10.1007/JHEP08(2018)117}{JHEP \textbf{08} (2018), 117}, \href{https://arxiv.org/abs/1806.00660}{arXiv:1806.00660 [hep-ph]}.
\bibitem{Dobrich:2018jyi}
B.~D\"obrich, F.~Ertas, F.~Kahlhoefer and T.~Spadaro,
{\it Model-independent bounds on light pseudoscalars from rare B-meson decays}, \href{https://doi.org/10.1016/j.physletb.2019.01.064}{Phys. Lett. B \textbf{790} (2019), 537-544}, \href{https://arxiv.org/abs/1810.11336}{arXiv:1810.11336 [hep-ph]}.
\bibitem{MartinCamalich:2020dfe}
J.~Martin Camalich, M.~Pospelov, P.~N.~H.~Vuong, R.~Ziegler and J.~Zupan,
{\it Quark Flavor Phenomenology of the QCD Axion}, \href{https://doi.org/10.1103/PhysRevD.102.015023}{Phys. Rev. D \textbf{102} (2020) no.1, 015023}, \href{https://arxiv.org/abs/2002.04623}{arXiv:2002.04623 [hep-ph]}.
\bibitem{Bandyopadhyay:2021wbb}
T.~Bandyopadhyay, S.~Ghosh and T.~S.~Roy,
{\it ALP-Pions generalized}, \href{https://doi.org/10.1103/PhysRevD.105.115039}{Phys. Rev. D \textbf{105} (2022) no.11, 115039}, \href{https://arxiv.org/abs/2112.13147}{arXiv:2112.13147 [hep-ph]}.
\bibitem{Bauer:2019gfk}
M.~Bauer, M.~Neubert, S.~Renner, M.~Schnubel and A.~Thamm,
{\it Axionlike Particles, Lepton-Flavor Violation, and a New Explanation of $a_\mu$ and $a_e$}, \href{https://doi.org/10.1103/PhysRevLett.124.211803}{Phys. Rev. Lett. \textbf{124} (2020) no.21, 211803}, \href{https://arxiv.org/abs/1908.00008}{arXiv:1908.00008 [hep-ph]}.
\bibitem{Calibbi:2020jvd}
L.~Calibbi, D.~Redigolo, R.~Ziegler and J.~Zupan,
{\it Looking forward to lepton-flavor-violating ALPs}, \href{https://doi.org/10.1007/JHEP09(2021)173}{JHEP \textbf{09} (2021), 173}, \href{https://arxiv.org/abs/2006.04795}{arXiv:2006.04795 [hep-ph]}.
\bibitem{Dolan:2017osp}
M.~J.~Dolan, T.~Ferber, C.~Hearty, F.~Kahlhoefer and K.~Schmidt-Hoberg,
{\it Revised constraints and Belle II sensitivity for visible and invisible axion-like particles}, \href{https://doi.org/10.1007/JHEP12(2017)094}{JHEP \textbf{03} (2021), 190}, \href{https://arxiv.org/abs/1709.00009}{arXiv:1709.00009 [hep-ph]}.
\bibitem{Acanfora:2023gzr}
F.~Acanfora, R.~Franceschini, A.~Mastroddi and D.~Redigolo,
{\it Fusing photons into nothing, a new search for invisible ALPs and Dark Matter at Belle II}, \href{https://arxiv.org/abs/2307.06369}{arXiv:2307.06369 [hep-ph]}.
\bibitem{Dobrich:2015jyk}
B.~D\"obrich, J.~Jaeckel, F.~Kahlhoefer, A.~Ringwald and K.~Schmidt-Hoberg,
{\it ALPtraum: ALP production in proton beam dump experiments}, \href{https://doi.org/10.1007/JHEP02(2016)018}{JHEP \textbf{02} (2016), 018}, \href{https://arxiv.org/abs/1512.03069}{arXiv:1512.03069 [hep-ph]}.
\bibitem{Dobrich:2019dxc}
B.~D\"obrich, J.~Jaeckel and T.~Spadaro,
{\it Light in the beam dump - ALP production from decay photons in proton beam-dumps}, \href{https://doi.org/10.1007/JHEP05(2019)213}{JHEP \textbf{05} (2019), 213}, \href{https://arxiv.org/abs/1904.02091}{arXiv:1904.02091 [hep-ph]}. [erratum: JHEP \textbf{10} (2020), 046]
\bibitem{Cadamuro:2011fd}
D.~Cadamuro and J.~Redondo,
{\it Cosmological bounds on pseudo Nambu-Goldstone bosons}, \href{https://doi.org/10.1088/1475-7516/2012/02/032}{JCAP \textbf{02} (2012), 032}, \href{https://arxiv.org/abs/1110.2895}{arXiv:1110.2895 [hep-ph]}. 
\bibitem{Millea:2015qra}
M.~Millea, L.~Knox and B.~Fields,
{\it New Bounds for Axions and Axion-Like Particles with keV-GeV Masses}, \href{https://doi.org/10.1103/PhysRevD.92.023010}{Phys. Rev. D \textbf{92} (2015) no.2, 023010}, \href{https://arxiv.org/abs/1501.04097}{arXiv:1501.04097 [astro-ph.CO]}. 
\bibitem{Depta:2020wmr}
P.~F.~Depta, M.~Hufnagel and K.~Schmidt-Hoberg,
{\it Robust cosmological constraints on axion-like particles}, \href{https://doi.org/10.1088/1475-7516/2020/05/009}{JCAP \textbf{05} (2020), 009}, \href{https://arxiv.org/abs/2002.08370}{arXiv:2002.08370 [hep-ph]}. 
\bibitem{Izaguirre:2016dfi} 
E.~Izaguirre, T.~Lin and B.~Shuve,
{\it Searching for Axionlike Particles in Flavor-Changing Neutral Current Processes}, \href{https://doi.org/10.1103/PhysRevLett.118.111802}{Phys. Rev. Lett. \textbf{118} (2017) no.11, 111802}, \href{https://arxiv.org/abs/1611.09355}{arXiv:1611.09355 [hep-ph]}.
\bibitem{Alonso-Alvarez:2018irt} 
G.~Alonso-Alvarez, M.~B.~Gavela and P.~Quilez,
{\it Axion couplings to electroweak gauge bosons}, \href{https://doi.org/10.1140/epjc/s10052-019-6732-5}{Eur. Phys. J. C \textbf{79} (2019) no.3, 223}, \href{https://arxiv.org/abs/1811.05466}{arXiv:1811.05466 [hep-ph]}.
\bibitem{Gavela:2019wzg} 
M.~B.~Gavela, R.~Houtz, P.~Quilez, R.~Del Rey and O.~Sumensari,
{\it Flavor-changing constraints on electroweak ALP couplings}, \href{https://doi.org/10.1140/epjc/s10052-019-6889-y}{Eur. Phys. J. C \textbf{79} (2019) no.5, 369}, \href{https://arxiv.org/abs/1901.02031}{arXiv:1901.02031 [hep-ph]}.
\bibitem{Ebadi:2019gij}
J.~Ebadi, S.~Khatibi and M.~Mohammadi Najafabadi,
{\it New probes for axionlike particles at hadron colliders}, \href{https://doi.org/10.1103/PhysRevD.100.015016}{Phys. Rev. D \textbf{100} (2019) no.1, 015016}, \href{https://arxiv.org/abs/1901.03061}{arXiv:1901.03061 [hep-ph]}.
\bibitem{Ertas:2020xcc}
F.~Ertas and F.~Kahlhoefer,
{\it On the interplay between astrophysical and laboratory probes of MeV-scale axion-like particles}, \href{https://doi.org/10.1007/JHEP07(2020)050}{JHEP \textbf{07} (2020), 050}, \href{https://arxiv.org/abs/2004.01193}{arXiv:2004.01193 [hep-ph]}.
\bibitem{Gori:2020xvq}
S.~Gori, G.~Perez and K.~Tobioka,
{\it KOTO vs. NA62 Dark Scalar Searches}, \href{https://doi.org/10.1007/JHEP08(2020)110}{JHEP \textbf{08} (2020), 110}, \href{https://arxiv.org/abs/2005.05170}{arXiv:2005.05170 [hep-ph]}.
\bibitem{Kelly:2020dda}
K.~J.~Kelly, S.~Kumar and Z.~Liu,
{\it Heavy axion opportunities at the DUNE near detector}, \href{https://doi.org/10.1103/PhysRevD.103.095002}{Phys. Rev. D \textbf{103} (2021) no.9, 095002}, \href{https://arxiv.org/abs/2011.05995}{arXiv:2011.05995 [hep-ph]}.
\bibitem{Galda:2021hbr}
A.~M.~Galda, M.~Neubert and S.~Renner,
{\it ALP \textemdash{} SMEFT interference}, \href{https://doi.org/10.1007/JHEP06(2021)135}{JHEP \textbf{06} (2021), 135}, \href{https://arxiv.org/abs/2105.01078}{arXiv:2105.01078 [hep-ph]}.
\bibitem{Bonilla:2021ufe}
J.~Bonilla, I.~Brivio, M.~B.~Gavela and V.~Sanz,
{\it One-loop corrections to ALP couplings}, \href{https://doi.org/10.1007/JHEP11(2021)168}{JHEP \textbf{11} (2021), 168}, \href{https://arxiv.org/abs/2107.11392}{arXiv:2107.11392 [hep-ph]}.
\bibitem{Bauer:2021mvw}
M.~Bauer, M.~Neubert, S.~Renner, M.~Schnubel and A.~Thamm,
{\it Flavor probes of axion-like particles}, \href{https://doi.org/10.1007/JHEP09(2022)056}{JHEP \textbf{09} (2022), 056}, \href{https://arxiv.org/abs/2110.10698}{arXiv:2110.10698 [hep-ph]}.
\bibitem{Wang:2021uyb}
D.~Wang, L.~Wu, J.~M.~Yang and M.~Zhang,
{\it Photon-jet events as a probe of axionlike particles at the LHC}, \href{https://doi.org/10.1103/PhysRevD.104.095016}{Phys. Rev. D \textbf{104} (2021) no.9, 095016}, \href{https://arxiv.org/abs/2102.01532}{arXiv:2102.01532 [hep-ph]}.
\bibitem{Craig:2018kne} 
N.~Craig, A.~Hook and S.~Kasko, {\it The Photophobic ALP}, \href{https://doi.org/10.1007/JHEP09(2018)028}{JHEP \textbf{09} (2018), 028}, \href{https://arxiv.org/abs/1805.06538}{arXiv:1805.06538 [hep-ph]}. 
\bibitem{Gavela:2019cmq}
M.~B.~Gavela, J.~M.~No, V.~Sanz and J.~F.~de Troc\'oniz, {\it Nonresonant Searches for Axionlike Particles at the LHC}, \href{https://doi.org/10.1103/PhysRevLett.124.051802}{Phys. Rev. Lett. \textbf{124} (2020) no.5, 051802}, \href{https://arxiv.org/abs/1905.12953}{arXiv:1905.12953 [hep-ph]}.
\bibitem{Carra:2021ycg}
S.~Carra, V.~Goumarre, R.~Gupta, S.~Heim, B.~Heinemann, J.~Kuechler, F.~Meloni, P.~Quilez and Y.~C.~Yap, {\it Constraining off-shell production of axionlike particles with Z\ensuremath{\gamma} and WW differential cross-section measurementsn}, \href{https://doi.org/10.1103/PhysRevD.104.092005}{Phys. Rev. D \textbf{104} (2021) no.9, 092005)}, \href{https://arxiv.org/abs/2106.10085}{arXiv:2106.10085 [hep-ex]}.
\bibitem{Bonilla:2022pxu}
J.~Bonilla, I.~Brivio, J.~Machado-Rodr\'\i{}guez and J.~F.~de Troc\'oniz, {\it Nonresonant searches for axion-like particles in vector boson scattering processes at the LHC}, \href{https://doi.org/10.1007/JHEP06(2022)113}{JHEP \textbf{06} (2022), 113}, \href{https://arxiv.org/abs/2202.03450}{arXiv:2202.03450 [hep-ph]}.
\bibitem{Buchmuller:1985jz}
W.~Buchmuller and D.~Wyler,
{\it Effective Lagrangian Analysis of New Interactions and Flavor Conservation}, \href{https://doi.org/10.1016/0550-3213(86)90262-2}{Nucl.\ Phys.\ B {\bf 268} (1986) 621-653}.
\bibitem{Grzadkowski:2010es}
.~Grzadkowski, M.~Iskrzynski, M.~Misiak and J.~Rosiek, {\it Dimension-Six Terms in the Standard Model Lagrangian}, \href{https://doi.org/10.1007/JHEP10(2010)085}{JHEP {\bf 1010} (2010) 085}, \href{https://arxiv.org/abs/1008.4884}{arXiv:1008.4884 [hep-ph]}.
\bibitem{Choi:1986zw}
K.~Choi, K.~Kang and J.~E.~Kim, {\it Effects of $\eta^\prime$ in Low-energy Axion Physics}, \href{https://doi.org/10.1016/0370-2693(86)91273-6}{Phys.\ Lett.\ B {\bf 181} (1986) 145}.
\bibitem{ATLAS:2019slw} 
The ATLAS collaboration [ATLAS Collaboration],{\it Combined measurements of Higgs boson production and decay using up to $80$ fb$^{-1}$ of proton--proton collision data at $\sqrt{s}=$ 13 TeV collected with the ATLAS experiment}, {ATLAS-CONF-2019-005 }
\bibitem{Alonso:2012px}
R.~Alonso, M.~B.~Gavela, L.~Merlo, S.~Rigolin and J.~Yepes,
{\it The Effective Chiral Lagrangian for a Light Dynamical "Higgs Particle"}, \href{https://doi.org/10.1016/j.physletb.2013.04.037}{Phys.\ Lett.\ B {\bf 722} (2013) 330}, Erratum: [Phys.\ Lett.\ B {\bf 726} (2013) 926], \href{https://arxiv.org/abs/1212.3305}{arXiv:1212.3305 [hep-ph]}.  
\bibitem{Feruglio:1992wf}
F.~Feruglio, {\it The Chiral approach to the electroweak interactions}, \href{https://doi.org/10.1142/S0217751X93001946}{Int.\ J.\ Mod.\ Phys.\ A {\bf 8} (1993) 4937}, \href{https://arxiv.org/abs/hep-ph/9301281}{hep-ph/9301281}.
\bibitem{Alonso:2012jc}
R.~Alonso, M.~B.~Gavela, L.~Merlo, S.~Rigolin and J.~Yepes,
{\it Minimal Flavour Violation with Strong Higgs Dynamics}, \href{https://doi.org/10.1007/JHEP06(2012)076}{JHEP {\bf 1206} (2012) 076}, \href{https://arxiv.org/abs/1201.1511}{arXiv:1201.1511 [hep-ph]}.
\bibitem{Azatov:2012bz}
A.~Azatov, R.~Contino and J.~Galloway, {\it Model-Independent Bounds on a Light Higgs}, \href{https://doi.org/10.1007/JHEP04(2012)127}{JHEP {\bf 1204} (2012) 127}, Erratum: [JHEP {\bf 1304} (2013) 140], \href{https://arxiv.org/abs/1202.3415}{ arXiv:1202.3415 [hep-ph]}.
\bibitem{Alonso:2012pz}
R.~Alonso, M.~B.~Gavela, L.~Merlo, S.~Rigolin and J.~Yepes, {\it Flavor with a light dynamical "Higgs particle"}, \href{https://doi.org/10.1103/PhysRevD.87.055019}{Phys.\ Rev.\ D {\bf 87} (2013) no.5,  055019}, \href{https://arxiv.org/abs/1212.3307}{arXiv:1212.3307 [hep-ph]}.
\bibitem{Buchalla:2013rka}
G.~Buchalla, O.~CatÃ  and C.~Krause, {\it Complete Electroweak Chiral Lagrangian with a Light Higgs at NLO}, \href{https://doi.org/10.1016/j.nuclphysb.2014.01.018}{Nucl.\ Phys.\ B {\bf 880} (2014) 552}, Erratum: [Nucl.\ Phys.\ B {\bf 913} (2016) 475], \href{https://arxiv.org/abs/1307.5017}{arXiv:1307.5017 [hep-ph]}.
\bibitem{Brivio:2013pma}
I.~Brivio, T.~Corbett, O.~J.~P.~\'Eboli, M.~B.~Gavela, J.~Gonzalez-Fraile, M.~C.~Gonzalez-Garcia, L.~Merlo and S.~Rigolin, {\it Disentangling a dynamical Higgs}, \href{https://doi.org/10.1007/JHEP03(2014)024}{JHEP {\bf 1403} (2014) 024}, \href{https://arxiv.org/abs/1311.1823}{arXiv:1311.1823 [hep-ph]}.
\bibitem{Cepeda:2019klc}
M.~Cepeda, S.~Gori, P.~Ilten, M.~Kado, F.~Riva, R.~Abdul Khalek, A.~Aboubrahim, J.~Alimena, S.~Alioli and A.~Alves, \textit{et al.}
{\it Report from Working Group 2: Higgs Physics at the HL-LHC and HE-LHC}, \href{https://doi.org/10.23731/CYRM-2019-007.221}{CERN Yellow Rep. Monogr. \textbf{7} (2019), 221-584}, \href{https://arxiv.org/abs/1902.00134}{arXiv:1902.00134 [hep-ph]}.
\bibitem{Biswas:2021qaf}
T.~Biswas, A.~Datta and B.~Mukhopadhyaya,
{\it Following the trail of new physics via the vector boson fusion Higgs boson signal at the Large Hadron Collider}, \href{https://doi.org/10.1103/PhysRevD.105.055028}{Phys. Rev. D \textbf{105} (2022) no.5, 055028}, \href{https://arxiv.org/abs/2107.05503}{arXiv:2107.05503 [hep-ph]}.
\bibitem{Biswas:2022fsr}
T.~Biswas and A.~Datta,
{\it Exploring Higgs-photon production at the LHC}, \href{https://doi.org/10.1007/JHEP05(2023)104}{JHEP \textbf{05} (2023), 104}, \href{https://arxiv.org/abs/2208.08432}{arXiv:2208.08432 [hep-ph]}.
\bibitem{Brivio:2016fzo}
I.~Brivio, J.~Gonzalez-Fraile, M.~C.~Gonzalez-Garcia and L.~Merlo,
{\it The complete HEFT Lagrangian after the LHC Run I}, \href{https://doi.org/10.1140/epjc/s10052-016-4211-9}{Eur. Phys. J. C \textbf{76} (2016) no.7, 416}, \href{https://arxiv.org/abs/1604.06801}{arXiv:1604.06801 [hep-ph]}.
\bibitem{Herrero:2020dtv}
M.~Herrero and R.~A.~Morales,
{\it Anatomy of Higgs boson decays into $\gamma \gamma$ and $\gamma Z$ within the electroweak chiral Lagrangian in the $R_\xi$ gauges}, \href{https://doi.org/10.1103/PhysRevD.102.075040}{Phys. Rev. D \textbf{102} (2020) no.7, 075040}, \href{https://arxiv.org/abs/2005.03537}{arXiv:2005.03537 [hep-ph]}.
\bibitem{Herrero:2021iqt}
M.~J.~Herrero and R.~A.~Morales,
{\it One-loop renormalization of vector boson scattering with the electroweak chiral Lagrangian in covariant gauges}, \href{https://doi.org/10.1103/PhysRevD.104.075013}{Phys. Rev. D \textbf{104} (2021) no.7, 075013}, \href{https://arxiv.org/abs/2107.07890}{arXiv:2107.07890 [hep-ph]}.
\bibitem{Alloul:2013bka} 
A.~Alloul, N.~D.~Christensen, C.~Degrande, C.~Duhr and B.~Fuks,
\emph{{FeynRules  2.0 - A complete toolbox for tree-level phenomenology}},
\href{https://doi.org/10.1016/j.cpc.2014.04.012}{\emph{Comput. \;Phys.\;Commun.}  {\bf 185}, 2250 (2014)},
\href{https://arxiv.org/abs/1310.1921}{{\ttfamily arXiv:1310.1921 [hep-ph]}}.
\bibitem{Alwall:2014hca}
J.~Alwall {\it et al.},
{\it The automated computation of tree-level and next-to-leading order differential cross sections, and their matching to parton shower simulations}, \href{https://doi.org/10.1007/JHEP07(2014)079}{JHEP {\bf 1407} (2014) 079}, \href{https://arxiv.org/abs/1405.0301}{arXiv:1405.0301 [hep-ph]}.
\bibitem{Sjostrand:2014zea}
T.~Sj\"ostrand, S.~Ask, J.~R.~Christiansen, R.~Corke, N.~Desai, P.~Ilten, S.~Mrenna, S.~Prestel, C.~O.~Rasmussen and P.~Z.~Skands, {\it An Introduction to PYTHIA 8.2}, \href{https://doi.org/10.1016/j.cpc.2015.01.024}{Comput.\ Phys.\ Commun.\  {\bf 191} (2015) 159}, \href{https://arxiv.org/abs/1410.3012}{arXiv:1410.3012 [hep-ph]}.
\bibitem{pdf}
R.~D.~Ball,  et al., \emph{Parton distributions with LHC data}, \href{https://doi.org/10.1016/j.nuclphysb.2012.10.003}{\emph{Nucl. Phys.} \textbf{B867}, 244 (2013)},
\href{https://arxiv.org/abs/1207.1303}{{\ttfamily arXiv:1207.1303 [hep-ph]}}. 
\bibitem{MLMMerging}
J.~Alwall, S.~Hoche, F.~Krauss, N.~Lavesson, L.~Lonnblad, F.~Maltoni, M.~L.~Mangano, M.~Moretti, C.~G.~Papadopoulos and F.~Piccinini, \textit{et al.}
\emph{{Comparative study of various algorithms for the merging of parton showers and matrix elements in hadronic collisions}}, 
\href{https://doi.org/10.1140/epjc/s10052-007-0490-5}{\emph{Eur. Phys. J.}  \textbf{C53}, 473-500 (2008)}, \href{https://arxiv.org/abs/0706.2569}{{\ttfamily arXiv:0706.2569 [hep-ph]}}.  
\bibitem{deFavereau:2013fsa}
{\scshape DELPHES 3} collaboration, J.~de~Favereau et~al., \emph{{DELPHES 3, A modular
		framework for fast simulation of a generic collider experiment}},
\href{https://doi.org/10.1007/JHEP02(2014)057}
{\emph{JHEP} {\bfseries 02} (2014) 057}, \href{https://arxiv.org/abs/1307.6346}{{\ttfamily arXiv:1307.6346 [hep-ph]}}.
\bibitem{Cacciari:2011ma}
M.~Cacciari, G.~P.~Salam and G.~Soyez,
\emph{FastJet User Manual},
\href{https://doi.org/10.1140/epjc/s10052-012-1896-2}{\emph{Eur. Phys. J.} \textbf{C72}, 1896 (2012)},
\href{https://arxiv.org/abs/1111.6097}{{\ttfamily arXiv:1111.6097 [hep-ph]}}.
\bibitem{ATLAS:2022enb}
G.~Aad \textit{et al.} [ATLAS],
{\it Search for heavy resonances decaying into a $Z$ or $W$ boson and a Higgs boson in final states with leptons and $b$-jets in $139~$fb$^{-1}$ of $pp$ collisions at $\sqrt{s}=13~$TeV with the ATLAS detector}, \href{https://doi.org/10.1007/JHEP06(2023)016023), 016}{JHEP \textbf{06} (2023), 016}, \href{https://arxiv.org/abs/2207.00230}{arXiv:2207.00230 [hep-ex]}.
\bibitem{ATLAS:2018sxj}
M.~Aaboud \textit{et al.} [ATLAS],
{\it Search for heavy resonances decaying to a photon and a hadronically decaying $Z/W/H$ boson in $pp$ collisions at $\sqrt{s}=13$ $\mathrm{TeV}$ with the ATLAS detector},\href{https://doi.org/10.1103/PhysRevD.98.032015}{
	Phys. Rev. D \textbf{98} (2018) no.3, 032015},\href{https://arxiv.org/abs/1805.01908}{arXiv:1805.01908 [hep-ex]}.
\bibitem{CMS:2019ppl}
A.~M.~Sirunyan \textit{et al.} [CMS],
{\it Search for anomalous triple gauge couplings in WW and WZ production in lepton + jet events in proton-proton collisions at $\sqrt{s} =$ 13 TeV}, \href{https://doi.org/10.1007/JHEP12(2019)062}{JHEP \textbf{12} (2019), 062}, \href{https://arxiv.org/abs/1907.08354}{arXiv:1907.08354 [hep-ex]}.
\bibitem{CMS:2023rcv}
A.~Hayrapetyan \textit{et al.} [CMS],
{\it Observation of WW$\gamma$ production and search for H$\gamma$ production in proton-proton collisions at $\sqrt{s}$ = 13 TeV},\href{https://arxiv.org/abs/2310.05164}{arXiv:2310.05164 [hep-ex]}.
\bibitem{Butterworth:2008iy}
J.~M.~Butterworth, A.~R.~Davison, M.~Rubin and G.~P.~Salam,
\emph{{Jet substructure as a new Higgs search channel at the LHC}},  \href{https://doi.org/10.1103/PhysRevLett.100.242001}{Phys. Rev. Lett. \textbf{100}, 242001 (2008)}, \href{https://arxiv.org/abs/0802.2470}{{\ttfamily arXiv:0802.2470 [hep-ph]}}.
\bibitem{CMS:2021xor}
A.~Tumasyan \textit{et al.} [CMS],
\emph{{Search for heavy resonances decaying to ZZ or ZW and axion-like particles mediating nonresonant ZZ or ZH production at $ \sqrt{s} $ = 13 TeV}}, 
\href{https://doi.org/10.1007/JHEP04(2022)087}{JHEP \textbf{04}, 087 (2022)}, \href{https://arxiv.org/abs/2111.13669}{{\ttfamily arXiv:2111.13669 [hep-ex]}}.
\bibitem{hepdata}
\url{https://www.hepdata.net/ }
\bibitem{Catani:2009sm}
S.~Catani, L.~Cieri, G.~Ferrera, D.~de Florian and M.~Grazzini,
{\it Vector boson production at hadron colliders: a fully exclusive QCD calculation at NNLO}, \href{https://doi.org/10.1103/PhysRevLett.103.082001}{\emph{Phys. Rev. Lett. \textbf{103} (2009), 082001}},
\href{https://arxiv.org/abs/0903.2120}{{\ttfamily arXiv:0903.2120 [hep-ph]}}.
\bibitem{Balossini:2009sa}
G.~Balossini, G.~Montagna, C.~M.~Carloni Calame, M.~Moretti, O.~Nicrosini, F.~Piccinini, M.~Treccani and A.~Vicini,
{\it Combination of electroweak and QCD corrections to single W production at the Fermilab Tevatron and the CERN LHC}, \href{https://doi.org/10.1007/JHEP01(2010)013}{\emph{JHEP \textbf{01} (2010), 013}},
\href{https://arxiv.org/abs/0907.0276}{{\ttfamily arXiv:0907.0276 [hep-ph]}}.
\bibitem{Kidonakis:2015nna}
N.~Kidonakis,
{\it Theoretical results for electroweak-boson and single-top production}, \href{https://doi.org/10.22323/1.247.0170}{\emph{PoS \textbf{DIS2015} (2015), 170}},
\href{https://arxiv.org/abs/1506.04072}{{\ttfamily arXiv:1506.04072 [hep-ph]}}.
\bibitem{Campbell:2011bn}
J.~M.~Campbell, R.~K.~Ellis and C.~Williams,
{\it Vector boson pair production at the LHC}, \href{https://doi.org/10.1007/JHEP07(2011)018}{\emph{JHEP \textbf{07} (2011), 018}},
\href{https://arxiv.org/abs/1105.0020}{{\ttfamily arXiv:1105.0020 [hep-ph]}}.
\bibitem{Muselli:2015kba}
C.~Muselli, M.~Bonvini, S.~Forte, S.~Marzani and G.~Ridolfi,
{\it Top Quark Pair Production beyond NNLO}, \href{https://doi.org/10.1007/JHEP08(2015)076}{\emph{JHEP \textbf{08} (2015), 076}},
\href{https://arxiv.org/abs/1505.02006}{{\ttfamily arXiv:1505.02006 [hep-ph]}}.
\bibitem{LHCWGxsec}
\href{https://twiki.cern.ch/twiki/bin/view/LHCPhysics/LHCHWG1HELHCXsecs}{https://twiki.cern.ch/twiki/bin/view/LHCPhysics/LHCHWG1HELHCXsecs},
\bibitem{Krause:2017nxq}
J.~Krause and F.~Siegert,
\emph{{NLO QCD predictions for $Z+\gamma$ + jets production with Sherpa}}, \href{https://doi.org/10.1140/epjc/s10052-018-5627-1}{\emph{Eur. Phys. J. C \textbf{78}, no.2, 161 (2018)}},
\href{https://arxiv.org/abs/1708.06283}{{\ttfamily arXiv:1708.06283 [hep-ph]}}.
\bibitem{Lombardi:2020wju}
D.~Lombardi, M.~Wiesemann and G.~Zanderighi,
\emph{{Advancing MiNNLO$_{\rm PS}$ to diboson processes: $Z\gamma$ production at NNLO+PS}}, \href{https://doi.org/10.1007/JHEP06(2021)095}{\emph{JHEP} {\bfseries 06}
	(2021) 095},
\href{https://arxiv.org/abs/2010.10478}{{\ttfamily arXiv:2010.10478 [hep-ph]}}.
\bibitem{Pagani:2021iwa}
D.~Pagani, H.~S.~Shao, I.~Tsinikos and M.~Zaro,
\emph{{Automated EW corrections with isolated photons: $t \bar t \gamma$, $t \bar t \gamma\gamma$ and $t \gamma j$ as case studies}},
\href{https://arxiv.org/abs/2106.02059}{{\ttfamily arXiv:2106.02059 [hep-ph]}}.
\bibitem{Gabrielli:2016mdd}
E.~Gabrielli, B.~Mele, F.~Piccinini and R.~Pittau,
\emph{{Asking for an extra photon in Higgs production at the LHC and beyond}}, \href{https://doi.org/10.1007/JHEP07(2016)003}{\emph{JHEP} \textbf{07}, 003 (2016)}, \href{https://arxiv.org/abs/1601.03635}{{\ttfamily arXiv:1601.03635 [hep-ph]}}.
\bibitem{Jaeckel:2015jla} 
J.~Jaeckel and M.~Spannowsky,
{\it Probing MeV to 90 GeV axion-like particles with LEP and LHC}, \href{https://doi.org/10.1016/j.physletb.2015.12.037}{Phys.\ Lett.\ B {\bf 753}, 482 (2016)}, \href{https://arxiv.org/abs/1509.00476}{arXiv:1509.00476 [hep-ph]}. 
\bibitem{Knapen:2016moh} 
S.~Knapen, T.~Lin, H.~K.~Lou and T.~Melia,
{\it Searching for Axionlike Particles with Ultraperipheral Heavy-Ion Collisions}, \href{https://doi.org/10.1103/PhysRevLett.118.171801}{Phys.\ Rev.\ Lett.\  {\bf 118}, no. 17, 171801 (2017)}, \href{https://arxiv.org/abs/1607.06083}{arXiv:1607.06083 [hep-ph]}. 
\bibitem{Mariotti:2017vtv} 
A.~Mariotti, D.~Redigolo, F.~Sala and K.~Tobioka,
{\it New LHC bound on low-mass diphoton resonances},\href{https://doi.org/10.1016/j.physletb.2015.12.037}{Phys.\ Lett.\ B {\bf 783}, 13 (2018)}, \href{https://arxiv.org/abs/1710.01743}{arXiv:1710.01743 [hep-ph]}.
\bibitem{Busoni:2013lha}
G.~Busoni, A.~De Simone, E.~Morgante and A.~Riotto,
\emph{On the Validity of the Effective Field Theory for Dark Matter Searches at the LHC},\href{http://dx.doi.org/10.1016/j.physletb.2013.11.069}{\emph Phys. Lett. B \textbf{728}, 412-421 (2014)}, \href{http://arxiv.org/abs/1307.2253}{{\tt arXiv:1307.2253 [hep-ph]}}.
\bibitem{Bhattacharya:2015vja}
S.~Bhattacharya and J.~Wudka,
\emph{Dimension-seven operators in the standard model with right handed neutrinos}, \href{http://dx.doi.org/10.1103/PhysRevD.94.055022}{\emph Phys. Rev. D \textbf{94}, no.5, 055022 (2016)} [erratum: \href{http://dx.doi.org/10.1103/PhysRevD.95.039904}{\emph Phys. Rev. D \textbf{95}, no.3, 039904 (2017)}] \href{http://arxiv.org/abs/1505.05264}{{\tt arXiv:1505.05264 [hep-ph]}}.
\bibitem{Thaler:2010tr}
J.~Thaler and K.~Van~Tilburg, \emph{{Identifying Boosted Objects with
		N-subjettiness}},
\href{http://dx.doi.org/10.1007/JHEP03(2011)015}{\emph{JHEP} {\bfseries 03}
	(2011) 015}, \href{https://arxiv.org/abs/1011.2268}{{\ttfamily 1011.2268 [hep-ph]}}.
\bibitem{Thaler:2011gf}
J.~Thaler and K.~Van~Tilburg, \emph{{Maximizing Boosted Top Identification by Minimizing N-subjettiness}},
\href{http://dx.doi.org/10.1007/JHEP02(2012)093}{\emph{JHEP} {\bfseries 02} (2012) 093}, \href{https://arxiv.org/abs/1108.2701}{{\ttfamily 1108.2701 [hep-ph]}}.
\bibitem{ATLAS:2021shl}
G.~Aad \textit{et al.} [ATLAS],
{\it Search for dark matter produced in association with a Standard Model Higgs boson decaying into b-quarks using the full Run 2 dataset from the ATLAS detector}, \href{https://doi.org/10.1007/JHEP11(2021)209}{JHEP \textbf{11} (2021), 209}, \href{https://arxiv.org/abs/2108.13391}{arXiv:2108.13391 [hep-ex]}. 
\bibitem{CMS:2017nxf}
A.~M.~Sirunyan \textit{et al.} [CMS],
{\it Search for new physics in events with a leptonically decaying Z boson and a large transverse momentum imbalance in proton\textendash{}proton collisions at $\sqrt{s} $ = 13 $\,\text {TeV}$}, \href{https://doi.org/10.1140/epjc/s10052-018-5740-1}{Eur. Phys. J. C \textbf{78} (2018) no.4, 291}, \href{https://arxiv.org/abs/1711.00431}{arXiv:1711.00431 [hep-ex]}. 
\bibitem{ATLAS:2019lsy}
G.~Aad \textit{et al.} [ATLAS],
{\it Search for a heavy charged boson in events with a charged lepton and missing transverse momentum from $pp$ collisions at $\sqrt{s} = 13$ TeV with the ATLAS detector}, \href{https://doi.org/10.1103/PhysRevD.100.052013}{Phys. Rev. D \textbf{100} (2019) no.5, 052013}, \href{https://arxiv.org/abs/1906.05609}{arXiv:1906.05609 [hep-ex]}. 
\bibitem{CMS:2022dwd}
A.~Tumasyan \textit{et al.} [CMS],
{\it A portrait of the Higgs boson by the CMS experiment ten years after the discovery}, \href{https://doi.org/10.1038/s41586-022-04892-x}{Nature \textbf{607} (2022) no.7917, 60-68}
Nature \textbf{607} (2022) no.7917, 60-68, \href{https://arxiv.org/abs/2207.00043}{arXiv:2207.00043 [hep-ex]}. 
\bibitem{Workman:2022ynf}
R.~L.~Workman \textit{et al.} [Particle Data Group],
{\it Review of Particle Physics}, \href{https://doi.org/10.1093/ptep/ptac097}{PTEP \textbf{2022}, 083C01 (2022)}.
\bibitem{Coadou:2022nsh}
Y.~Coadou,
{\it Boosted decision trees}, \href{https://arxiv.org/abs/2206.09645}{arXiv:2206.09645 [physics.data-an]}. 
\bibitem{Baldi:2014kfa}
P.~Baldi, P.~Sadowski and D.~Whiteson,
{\it Searching for Exotic Particles in High-Energy Physics with Deep Learning}, \href{https://doi.org/10.1038/ncomms5308}{Nature Commun. \textbf{5} (2014), 4308}, \href{https://arxiv.org/abs/1402.4735}{arXiv:1402.4735 [hep-ph]}. 
\bibitem{Oyulmaz:2019jqr}
K.~Y.~Oyulmaz, A.~Senol, H.~Denizli and O.~Cakir,
{\it Top quark anomalous FCNC production via $tqg$ couplings at FCC-hh}, \href{https://doi.org/10.1103/PhysRevD.99.115023}{Phys. Rev. D \textbf{99} (2019) no.11, 115023}, \href{https://arxiv.org/abs/1902.03037}{arXiv:1902.03037 [hep-ph]}. 
\bibitem{Bakhet:2015uca}
N.~Bakhet, M.~Y.~Khlopov and T.~Hussein,
{\it Neural Networks Search for Charged Higgs Boson of Two Doublet Higgs Model at the Hadrons Colliders}, \href{https://arxiv.org/abs/1507.06547}{arXiv:1507.06547 [hep-ph]}. 
\bibitem{Lasocha:2020ctd}
K.~Lasocha, E.~Richter-Was, M.~Sadowski and Z.~Was,
{\it Deep neural network application: Higgs boson $CP$ state mixing angle in $H \to \tau\tau$ decay and at the LHC}, \href{https://doi.org/10.1103/PhysRevD.103.036003}{Phys. Rev. D \textbf{103} (2021) no.3, 036003}, \href{https://arxiv.org/abs/2001.00455}{arXiv:2001.00455 [hep-ph]}.
\bibitem{Aiko:2023trb}
M.~Aiko and M.~Endo,
{\it Electroweak precision test of axion-like particles}, \href{https://doi.org/10.1007/JHEP05(2023)147}{JHEP \textbf{05} (2023), 147}, 
\href{https://arxiv.org/abs/2302.11377}{arXiv:2302.11377 [hep-ph]}.
\bibitem{ATLAS:2020uiq}
G.~Aad \textit{et al.} [ATLAS],
{\it Search for dark matter in association with an energetic photon in $pp$ collisions at $\sqrt{s}$ = 13 TeV with the ATLAS detector}, \href{https://doi.org/10.1007/JHEP02(2021)226}{JHEP \textbf{02} (2021), 226}, \href{https://arxiv.org/abs/2011.05259}{arXiv:2011.05259 [hep-ph]}.
\bibitem{L3:1992kcg}
O.~Adriani \textit{et al.} [L3],
{\it Isolated hard photon emission in hadronic Z0 decays}, \href{https://doi.org/10.1016/0370-2693(92)91205-N}{Phys. Lett. B \textbf{292} (1992), 472-484}.
\bibitem{NA62:2021zjw}
E.~Cortina Gil \textit{et al.} [NA62],
{\it Measurement of the very rare K$^{+}\rightarrow{} {\pi}^{+}\nu \overline{\nu} $ decay
}, \href{https://doi.org/10.1007/JHEP06(2021)093}{JHEP \textbf{06} (2021), 093}, \href{https://arxiv.org/abs/2103.15389}{arXiv:2103.15389 [hep-ex]}.
\bibitem{BaBar:2013npw}
J.~P.~Lees \textit{et al.} [BaBar],
{\it Search for $B \to K^{(*)} \nu \overline \nu$ and invisible quarkonium decays
}, \href{https://doi.org/10.1103/PhysRevD.87.112005}{Phys. Rev. D \textbf{87} (2013) no.11, 112005}, \href{https://arxiv.org/abs/1303.7465}{arXiv:1303.7465 [hep-ex]}.
\bibitem{Kurz:2022rsg}
S.~Kurz [Belle II],
{\it Search for $B \rightarrow K \nu \bar{\nu}$ and other electroweak/radiative penguin processes at Belle II
}, \href{https://doi.org/10.22323/1.398.0554}{PoS \textbf{EPS-HEP2021} (2022), 554}
\bibitem{Belle-II:2022cgf}
L.~Aggarwal \textit{et al.} [Belle-II],
{\it Snowmass White Paper: Belle II physics reach and plans for the next decade and beyond}, \href{https://arxiv.org/abs/2207.06307}{arXiv:2207.06307 [hep-ex]}.
\bibitem{Riordan:1987aw}
E.~M.~Riordan, M.~W.~Krasny, K.~Lang, P.~De Barbaro, A.~Bodek, S.~Dasu, N.~Varelas, X.~Wang, R.~G.~Arnold and D.~Benton, \textit{et al.}
{\it A Search for Short Lived Axions in an Electron Beam Dump Experiment}, \href{https://doi.org/10.1103/PhysRevLett.59.755}{Phys. Rev. Lett. \textbf{59} (1987), 755}.
\bibitem{Bjorken:1988as}
J.~D.~Bjorken, S.~Ecklund, W.~R.~Nelson, A.~Abashian, C.~Church, B.~Lu, L.~W.~Mo, T.~A.~Nunamaker and P.~Rassmann,
{\it Search for Neutral Metastable Penetrating Particles Produced in the SLAC Beam Dump}, \href{https://doi.org/10.1103/PhysRevD.38.3375}{Phys. Rev. D \textbf{38} (1988), 3375}.
\bibitem{Blumlein:2013cua}
J.~Bl\"umlein and J.~Brunner,
{\it New Exclusion Limits on Dark Gauge Forces from Proton Bremsstrahlung in Beam-Dump Data}, \href{https://doi.org/10.1016/j.physletb.2014.02.029}{Phys. Lett. B \textbf{731} (2014), 320-326}, \href{https://arxiv.org/abs/1311.3870}{arXiv:1311.3870 [hep-ph]}.
\bibitem{Payez:2014xsa} 
A.~Payez, C.~Evoli, T.~Fischer, M.~Giannotti, A.~Mirizzi and A.~Ringwald,
{\it Revisiting the SN1987A gamma-ray limit on ultralight axion-like particles}, \href{https://doi.org/10.1088/1475-7516/2015/02/006}{JCAP {\bf 1502}, no. 02, 006 (2015)}, \href{https://arxiv.org/abs/1410.3747}{arXiv:1410.3747 [astro-ph.HE]}.
\bibitem{Jaeckel:2017tud} 
J.~Jaeckel, P.~C.~Malta and J.~Redondo,
{\it Decay photons from the axionlike particles burst of type II supernovae}, \href{https://doi.org/10.1103/PhysRevD.98.055032}{Phys.\ Rev.\ D {\bf 98}, no. 5, 055032 (2018)}, \href{https://arxiv.org/abs/1702.02964}{arXiv:1702.02964 [hep-ph]}.
\end{thebibliography}
\end{document}